\documentclass[fleqn,usenatbib]{mnras}

\usepackage{newtxtext,newtxmath}

\usepackage[T1]{fontenc}
\usepackage{ae,aecompl}

\usepackage{graphicx}	
\usepackage{amsmath}	
\usepackage{amssymb}	
\usepackage{subfig}
\usepackage{booktabs}
\usepackage{afterpage}
\usepackage{longtable}
\usepackage{threeparttable}
\usepackage{threeparttablex}
\usepackage[section]{placeins}

\usepackage{mathtools}
\usepackage{footnotehyper}
\makesavenoteenv{longtable}
\DeclarePairedDelimiter{\ceil}{\lceil}{\rceil}

\newcommand{\angstrom}{\mbox{\normalfont\AA}}

\title[$\gamma$-ray deposition histories]{The $\gamma$-ray deposition histories of core-collapse supernovae}

\author[Sharon \& Kushnir]{
Amir Sharon$^{1}$\thanks{E-mail: amir.sharon@weizmann.ac.il}
and Doron Kushnir$^{1}$
\\
$^{1}$Dept.of Particle Phys. \& Astrophys., Weizmann Institute of Science, Rehovot 76100, Israel\\
}

\date{Accepted XXX. Received YYY; in original form ZZZ}

\pubyear{2020}

\begin{document}
\label{firstpage}
\pagerange{\pageref{firstpage}--\pageref{lastpage}}
\maketitle

\begin{abstract}
The $\gamma$-ray deposition history in an expanding supernova (SN) ejecta has been mostly used to constrain models for Type Ia SN. Here we expand this methodology to core-collapse SNe, including stripped envelope (SE; Type Ib/Ic/IIb) and Type IIP SNe. We construct bolometric light curves using photometry from the literature and we use the Katz integral to extract the $\gamma$-ray deposition history. We recover the tight range of $\gamma$-ray escape times, $t_0\approx30-45\,\textrm{d}$, for Type Ia SNe, and we find a new tight range $t_0\approx80-140\,\textrm{d}$, for SE SNe. Type IIP SNe are clearly separated from other SNe types with $t_0\gtrsim400\,\textrm{d}$, and there is a possible negative correlation between $t_0$ and the synthesized $^{56}$Ni mass. We find that the typical masses of the synthesized $^{56}$Ni in SE SNe are larger than those in Type IIP SNe, in agreement with the results of Kushnir. This disfavours progenitors with the same initial mass range for these explosions. We recover the observed values of $ET$, the time-weighted integrated luminosity from cooling emission, for Type IIP, and we find hints of non-zero $ET$ values in some SE SNe. We apply a simple $ \gamma$-ray radiation transfer code to calculate the $\gamma$-ray deposition histories of models from the literature, and we show that the observed histories are a powerful tool for constraining models.
\end{abstract}





\section{Introduction}
\label{sec:intro}

It is widely accepted that the light curves of both Type Ia supernovae (SNe) and, at least for late times, core-collapse SNe, including stripped envelope (SE; Type Ib/Ic/IIb) and Type IIP SNe, are powered by the decay of radionuclides synthesized in the explosion. The most important power source is the decay chain \citep{pankey1962,colgate1969early}
\begin{eqnarray}\label{eq:56 chain}
^{56}\textrm{Ni}\xrightarrow{t_{1/2}=6.07\,\textrm{d}}{}^{56}\textrm{Co}\xrightarrow{t_{1/2}=77.2\,\textrm{d}}{}^{56}\textrm{Fe}.
\end{eqnarray}
The decay products consist of $\gamma $-rays and positrons, which transfer their energy to the expanding ejecta and heat it. This thermal energy is emitted as photons ranging from the infrared (IR) to ultraviolet (UV) that produce the bolometric luminosity of the SN, $L(t)$, where $t$ is the time since explosion. The $\gamma$-ray optical depth is high shortly following the explosion, such that all the $\gamma$-ray energy is deposited within the ejecta. As the ejecta expands, the $\gamma$-ray optical depth decreases, and some $\gamma$-rays only partially deposit their energy, or just escape without interacting with the ejecta \citep{jeffery1999radioactive}. These $\gamma$-ray photons are usually not observed, with the exceptions of a handful of nearby SNe \citep{matz1988gamma,churazov2014cobalt}. As the ejecta becomes optically thin, the bolometric luminosity equals the deposited energy,
\begin{equation}\label{eq:key1}
L(t) = Q_\text{dep}(t),
\end{equation}
such that the fraction of deposited $\gamma$-ray energy has a significant impact on the shape of the bolometric light curve. The deposited energy is given by
\begin{equation}\label{eq:q_total}
Q_\text{dep}(t) = Q_\gamma(t) f_\text{dep}(t)+Q_\text{pos}(t),
\end{equation}
where $Q_\gamma(t)$ and $Q_\text{pos}(t)$ are the radioactive energy generated from $\gamma$-ray photons and the kinetic energy of positrons, respectively. The $\gamma$-ray deposition function, $f_\text{dep}(t)$, describes the fraction of the generated $\gamma$-ray energy deposited in the ejecta. We limit our analysis to times for which the kinetic energy of the positrons is deposited locally and instantaneously \citep{Colgate1980,Ruiz1998,Milne1999,kushnir2020constraints} and for which the contribution from other decay channels \citep{Seitenzahl2009} is negligible. The rest-mass energy of the positrons is emitted as $\gamma$-ray photons, which are included in $Q_\gamma(t)$. The energy generation rates of the $\gamma$-rays and of the positrons are \citep{Swartz1995,Junde1999}:
\begin{equation}\label{eq:key333}
Q_\gamma (t) = \frac{M_\text{Ni56}}{M_\odot}\left[6.54\,\textrm{e}^{-\frac{t}{8.76\text{d}}}+1.38\,\textrm{e}^{-\frac{t}{111.4\text{d}}}\right]\times 10^{43}\,\text{erg}\,\text{s}^{-1}.
\end{equation}
and
\begin{equation}\label{eq:key222}
Q_\text{pos} (t) = 4.64\frac{M_\text{Ni56}}{M_\odot}\left[\textrm{e}^{-\frac{t}{111.4\text{d}}}-\textrm{e}^{-\frac{t}{8.76\text{d}}}\right]\times 10^{41}\,\text{erg}\,\text{s}^{-1},
\end{equation}
where $M_{\rm{Ni}56}$ is the mass of $^{56}$Ni and all its radioactive parents at the time of the explosion. For a small enough $\gamma$-ray optical depth, each $\gamma$-ray photon has a small chance of colliding with matter from the ejecta (and a negligible chance for additional collisions), such that the deposition function is proportional to the column density, which scales as $ t^{-2} $, and is given by \citep{jeffery1999radioactive}:
\begin{equation}\label{eq:dep_late}
f_\text{dep}(t) = \frac{t_0^2}{t^2},\;\;\;\text{for}\;\;f_\text{dep}\ll 1,
\end{equation}
where $t_0$ is the $\gamma$-ray escape time. The time $t_c$, for which the $\gamma$-rays deposited energy equals the positrons deposited energy, $Q_{\gamma}f_{\text{dep}}=Q_{\textrm{pos}}$, is given by Equations~(\ref{eq:key333}-\ref{eq:dep_late}) as $t_c\approx5t_0$. For times when $t> t_c$, the main heating source of the ejecta is the kinetic energy loss of the positrons \citep{Arnett1979,Axelrod1980}.  

The bolometric or pseudo-bolometric light curves of Type Ia SNe have been extensively studied in the past \citep[see e.g.][]{branch1992type,milne2001late}. A common method to infer the ejecta properties of $^{56}$Ni powered SNe is the analytical model of Arnett \citep{Arnett1979,Arnett1982ApJ...253..785A}. In this model, the peak bolometric luminosity equals the instantaneous energy deposition rate at the peak time. This method is easy to implement and allows an estimate of the ejecta properties. However, the derivation of this `Arnett's rule' includes some simplifying assumptions, such as constant opacity and a uniform-heating-to-energy-density ratio, which do not hold for some ejecta profiles \citep{dessart2016inferring,khatami2019physics}. As a result, the uncertainty in the estimated properties using `Arnett's rule' is hard to quantify.  

The $ \gamma $-ray escape time $t_0$ has been measured for Type Ia SNe \citep{stritzinger2006,scalzo2014type}, by fitting the luminosity at late times under the assumption $L(t)=Q_{\text{dep}}(t)$ (\textit{the direct method}). In order to include a transition between the $\gamma$-ray optically thick and thin regions, an interpolating function was used for $f_\text{dep}(t)$ \citep{jeffery1999radioactive}: 
\begin{equation}\label{eq:dep_exp}
f_\text{dep}(t) = 1-\textrm{e}^{-t_0^2/t^2},
\end{equation}
which provides the correct expressions for both regions. However, since the assumption $L(t)=Q_{\text{dep}}(t)$ is most valid for times where $f_{\rm{dep}}\approx(t_0/t)^2$, then $Q_{\gamma}\propto M_{\rm{Ni}56}t_0^2$, and there is a degeneracy between $M_{\rm{Ni}56}$ and $t_0$. In order to remove this degeneracy, observations with $t\gtrsim t_c$, for which the positrons contribution is significant, are required. 

A different approach to measure $t_0$ is based on the Katz integral \citep{Katz2013integral}, 
\begin{equation}\label{eq:integral1}
\int_0^t Q_\text{dep}(t')t'\textrm{d}t'=\int_0^t L(t')t'\textrm{d}t'+\left.t'E(t')\right\vert_{0}^{t},
\end{equation}
where $E(t)$ is the (radiation-dominated) thermal energy of the ejecta gas. This relation is accurate for a non-relativistic expanding ejecta, independent of the assumption $L(t)=Q_{\text{dep}}(t)$. For Type Ia SNe, the term $tE(t)$ is small for both early and late times, such that for times when $L(t)=Q_{\text{dep}}(t)$, the relation
\begin{equation}\label{eq:integral_ratio1}
\frac{L(t)}{\int_0^t L(t')t'\textrm{d}t'}=\frac{Q_\text{dep}(t)}{\int_0^t Q_\text{dep}(t')t'\textrm{d}t'},
\end{equation}
holds \citep{kushnir2013head}. Using Equations~(\ref{eq:dep_exp},\ref{eq:integral_ratio1}), \cite{WygodaI2019} performed a one parameter fit to find $t_0$ values that best match the late-time light curve (\textit{the integral method}) of well observed Type Ia SNe. This method requires the measurements of the bolometric luminosity from early times until $t\sim\textrm{few}\times t_0$ (although early times are less significant because the integral is time weighted). The fit, however, is ensured to maintain energy conservation. The value of $t_0$ can be used to find $M_{\rm{Ni}56}$ by comparing the late-time light curve to the deposited energy given by Equations~(\ref{eq:q_total}-\ref{eq:key222}) with an estimated distance to the SN. \cite{WygodaI2019} assumed the interpolating function of Equation~\eqref{eq:dep_exp} and found a tight range of $\gamma$-ray escape times, $t_0\approx35-40\,\textrm{d}$, over the entire range of observed $M_{\rm{Ni}56}$. These findings were used to constrain different Type Ia models. Note that \cite{childress2015} found a much larger range of $t_0\approx13-48\,\textrm{d}$, and specifically a very low value of $t_0\approx16\,\textrm{d}$ for SN 2011fe. As we show in Section~\ref{sec:results}, their value of $t_0$ for SN 2011fe is clearly inconsistent with the observations, and we also suggest there what was the source for this discrepancy. 

\cite{Shussman2016} and \cite{Nakar2016} used Equation~\eqref{eq:integral1} to study Type IIP SNe, where the term $tE(t)$ cannot be neglected at early times. In this case, Equation~\eqref{eq:integral1} can be written (for times where $tE(t)$ is small) as
\begin{equation}\label{eq:integral2}
\int_0^t Q_\text{dep}(t')t'\textrm{d}t'=\int_0^t L(t')t'\textrm{d}t'-ET,
\end{equation}
where $ET$ is the integrated time-weighted luminosity that would be emitted if no $^{56}$Ni was produced. The procedure was to find  $M_{\rm{Ni}56}$ from the nickel tail (assuming full deposition), and to derive $ET$ from Equation~\eqref{eq:integral2}. They found $ET$ values of $\sim10^{55}\text{erg}\,\text{s}$, which is a substantial fraction of the total time-weighted luminosity.

\begin{figure}
	\includegraphics[width=\columnwidth]{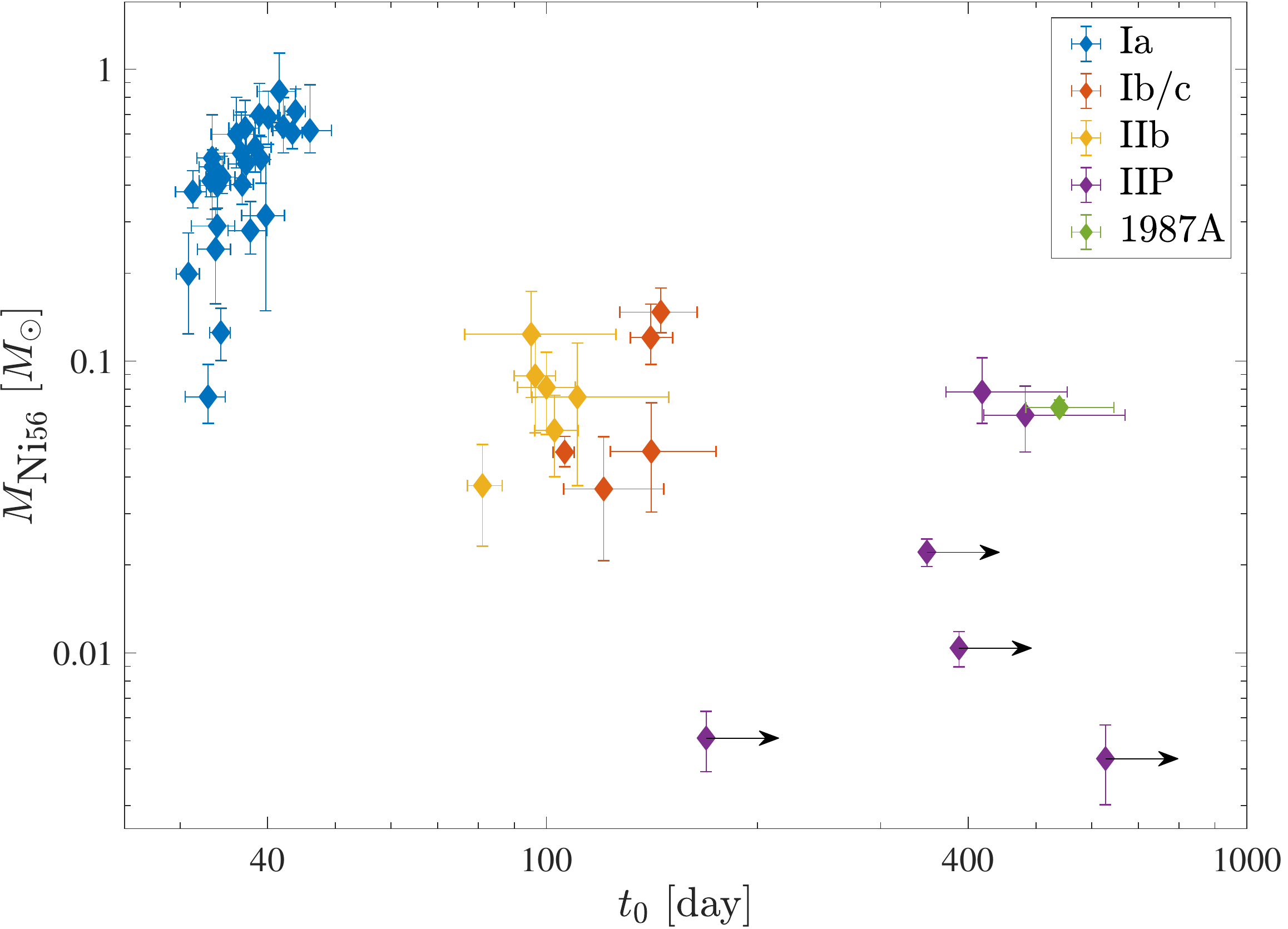}
	\caption{$M_{\rm{Ni}56}-t_0$ distribution of the SNe sample. Different types of SNe, marked by different colours, are clustered around distinct typical $ t_0 $ values, while the synthesized $^{56} $Ni amount overlaps among the different types. The tight range of $\gamma$-ray escape times, $t_0\approx30-45\,\textrm{d}$, of Type Ia SNe (blue) is recovered, and we find a new tight range, $t_0\approx80-140\,\textrm{d}$, for SE SNe (red: Type Ib/c, orange: Type IIb). Type IIP SNe (purple, SN 1987A in green) are clearly separated from other SNe types with $t_0\gtrsim400\,\textrm{d}$, and there is a possible negative correlation between $t_0$ and $M_{\rm{Ni}56}$. The error-bars correspond to $68$ per cent confidence level around the median values (see the text for details).}
	\label{fig:t0Ni}
\end{figure}

Bolometric light curves of SE SNe have been constructed and studied in the past \citep[e.g., ][]{lyman2013bolometric,wheeler2015analysis,lyman2016bolometric,prentice2018investigating,meza2020strippedenvelope}. The values of $t_0$ were estimated with two different methods by \cite{wheeler2015analysis}. They either used Arnett's rule with some extensions, or a fit to the late-time light curve, by using the exponential interpolating function, Equation~\eqref{eq:dep_exp}, assuming positrons are not fully trapped \citep[which is unrealistic, see][]{kushnir2020constraints}. Using the first method, they found $t_0\approx50-90\,\textrm{d}$, while the second method yielded $t_0\approx100-300\,\textrm{d}$ (with some exceptions).

In this paper, we expand the use of the Katz integral to study the $\gamma$-ray deposition history of core-collapse SNe. The construction of bolometric light curves for a sample of SNe is described in Section~\ref{sec:bolometric}. We find that throughout most of the available measurements, the ejecta is not sufficiently optically thin for the approximation of Equation~\eqref{eq:dep_late} to hold. As a result, the interpolating function has a significant effect on the results, which forces us to introduce an additional "smoothness" parameter, $n$, to the interpolating function of Equation~\eqref{eq:dep_exp}. We therefore fit in Section~\ref{sec:fit} for four ejecta parameters: $t_0$, $n$, $M_{\rm{Ni}56}$ and $ET$. We use a sample of Type Ia SNe as a control sample, and we are able to recover the results of \citet{WygodaI2019} with our method. The use of the additional parameter $n$ allows us to reduce a $\sim10$ per cent systematic error that was introduced by the interpolation function of Equation~\eqref{eq:dep_exp} to the $t_0$ values of Type Ia SNe. 

Our results are presented in Section~\ref{sec:results}. The main result is presented in Figure~\ref{fig:t0Ni}, where the distribution of $t_0$ as a function of  $M_{\rm{Ni}56}$ for various types of SNe is shown. As can be seen in the figure, we recover the tight range of $\gamma$-ray escape times, $t_0\approx30-45\,\textrm{d}$, for Type Ia SNe, and we find a new tight range, $t_0\approx80-140\,\textrm{d}$, for SE SNe. Type IIP SNe are clearly separated from other SNe types with $t_0\gtrsim400\,\textrm{d}$, and there is a possible negative correlation between $t_0$ and $M_{\rm{Ni}56}$.

\citet{Kushnir2015progenitors} used a compilation from the literature to show that the typical $M_{\rm{Ni}56}$ in SE SNe are larger than those in Type IIP SNe \footnote{This result was later reproduced by \citet{Anderson2019A&A...628A...7A}, using similar methods.}. As can be seen in Figure~\ref{fig:t0Ni}, the result of \citet{Kushnir2015progenitors} still holds for our (smaller) sample, where the $M_{\rm{Ni}56}$ are determined in a consistent way with systematic uncertainties under control. \citet{meza2020strippedenvelope} also reached the same conclusion, but their methods suffers from systematic effects that are hard to quantify (see discussion in Section~\ref{sec:results}). The larger typical values of $M_{\rm{Ni}56}$ in SE SNe, as compared to $M_{\rm{Ni}56}$ in Type IIP SNe, disfavours progenitors with the same initial mass range for these explosions \citep[see detailed discussion in][]{Kushnir2015progenitors}. 

In section \ref{sec:simulations}, we apply a simple $ \gamma$-ray radiation transfer code to calculate the $\gamma$-ray deposition histories of models from the literature, and we show that the observed histories are a powerful tool to constrain models. We discuss our results and conclude in Section~\ref{sec:discussion}.


\section{Bolometric light curves sample}
\label{sec:bolometric}

Our analysis requires the acquisition of bolometric light curves for several types of well-observed SNe. These were constructed using published photometry and estimated reddening values and distances \citep[except for SN 1987A, for which we used the bolometric light curve of][]{suntzeff1990bolometric}. The SN sample is described in Section~\ref{sec:sample}, and the bolometric light curve construction method is described in Section~\ref{sec:construction}.

\subsection{SNe sample}
\label{sec:sample}

Our sample of well-observed SNe is composed of 27 Type Ia SNe, 11 SE SNe (5 Type Ib/c, 6 Type IIb), 7 Type IIP SNe and SN 1987A.
\begin{itemize}
	\item \textbf{Type Ia SNe} --Photometry was taken from the Open Supernova Catalogue \citep{guillochon2017open}, the CSP data release 3 \citep{krisciunas2017carnegie,csphubbleApJ...869...56B}, the Berkeley supernova Ia program \citep{berkeley2012MNRAS.425.1789S}, the Lick Observatory Supernova Search Follow-up Photometry Program \citep{lick2010ApJS..190..418G}, the Sternberg Astronomical Institute Supernova Light Curve Catalogue \citep{tsvetkov2004sai}, the Harvard-Smithsonian Center for Astrophysics (CfA) \citep{cfaApJ...700..331H,cfair2ApJS..220....9F}, and the Swift Optical/Ultraviolet Supernova Archive \citep{sousa2014Ap&SS.354...89B}. Additional photometry was taken from the individual SN studies of SN 2003du \citep{2003duA&A...469..645S}, SN 2004eo \citep{2004eoMNRAS.377.1531P}, SN 2005cf \citep{2005cfMNRAS.376.1301P}, SN 2011fe \citep{Matheson2012ApJ...754...19M,Munari2013NewA...20...30M,Tsvetkov2013CoSka..43...94T,Firth2015MNRAS.446.3895F,Graham2017MNRAS.472.3437G}, SN 2012fr \citep{2012fr2AJ....148....1Z,2012fr1ApJ...859...24C}, and SN 2013dy \citep{2013dy1.452.4307P,2013dy2....151..125Z}.
	\item \textbf{SE SNe} --The sample consists of three SNe from the CSP-I SE sample photometry \citep{carData} with the distance and reddening estimation taken from \citep{carAnalysis,carMethods}, SN 1993J \citep{1993JPhoto,1993J_IR}\footnote{Since photometric errors are not given, we estimate $0.1$ mag error for all measurements.}, SN 2002ap \citep{yoshii2003optical,tomita2006optical}, SN 2007gr \citep{hunter2009extensive,Chen_2014}, SN 2008ax \citep{pastorello2008type,taubenberger2011he,bianco2014multi}, SN 2009jf \citep{valenti2011sn}, SN 2010as \citep{folatelli2014supernova}, SN 2011dh \citep{ergon2015type}\footnote{Data up to 150 days from the explosion was used, see Appendix~\ref{app:2011dh} for details.}, SN 2016coi \citep{Prentice2018MNRAS.478.4162P,Terreran2019ApJ...883..147T}.
	\item \textbf{Type IIP SNe} --The sample consists of SN 2004et \citep{sahu2006photometric,maguire2010optical}, SN 2005cs \citep{pastorello2009sn}, SN 2009N \citep{takats2013sn}, SN 2009md \citep{fraser2011sn}, SN 2012A \citep{tomasella2013comparison}, SN 2013ej \citep{yuan2016450}\footnote{This SN is generally regarded as Type IIL, but we do not make this distinction here.}, and SN 2017eaw \citep{rho2018near,szalai2019type}. Note that SN 2004et and SN 2017eaw occurred at the same host galaxy, so the distance estimate for both SNe was taken from \cite{szalai2019type}.
	\item \textbf{SN 1987A} --The bolometric light curve was taken from \cite{suntzeff1990bolometric}.
\end{itemize}

\subsection{Constructing bolometric light curves}
\label{sec:construction}

\begin{figure}
	\includegraphics[width=\columnwidth]{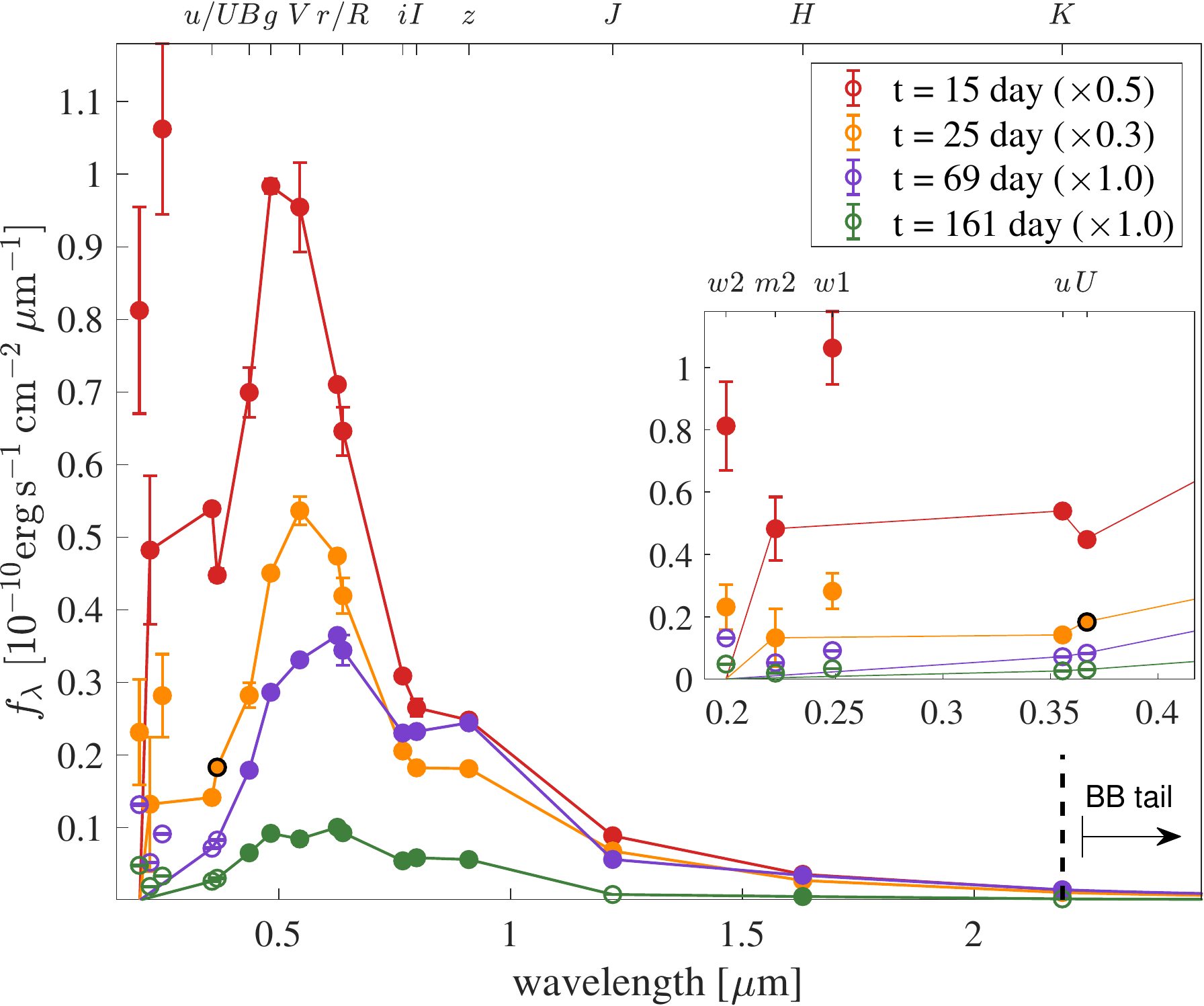}
	\caption{The obtained SED of SN 2016coi at several phases after explosion. The filled circles indicate the de-reddened flux intensity at the effective wavelength of each band, and the open circles correspond to the extrapolated values of the UV filters and the $J$ and $K$ bands at late times. The filled circles with black edges mark values interpolated with an auxiliary band. The solid lines represent the flux that is integrated for the calculation of the luminosity at each epoch. For clarity, the SED at $t=15(25)\,\textrm{d}$ is scaled by $0.5(0.3)$. A magnification of the UV region shows the large flux excess of the \textit{Swift} uvw1 and uvw2 filters, which are not taken into account for the construction of the SED. The BB region of the SED, covering wavelengths higher than the $K$ band, is indicated with a dashed black line, and includes a small fraction of the SED at all epochs.}
	\label{fig:SED}
\end{figure}

In this section, we describe the method we use to construct bolometric light curves from the photometric data. Our sample only includes SNe with photometry that covers the near-UV ($U$ or $u'$ band), optical, and near-infrared (IR) wavelengths for some phases of the SN. We require IR observations, since a significant fraction of the total flux of SE SNe is emitted in the IR wavelengths \citep{lyman2013bolometric,prentice2016bolometric}, reaching up to $\sim40$ per cent at some phases. We find a similar IR fraction for our Type IIP sample, and a slightly lower fraction, $\sim30$ per cent, for Type Ia SNe at some phases. The IR fraction has some non-trivial time evolution, and omitting the IR contribution changes our results significantly. Near UV measurements from \textit{Swift} UVOT were also available sometimes. However, we have omitted data from the uvw1 and uvw2 bands, as they were inconsistent with the flux observed at neighbouring filters (see Figure~\ref{fig:SED}). The motivation for this omission is the red leak \citep[see][for details]{brown2016interpreting}. We have also omitted the data from the uvm2 band for phases later than $30\,\textrm{d}$ since the explosion, because the UV contribution in these phases is quite small. 

The bolometric light curve construction procedure begins with identification of missing data. For each considered band, we find all pairs of adjacent measurements with a magnitude difference larger than $0.5$ magnitude and a time difference larger than $15$ d. The missing data between each pair is estimated by using an adjacent auxiliary band, provided it has more complete data during these times. This is done by calculating the color between the two bands at the beginning and at the end of the considered time range, and linearly interpolating the color at the intermediate phases. This procedure is important in cases where the light curve changes fast, for example, at the transition from the plateau to the nickel tail of Type IIP SNe.

We next extrapolate for bands that lack observations at the beginning and at the end the light curve. This is done by assuming a constant color between the band with the missing data and the nearest band with data over the relevant time. This method introduces large uncertainties and should be done carefully, but since IR measurements do not always cover the time range of the optical measurements, extrapolating their magnitudes is preferable over omitting them. The errors introduced by this method increase as the time from the last measurement increases. 

The light curves are then corrected for extinction using the methods of \cite{fitzpatrick1999correcting}, with the reddening values taken from the literature. The extinction-corrected light curves are then converted to flux densities at the effective wavelength of each band, creating a spectral energy distribution (SED) for each phase. The flux densities between the filters are linearly interpolated. To account for the missing UV contribution to the SED, we linearly extrapolate to zero flux at $2000\,\angstrom$. The flux density for wavelengths longer than the longest effective wavelength band are estimated with a blackbody (BB) fit, calculated using the flux densities of bands with an effective wavelength higher than $5000\,\angstrom$. Since all SNe in our sample contain IR measurements, the BB Rayleigh Jeans tail does not contribute more than a few per cents to the total flux (see also Figure~\ref{fig:SED}). We do not consider epochs with three or less measured bands (without extrapolation). For example, if following some epoch, only the $B$, $V$ and $R$ bands were measured (and the rest of the bands were extrapolated), then these late epochs will not be included in the bolometric light curve. The obtained SED for a few phases of the Type Ic SN 2016coi are presented in Figure~\ref{fig:SED}. The flux densities of the measured bands are marked with filled circles (black edges indicate interpolated values from an auxiliary band), and extrapolated values are marked with open circles. The large flux excess of the uvw1 and uvw2 bands, and the small contribution of the BB tail, are seen in the figure.

Finally, the total flux is calculated by trapezoidal integration of the SED and is converted to the luminosity using the luminosity distance. We also correct the time to the source frame, although this is a small correction for the low redshifts of our SNe sample ($z<0.03$ for almost all SNe). 

The uncertainty of the bolometric luminosity includes the given observed photometry (statistical) errors, which are propagated in the standard manner, and several sources of systematic errors. The systematic error includes the error due to the UV extrapolation, which we estimate by fixing zero flux at $\pm500\,\angstrom $ away from the default $2000\,\angstrom$, and the error of the BB extrapolation, which we estimate to be $25$ per cent of the flux obtained in the BB regime. An additional $5$ per cent of the total luminosity is added (as quadrature to the total error budget), accounting for additional unknown systematic errors. We (conservatively) treat this systematic error estimate as one standard deviation that is added to the (photometry) statistical uncertainty. As a result of this conservative error estimate, the total error is dominated by the significant systematic error uncertainty (see~Figure \ref{fig:fits}), and statistical estimates for the model goodness-of-fit are not very informative (e.g. the $\chi^{2}$ of the best-fitting models is significantly smaller than the number of degrees of freedom). Nevertheless, the uncertainty of the model parameters reflects the assumed systematic errors, and provides reliable estimate for possible models that can describe the observations. As we show in the following sections, this procedure allows us to draw strong conclusions, despite the large systematic errors. The treatment of the distance and extinction uncertainties is described in the next section.

Files containing the observed magnitudes, the processed magnitudes (after interpolation, extrapolation and de-reddening), and the bolometric luminosity are included in the supplementary materials. The processed magnitudes file also contains the extinction coefficients, the distance and the explosion epoch that we use. 


\section{Deducing four ejecta properties from the bolometric light curve}
\label{sec:fit}

\begin{figure}
	\includegraphics[width=\columnwidth]{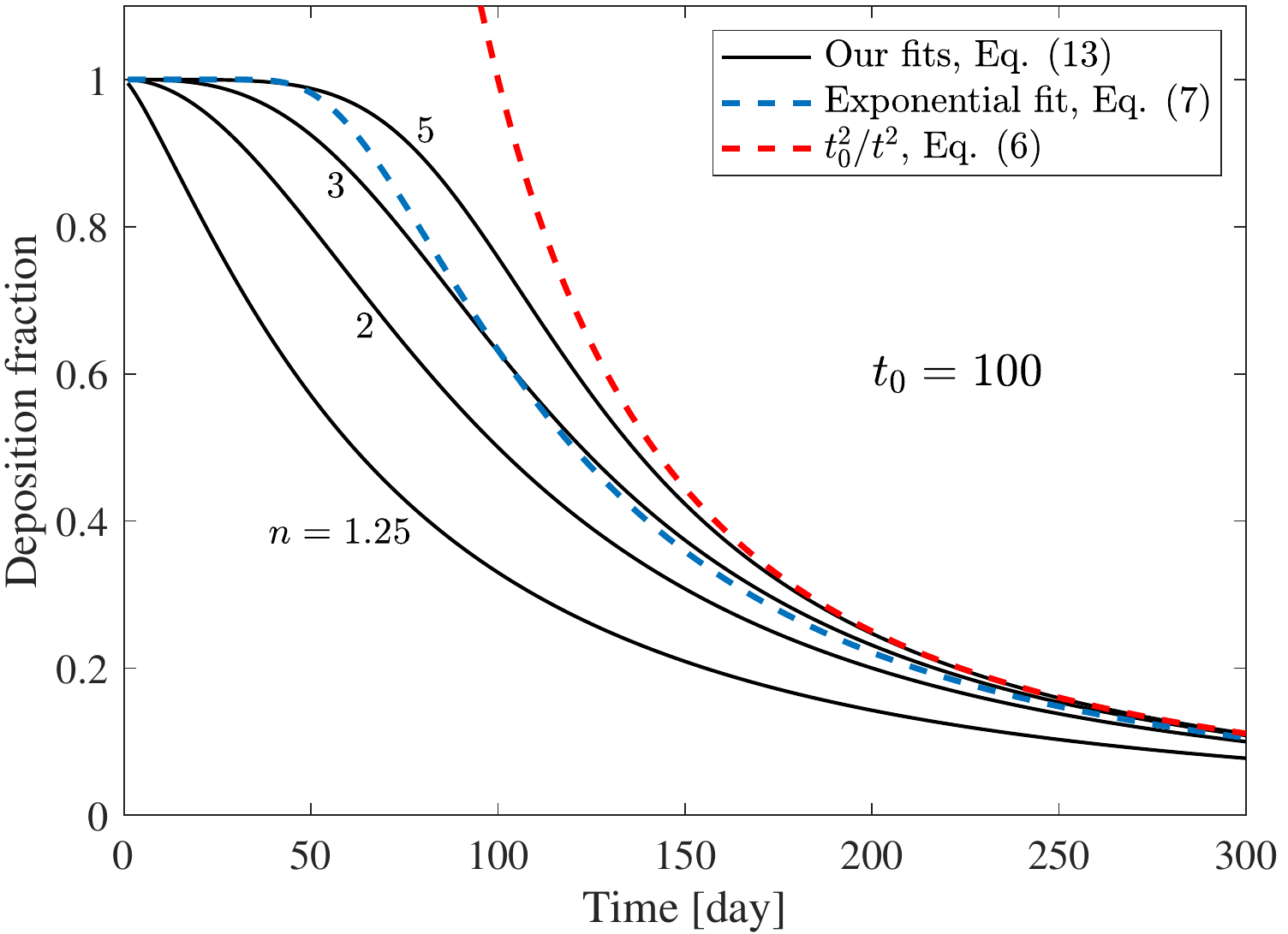}
	\caption{Deposition fraction as a function of time for $t_0=100\,\textrm{d}$. The black solid lines represent the functions used in this analysis, Equation~\eqref{eq:deposition}, for $n=1.25,2,3,5$ (bottom to top). The exponential function, Equation~\eqref{eq:dep_exp}, is shown in the blue dashed line, and the late time behavior, Equation~\eqref{eq:dep_late} is shown in the red dashed line. The parameter $n$ controls the sharpness of the transition between optically thin and thick regimes. The larger $n$ is, the sharper the transition is, and in the limit $n\rightarrow \infty$ the deposition fraction turns from being unity to following the behavior of Equation~\eqref{eq:dep_late} at $t = t_0$.} 
	\label{fig:deposition}
\end{figure}

In this section, we expand the methods of \cite{WygodaI2019} and \cite{Nakar2016} to extract various ejecta properties from the bolometric light curves. The case of SE SNe is more challenging to treat than the Type Ia SNe case for a number of reasons. The first is that $ET$, while not being as significant as in Type IIP SNe, is found to be non-negligible for most Type Ib/c SNe and for some Type IIb SNe. In order to include this parameter in our analysis, we use Equation~\eqref{eq:integral2} to write for times where $L=Q_{\text{dep}}$:
\begin{equation}\label{eq:integral_ratio3}
\frac{L(t)}{LT(t)}=\frac{\widetilde{Q}_\text{dep}(t)}{\widetilde{QT}(t)+\widetilde{ET}},
\end{equation}
where
\begin{equation}\label{eq:definitions1}
LT(t) \equiv \int_0^t L(t')t'\textrm{d}t' ,\;\;\;QT(t) \equiv \int_0^t Q_\text{dep}(t')t'\textrm{d}t', 
\end{equation}
and tilde stands for $^{56}$Ni normalized quantities (for example, $\widetilde{ET}=ET/M_{\text{Ni}56}$). The second difficulty is that in most of the available measurements of SE SNe, the ejecta is not sufficiently optically thin to $\gamma$-rays for the approximation of Equation~\eqref{eq:dep_late} to hold. As a result, the shape of the interpolating function has a significant effect on the derived ejecta properties, forcing us to use a more versatile $\gamma$-ray deposition function than Equation~\eqref{eq:dep_exp}, which we found to be inappropriate both from fitting the data and from $\gamma$-ray transfer Monte Carlo simulations (see Section~\ref{sec:simulations}). We therefore introduce an additional parameter $n$ to the interpolating function and replaced it with:
\begin{equation}\label{eq:deposition}
f_\text{dep}(t)=\frac{1}{\left(1+\left(t/t_0\right)^n\right)^{\frac{2}{n}}}.
\end{equation} 
The parameter $n$ controls the sharpness of the transition between optically thin and thick regimes. The larger $n$ is, the sharper the transition is, and in the limit $n\rightarrow \infty$, the deposition fraction changes instantaneously from being unity to following the behavior of Equation~\eqref{eq:dep_late} at $t = t_0$. Deposition functions for a few values of $n$ and for a typical $t_0=100\,\textrm{d}$ value of SE SNe are shown in Figure~\ref{fig:deposition}, along with the exponential interpolating function, Equation~\eqref{eq:dep_exp}. As can be seen in the figure, there is a large effect on the shape of the interpolating function up to $t\lesssim200\,\textrm{d}$ (where most of the observed data is given). In addition, the effect of the interpolating function is enhanced, since deviations accumulate in the integrated luminosity $LT$. 

The integral method is a three-parameter fit to Equation~\eqref{eq:integral_ratio3}, in which we find the values of the triplet $\{ t_0,\widetilde{ET},n\}$ that minimize the expression
\begin{equation}\label{eq:likelihood}
\frac{N_\text{bins}}{N_\text{obs}}\sum_{t_i\in t_{L=Q}} \left[\left(\frac{L(t_i)}{LT(t_i)}-\frac{\widetilde{Q}_{\text{dep}}(t_i)}{\widetilde{QT}(t_i)+\widetilde{ET}}\right)\frac{LT(t_i)}{L_{err}(t_i)}\right]^2.
\end{equation}
The time range $t_{L=Q}$ accounts for the times where the assumption $L=Q_{\text{dep}}$ is valid (see below). The second factor in the squared parentheses accounts for the measurements errors of the ratio $L/LT$, assumed to be Gaussian, where we have neglected the error in $LT$. This is because $LT$ is the sum of $N_i$ independent measurements until $t_i$ (with time weighting), so it is roughly suppressed by a factor $\sqrt{N_i}$. The $N_\text{bins}/N_\text{obs}$ factor is relevant for the estimation of the parameters uncertainty, which is discussed below, and does not affect the best-fitting values. This three-parameter fit is equivalent to maximizing the likelihood of the triplet $\{ t_0,\widetilde{ET},n\}$ to satisfy Equation~\eqref{eq:integral_ratio3}, given the measurements errors. Note that this procedure is independent of $M_{\rm{Ni}56}$ and of the distance to the SN. $M_{\rm{Ni}56}$ is found by comparing the luminosity in the fitted range to the deposited radioactive energy. 

\begin{figure}
	\includegraphics[width=\columnwidth]{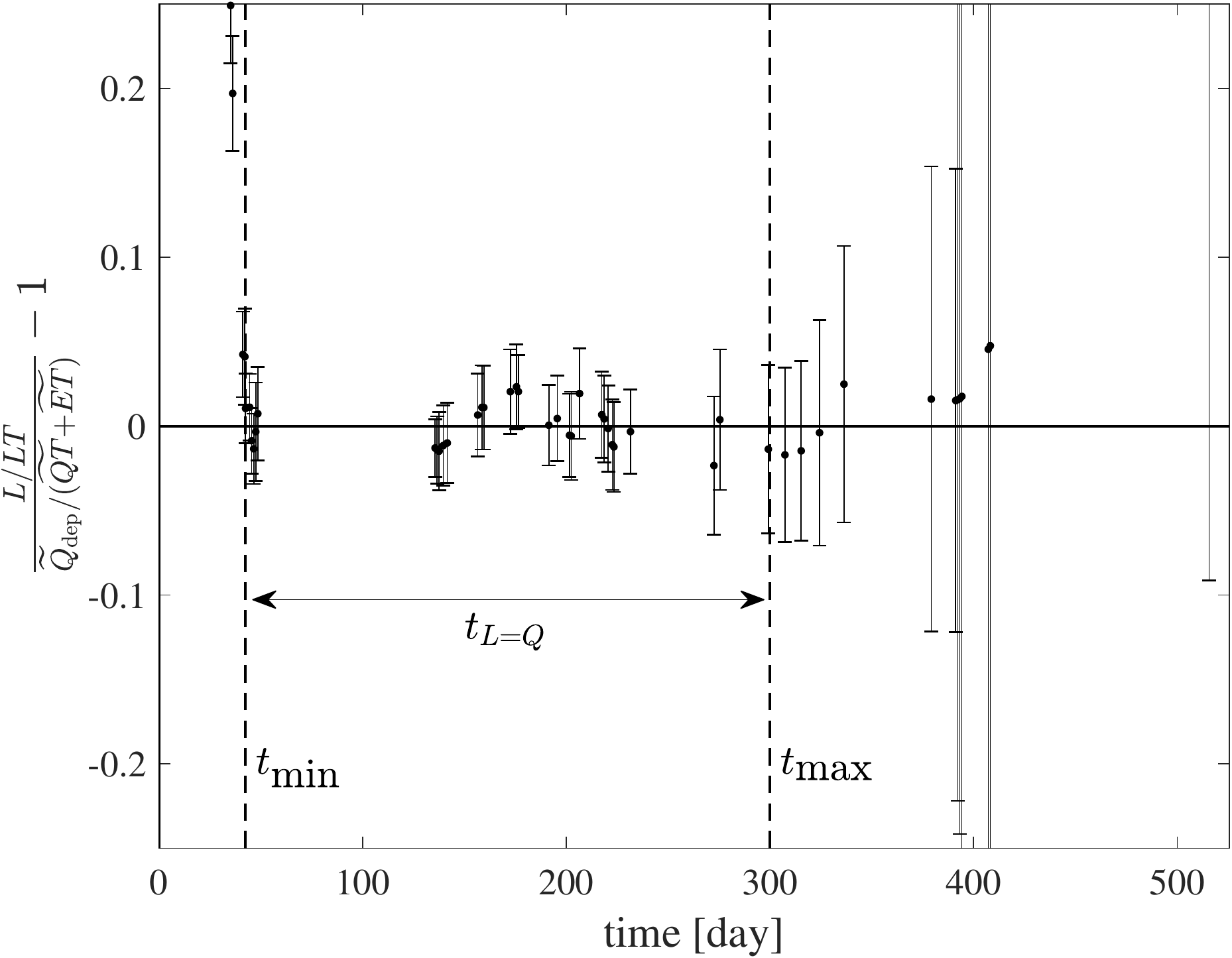}
	\caption{Relative error of the fit for SN 2002ap as a function of phase, with the error-bars being the statistical errors (the total errors are not presented for clarity). The vertical lines indicate the time range $ t_{\min}=42.5\,\textrm{d}\le t_{L=Q}\le t_{\max}=300\,\textrm{d}$, over which the fit is being performed. The relative errors are distributed around zero during the calibrated $t_{L=Q}$ (this remains true until $\sim400\,\textrm{d}$). As a result, increasing $t_\text{min}$ would not change significantly the best-fitting parameters, and the relative errors during the original $t_{L=Q}$ would still be distributed around zero.} 
	\label{fig:fitting}
\end{figure}

We determine the time range $t_\text{min}\le t_{L=Q}\le t_\text{max}$, by fixing $t_\text{max}$ for each SN type and by determining $t_\text{min}$ self-consistently for each SN. A good choice for $t_{\max}$ is $t_\text{max}\sim\textrm{few}\times t_0$, since it allows us to determine $t_0$ with a reasonable accuracy, while observations at later time are usually at lower quality and includes a significant contribution from positrons (recall that following $t_c\approx5t_0$ the energy deposition is dominated by positrons, such that it is more difficult to extract $t_0$). We fix $t_\text{max}=120\,\textrm{d}$ for Type Ia SNe, which is also roughly the median value of the last observational phase of our sample, and $t_\text{max}=300\,\textrm{d}$ for SE SNe. The available observations of Type IIP SNe force us to use $t_{\max}\sim500\,\textrm{d}$ (ideally we would like to use $t_{\max}\sim1500\,\textrm{d}$), with slight variations for SN 2017eaw and 1987A. In cases that the phase of the last available data is smaller than $t_\text{max}$, we fit until the last available phase. For most Type Ia and SE SNe that have measurements beyond $t_\text{max}$, our fits provide good matches for $t>t_{\max}$, in some cases until the last available observation (see Appendix~\ref{app:deposition plots}). The value of $t_\text{min}$ is harder to determine, since the data does not provide a clear indication for the times in which the assumption $L=Q_{\text{dep}}$ is valid. Specifically, too small value for $t_\text{min}$, with phases in which the assumption $L=Q_{\text{dep}}$ is not valid, would significantly bias the obtained fit parameters. In order to find self-consistently $t_\text{min}$, we inspect the relative error of the fit, given by the deviation of $(L/LT)/(\widetilde{Q}_{\text{dep}}/(\widetilde{QT}+\widetilde{ET}))$ from unity, during $t_{L=Q}$. For a self-consistently $t_\text{min}$, the relative errors should be distributed around zero during $t_{L=Q}$. In this case, increasing $t_\text{min}$ would not change the best-fitting parameters significantly, and the relative errors during the original $t_{L=Q}$ would still be distributed around zero. We demonstrate this procedure in Figure~\ref{fig:fitting}, where the relative errors of the SN 2002ap fit are shown. The time range $t_{L=Q}$ of each SN is given in the bolometric luminosity file of the supplementary material.

An alternative method (the direct method) for deriving the four parameters, is to directly compare $L(t)$ during $t\in t_{L=Q}$ to the radioactive energy deposition rate. This method uses only the instantaneous luminosity, so $ET$ is not a part of the calculation, but it requires $M_{\rm{Ni}56}$, such that the number of parameters remains three: $t_0,n$, and $M_{\rm{Ni}56}$. $ET$ can be calculated by taking the difference $LT-QT$ at late times. As explained in Section~\ref{sec:intro}, for this method, observations with $t\gtrsim t_c$ are required in order to remove the degeneracy between $M_{\rm{Ni}56}$ and $t_0$\footnote{The transition between the $\gamma$-ray optically thick and thin regimes also breaks this degeneracy. However, one has to trust that the interpolating function accurately describes the deposition during this transition.}. We find that, particularly for Type Ia SNe, the direct method is less stable than the integral method since it is more sensitive to the estimated time of explosion and to the values of $t_{\min}$ and $t_{\max}$. It also has the disadvantage that the fit may result in a negative $ET$, which is not physical. However, it is still useful to apply both methods for validation and cross-checking. For most of the SNe in our sample, the variations of $M_{\rm{Ni}56}$ and $t_0$ between the two methods are smaller than $10$ per cent. For several Type Ia SNe, because of the degeneracy between $ t_0 $ and $ M_{\rm{Ni}56} $, the direct method finds considerably lower $t_0$ values than the integral method with large negative values of $ET$. We conclude that the values obtained by the integral method are more reliable.

Another advantage of the integral method for the analysis of Type Ia SNe, is that the $ET$ parameter can be set to zero \citep[as was done in][]{WygodaI2019}, as expected for a white dwarf progenitor. To justify this simplification, we compare the results for Type Ia SNe with $ET$ as a free positive parameter and without it. The results for most SNe remain the same, where the calibrated $ET$ is negligibly small. For a few SNe, the $t_0$ values change by up to $7$ per cent and gain an $ET$ value of up $3.5$ per cent of the total time-weighted luminosity $LT$ at infinity. This exercise allows us to identify suspicious light curves, where more detailed inspection is required.

The missing bolometric luminosity from explosion until the first observed epoch, have to be estimated for the calculation of $LT$. \citet{WygodaI2019} suggested to interpolate linearly between the first epoch and $L=0$ at $t=0$. Since some Type Ia and SE SNe in our sample lack observation from very early times, we use instead a template of SN 2011fe (observed from hours since explosion) that is scaled to the first observed epoch. We found that the difference of the inferred parameters between the two methods is negligible. For Type IIP SNe, we assumed that the luminosity is constant from $t=0$ up to the first epoch, and equals to the luminosity of the first epoch. Not extending the bolometric luminosity to $t=0$ usually has a negligible effect, except for a few Type Ia SNe with relatively late first epoch measurements, and for a small influence on the ET value of core-collapse SNe. 

\begin{figure}
	\includegraphics[width=\columnwidth]{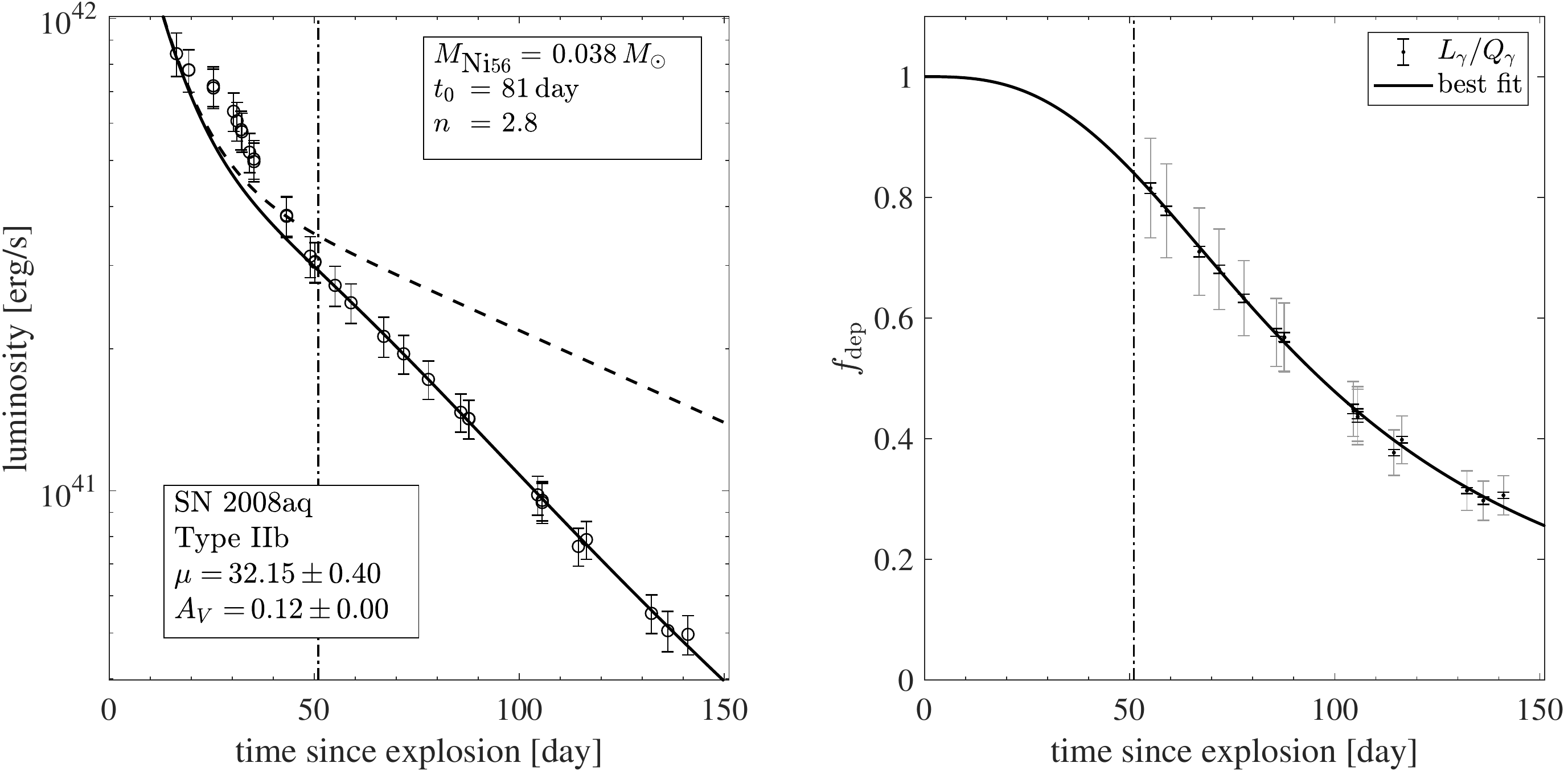}
	\includegraphics[width=\columnwidth]{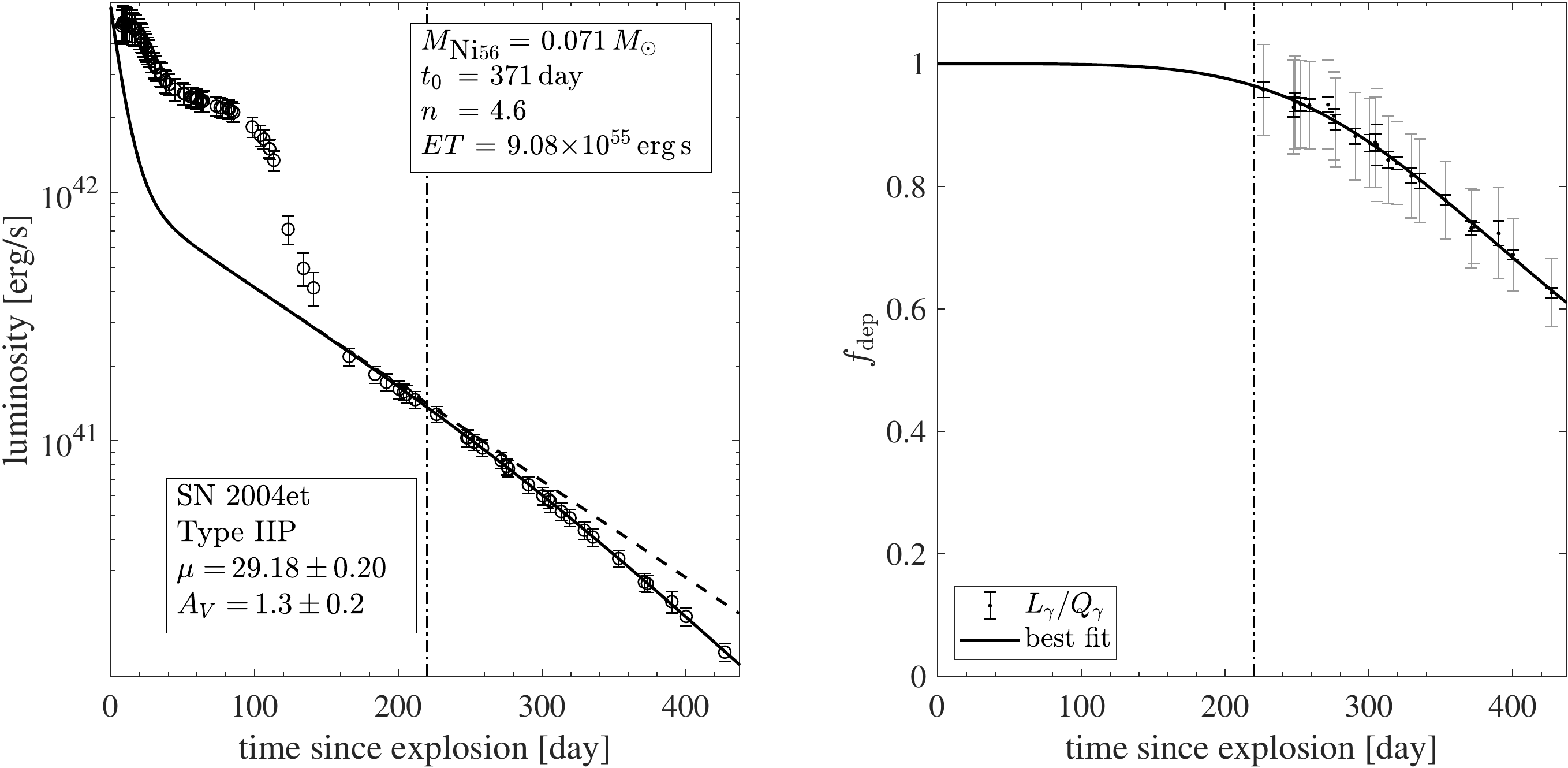}
	\includegraphics[width=\columnwidth]{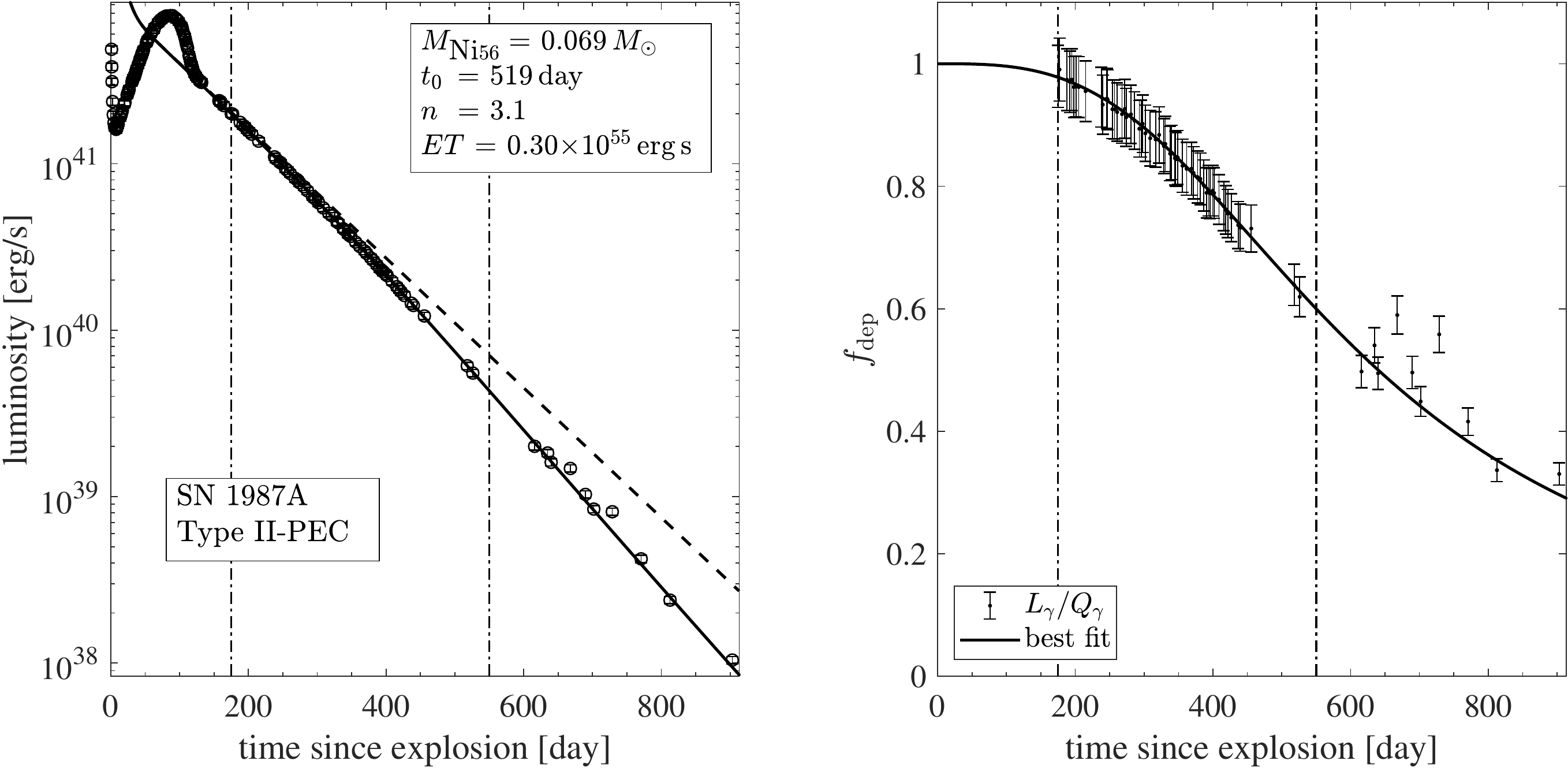}
	\caption{The best-fitting results of the integral method are demonstrated for SN 2008aq (Type IIb), SN 2004et (Type IIP), and SN 1987A (Type II-pec). The rest of the SNe in our sample are presented in Appendix~\ref{app:deposition plots}. The median values and 68\% confidence level of the parameters are presented in Table~\ref{tab:results}. left-hand side panels: the observed bolometric light curves are compared to the best-fitting models (with the parameters given in the boxes, solid lines) and to the radioactive energy generation rates (same as assuming $f_{\text{dep}}=1$, dashed lines). The distance and extinction estimates are given in the boxes as well. The errors represent the total errors (statistical and systematic). The bolometric light curve of SN 1987A was taken from the literature, so the errors (not shown) were set to 5\% of the total luminosity, and the distance and extinction were not used. right-hand side panels: The deposition functions, $f_{\text{dep}}$, that correspond to the best-fitting models (solid lines) are compared to the ratio $(L-Q_{\text{pos}})/Q_{\gamma}$. This ratio corresponds to $L_{\gamma}/Q_{\gamma}$ for $t\in t_{L=Q}$, where we use the observed $L$ and the derived $Q_{\text{pos}},Q_{\gamma}$. The total errors (statistical and systematic) are indicated by grey bars, while the (photometric) statistical errors are indicated by black bars. The treatment of the large systematic errors is discussed in the text. In both panels, the epochs of $t_\text{min}$ and $t_\text{max}$ (if different from the last phase) are indicated by vertical dashed-dotted lines.}
	\label{fig:fits}
\end{figure}

The best-fitting results obtained with the integral method are demonstrated in Figure~\ref{fig:fits} for SN 2008aq (Type IIb), SN 2004et (Type IIP), and SN 1987A (Type II-pec). The entire sample is presented in Appendix~\ref{app:deposition plots}. The observed bolometric light curves are compared to the best-fitting models (solid lines) and to the radioactive energy generation rates (same as assuming $f_{\text{dep}}=1$, dashed lines) in the left-hand side panels. In the right-hand side panels, the deposition functions $f_{\text{dep}}$ that correspond to the best-fitting models (solid lines) are compared to the ratio $(L-Q_{\text{pos}})/Q_{\gamma}$. This ratio corresponds to $L_{\gamma}/Q_{\gamma}$ for $t\in t_{L=Q}$, where we use the observed $L$ and the derived $Q_{\text{pos}}$ and $Q_{\gamma}$. The vertical dashed-dotted lines indicate the time range $t_{L=Q}$. As can be seen in the figure, the ejecta is not at the optically thin regime for almost the entire time-span of the observations, and the shape of the interpolating function has a large effect. Additionally, for both SN 2004et (Type IIP) and SN 1987A, the $\gamma$-rays are not fully trapped at late times.

As discussed in Section~\ref{sec:bolometric}, the total error is dominated by the significant systematic error uncertainty, and statistical estimates for the model goodness-of-fit are not very informative (e.g. the $\chi^{2}$ of the best-fitting models is significantly smaller than the number of degrees of freedom). We estimate in the following section the uncertainty of the model parameters, which will reflect the assumed systematic errors. This estimate provides reliable estimate for possible models that can describe the observations. 

\subsection{The uncertainty of the parameters}
\label{sec:fit uncertainty}

The uncertainty of the observed bolometric light curve introduces errors to the fitted parameters. We estimate the median and the 68 per cent confidence levels by performing a Markov Chain Monte Carlo (MCMC) algorithm using the \textsc{mcmcstat} \textsc{matlab} package\footnote{https://mjlaine.github.io/mcmcstat/}, where the likelihood function is Equation~\eqref{eq:likelihood} and the priors are uniformly distributed over reasonable domains. The priors for $t_0$ and $\widetilde{ET}$ are $[0,\infty)$. The prior for $n$ is [1,5), since below $n=1$ the derivative of Equation \eqref{eq:deposition} diverges at $ t\rightarrow0$ and creating a profile with $n\ge5$ using our $\gamma$-ray transfer Monte Carlo code required extremely centred $ ^{56} $Ni distribution (and for all SNe in our sample we find $n<5$).

The $N_\text{bins}/N_\text{obs}$ factor in the likelihood function, Equation~\eqref{eq:likelihood}, represents the ratio between the number of independent time bins, $N_\text{bins}$, to the number of observations, $N_\text{obs}$, and is always $\le1$. We introduce this factor since the uncertainty is dominated by systematic errors, such that one cannot assume that the observations are statistically independent. Instead, since the typical systematic error is $\sim10$ per cent, we estimate $N_\text{bins}$ as the number of times that $Q_\text{dep}$ changes by $10$ per cent over the time range of each SN. For Type Ia SNe, with the longest duration of $40\,\textrm{d}\le t_{L=Q}\le 120\,\textrm{d}$ and typical $t_0$ values, we find at most $N_\text{bins,max}\approx24$, such that the average duration of each independent bin is $\approx3.5\,\textrm{d}$. For SE SNe, with the longest duration of $50\,\textrm{d}\le t_{L=Q}\le 300\,\textrm{d}$, we find at most $N_\text{bins,max}\approx40$, and for Type IIP, with the longest duration of $130\,\textrm{d}\le t_{L=Q}\le 500\,\textrm{d}$, we find $N_\text{bins,max}\approx43$. We determine $N_\text{bins}$ for each SN as the ceiling of the maximal number of bins of that type multiplied with the ratio of $t_{L=Q}$ to the maximal duration. For example, the time range of the Type Ib/c SN 2007gr is $ 87\le t_{L=Q}\le183 $, so we find
\begin{equation}\label{key}
N_\text{bins}= \ceil[\Big]{\frac{183-87}{300-50}\times 40} = 16.
\end{equation}

As explained above, the values of the triplet $\{ t_0,\widetilde{ET},n\}$ are independent of the distance to the SN. However, the values of $M_{\rm{Ni}56}$ and $ET$ depend on the distance to the SN, and as a result, we propagate the distance uncertainty to the errors of these parameters. 

In order to estimate the effect of the uncertainty of the extinction, we construct two additional light curves for each SN by using the upper and lower $ 1\sigma $ uncertainty of the extinction. We find the four parameters for these light curves using the same procedure. The error due to extinction uncertainty is determined as half the difference between the obtained parameters of the high and low extinction light curves, and is propagated to the errors as well. The extinction uncertainty mainly affects the errors estimate of $M_{\rm{Ni}56}$ and $ET$, and can only lead to a few percent uncertainty for $ t_0 $ and $ n $ in the case of high extinction uncertainties ($ \delta E(B-V)\gtrsim 0.1$). In two Type IIb SNe with low $ ET $ values (SN 1993J and SN 2008ax), the lower $ET$ limits become negative after this process, so we fixed them to zero in order to remain consistent with our priors. 


\section{The ejecta properties of the SNe sample}
\label{sec:results}

The bolometric light curve parameters, derived using the integral method, for the SNe sample, are given in Table~\ref{tab:results}. The values in Table~\ref{tab:results} are the median values of the posterior distribution of the parameters, together with the $68$ per cent confidence level (note that the values presented in Appendix~\ref{app:deposition plots} are the best-fitting values). Except for Type IIP SNe, the best-fitting and median $t_0$ values of most SNe agree to within $1$ per cent, while some differ by up to $\sim5$ per cent. The relative uncertainty of the $n$ parameter is larger than the relative uncertainty of $t_0$, since $n$ is largely dependent on the deposition function at early times, typically not satisfying the condition $L=Q_{\text{dep}}$. 

The main result is presented in Figure~\ref{fig:t0Ni}, where the distribution of $t_0$ as a function of $M_{\rm{Ni}56}$ for various types of SNe is shown (plotted are the median values). As can be seen in the figure, we were able to recover the tight range of $\gamma$-ray escape time, $t_0\approx30-45\,\textrm{d}$, for Type Ia SNe \citep{WygodaI2019}. The Type Ia SNe $M_{\rm{Ni}56}$ values range from $0.075\pm0.029\,M_\odot$ for the faint SN 2007N to $\sim0.8\,M_\odot$, with an average value of $\approx0.5\,M_\odot$. Our $ t_0 $ values are within $ 10\% $ of the results of \cite{WygodaI2019}, with the differences either due to the more careful construction of the bolometric light curve or due to the more versatile deposition interpolating function. We find a new tight range, $t_0\approx80-140\,\textrm{d}$, for SE SNe, with Type Ib/c having somewhat higher values than Type IIb ($107-140\,\textrm{d}$ for Type Ib/c and $80-110\,\textrm{d}$ for Type IIb). The average $M_{\rm{Ni}56}$ of SE SNe, $\approx 0.08\,M_\odot$, is lower than the average value of Type Ia SNe, with almost the same values for Type Ib/c and Type IIb SNe. The $t_0$ values of Type IIP are much larger, and it is therefore much harder to evaluate their deposition functions. We are only able to determine the deposition function for three SNe (SN 2004et, SN 2017eaw and SN 1987A), while for the rest of the Type II SNe we can only able to provide a lower limit for $t_0$ (although $M_{\rm{Ni}56}$ and $ET$ are measured to 20\% accuracy, with the distance and extinction uncertainties being the main cause for the error). The lower limit was obtained by finding the $t_0$ value that would lead to a deposition equal to the mean value of the last three available phases, using the measured value of $M_{\rm{Ni}56}$ and $n=5$ (which is higher than any value in the sample, with a lower $n$ leading to a higher $t_0$). This lower limit is roughly the time of the last measurement. Type IIP SNe are clearly separated from other SNe types with $t_0\gtrsim400\,\textrm{d}$, and have a possible negative correlation between $t_0$ and $M_{\rm{Ni}56}$. We find that the typical $M_{\rm{Ni}56}$ of SE SNe are larger than those of Type IIP SNe, in agreement with \citet{Kushnir2015progenitors}.

\begin{table*}
		\vspace{-0.25 cm}
	\begin{threeparttable}
		\caption{The bolometric light-curve parameters, derived using the integral method. The values of the derived parameters are the median values of the posterior distribution, together with the $68$ per cent confidence levels.}
		\renewcommand{\arraystretch}{1.3}
	\begin{tabular}{llccllllcc}
Type	& Name        &  $ \mu $\tnote{a}  & $ E(B-V)_\text{MW} $\tnote{b}  &$ E(B-V)_\text{host} $\tnote{c}& $ R_V^\text{host} $,\tnote{d}& $M_{\text{Ni}56}$ & $t_0$	& $ET$ & $n$  \\[-0.15cm]
&           &           &    &    &   & $(M_\odot)$& (day)	& $ (10^{55}\,\text{erg}\,\text{s})$ &     \\\midrule 
Ia     & 2003du     &  32.79$\,\pm\,$0.04  &  0.01 &  0.00$\,\pm\,$0.05   &  3.1 & $   0.60^{+0.20}_{-0.14}$ & $ 36.1^{+2.1}_{-2.9}$ &	0 & $ 3.0^{+1.3}_{-1.1}$\\
& 2004eo     &  34.12$\,\pm\,$0.10  &  0.11 &  0.00$\,\pm\,$0.01   &  3.1 & $   0.47^{+0.14}_{-0.08}$ & $ 37.3^{+1.8}_{-2.1}$ &	0 & $ 2.6^{+1.3}_{-1.0}$\\
& 2004gs     &  35.49$\,\pm\,$0.05  &  0.03 &  0.19$\,\pm\,$0.01   &  1.9 & $   0.40^{+0.09}_{-0.07}$ & $ 33.9^{+1.5}_{-1.9}$ &	0 & $ 2.0^{+0.7}_{-0.5}$\\
& 2005cf     &  32.29$\,\pm\,$0.10  &  0.10 &  0.09$\,\pm\,$0.03   &  3.1 & $   0.63^{+0.15}_{-0.12}$ & $ 37.2^{+1.7}_{-1.9}$ &	0 & $ 2.7^{+1.2}_{-0.8}$\\
& 2005el     &  34.04$\,\pm\,$0.40  &  0.10 &  0.01$\,\pm\,$0.01   &  3.1 & $   0.50^{+0.20}_{-0.19}$ & $ 33.3^{+1.4}_{-1.6}$ &	0 & $ 3.3^{+1.1}_{-1.0}$\\
& 2005eq     &  35.40$\,\pm\,$0.04  &  0.06 &  0.11$\,\pm\,$0.02   &  2.4 & $   0.84^{+0.30}_{-0.18}$ & $ 41.6^{+2.3}_{-2.9}$ &	0 & $ 1.9^{+1.1}_{-0.6}$\\
& 2005ke     &  31.86$\,\pm\,$0.19  &  0.02 &  0.32$\,\pm\,$0.03   &    1 & $   0.13^{+0.03}_{-0.02}$ & $ 34.3^{+1.1}_{-1.2}$ &	0 & $ 3.3^{+0.8}_{-0.6}$\\
& 2005ki     &  34.67$\,\pm\,$0.05  &  0.03 &  0.02$\,\pm\,$0.01   &  1.4 & $   0.46^{+0.07}_{-0.05}$ & $ 33.4^{+1.2}_{-1.5}$ &	0 & $ 3.2^{+0.9}_{-0.7}$\\
& 2006D      &  32.97$\,\pm\,$0.10  &  0.04 &  0.14$\,\pm\,$0.01   &  1.6 & $   0.41^{+0.06}_{-0.05}$ & $ 33.2^{+1.1}_{-1.2}$ &	0 & $ 3.8^{+0.8}_{-0.8}$\\
& 2006hb     &  34.08$\,\pm\,$0.07  &  0.02 &  0.09$\,\pm\,$0.02   &  1.8 & $   0.28^{+0.07}_{-0.05}$ & $ 37.8^{+2.1}_{-2.7}$ &	0 & $ 2.5^{+1.1}_{-0.7}$\\
& 2006is     &  35.40$\,\pm\,$0.06  &  0.03 &  0.01$\,\pm\,$0.01   &  1.6 & $   0.62^{+0.27}_{-0.10}$ & $ 46.0^{+3.3}_{-5.3}$ &	0 & $ 2.8^{+1.4}_{-1.2}$\\
& 2006kf     &  34.80$\,\pm\,$0.05  &  0.21 &  0.04$\,\pm\,$0.02   &  1.8 & $   0.38^{+0.07}_{-0.05}$ & $ 31.3^{+1.4}_{-1.7}$ &	0 & $ 3.4^{+1.1}_{-1.0}$\\
& 2007N      &  33.93$\,\pm\,$0.09  &  0.03 &  0.44$\,\pm\,$0.04   &  1.3 & $   0.08^{+0.02}_{-0.01}$ & $ 32.9^{+1.9}_{-2.4}$ &	0 & $ 2.3^{+1.1}_{-0.7}$\\
& 2007af     &  31.72$\,\pm\,$0.07  &  0.01 &  0.13$\,\pm\,$0.02   &  3.1 & $   0.40^{+0.07}_{-0.06}$ & $ 36.8^{+1.4}_{-1.5}$ &	0 & $ 2.9^{+1.0}_{-0.7}$\\
& 2007as     &  34.44$\,\pm\,$0.06  &  0.12 &  0.12$\,\pm\,$0.03   &  1.4 & $   0.52^{+0.20}_{-0.09}$ & $ 36.7^{+2.1}_{-2.9}$ &	0 & $ 2.5^{+1.5}_{-1.0}$\\
& 2007hj     &  33.94$\,\pm\,$0.08  &  0.07 &  0.12$\,\pm\,$0.02   &  1.4 & $   0.29^{+0.09}_{-0.05}$ & $ 33.9^{+2.0}_{-2.8}$ &	0 & $ 2.5^{+1.3}_{-0.8}$\\
& 2007le     &  32.23$\,\pm\,$0.16  &  0.03 &  0.38$\,\pm\,$0.02   &  1.9 & $   0.64^{+0.16}_{-0.12}$ & $ 42.1^{+1.6}_{-1.6}$ &	0 & $ 2.9^{+1.3}_{-1.0}$\\
& 2007on     &  31.28$\,\pm\,$0.36  &  0.01 &  0.00$\,\pm\,$0.05   &  3.1 & $   0.20^{+0.08}_{-0.07}$ & $ 30.8^{+1.1}_{-1.2}$ &	0 & $ 3.9^{+0.7}_{-0.8}$\\
& 2008bc     &  34.15$\,\pm\,$0.06  &  0.23 &  0.02$\,\pm\,$0.02   &  1.5 & $   0.70^{+0.20}_{-0.10}$ & $ 38.9^{+2.4}_{-3.2}$ &	0 & $ 3.0^{+1.3}_{-1.1}$\\
& 2008fp     &  31.72$\,\pm\,$0.14  &  0.17 &  0.52$\,\pm\,$0.03   &  2.2 & $   0.68^{+0.16}_{-0.13}$ & $ 40.1^{+1.5}_{-1.7}$ &	0 & $ 3.3^{+1.1}_{-1.0}$\\
& 2008hv     &  33.76$\,\pm\,$0.08  &  0.03 &  0.01$\,\pm\,$0.01   &  1.1 & $   0.43^{+0.08}_{-0.05}$ & $ 34.5^{+1.5}_{-1.8}$ &	0 & $ 3.3^{+1.1}_{-0.9}$\\
& 2009Y      &  33.13$\,\pm\,$0.09  &  0.09 &  0.15$\,\pm\,$0.03   &  1.3 & $   0.72^{+0.14}_{-0.11}$ & $ 43.8^{+1.4}_{-1.5}$ &	0 & $ 2.3^{+0.8}_{-0.5}$\\
& 2011fe     &  29.03$\,\pm\,$0.17  &  0.01 &  0.03$\,\pm\,$0.01   &  3.1 & $   0.49^{+0.09}_{-0.09}$ & $ 39.1^{+1.1}_{-1.2}$ &	0 & $ 3.4^{+0.9}_{-0.8}$\\
& 2012fr     &  31.27$\,\pm\,$0.05  &  0.02 &  0.03$\,\pm\,$0.01   &  3.1 & $   0.61^{+0.09}_{-0.07}$ & $ 43.4^{+1.4}_{-1.5}$ &	0 & $ 2.8^{+1.0}_{-0.7}$\\
& 2012ht     &  31.50$\,\pm\,$0.40  &  0.02 &  0.00$\,\pm\,$0.01   &  3.1 & $   0.24^{+0.09}_{-0.08}$ & $ 33.7^{+1.7}_{-2.0}$ &	0 & $ 3.7^{+0.9}_{-1.0}$\\
& 2013dy     &  30.68$\,\pm\,$0.48  &  0.14 &  0.15$\,\pm\,$0.06   &  3.1 & $   0.31^{+0.17}_{-0.17}$ & $ 39.8^{+2.5}_{-3.0}$ &	0 & $ 3.0^{+1.2}_{-0.9}$\\
& 2015F      &  31.89$\,\pm\,$0.04  &  0.18 &  0.04$\,\pm\,$0.03   &  3.1 & $   0.54^{+0.15}_{-0.10}$ & $ 38.4^{+2.1}_{-2.3}$ &	0 & $ 3.0^{+1.2}_{-1.1}$\\\midrule
Ib/c    & 2002ap     &  29.50$\,\pm\,$0.10  &  0.09 &  0.00$\,\pm\,$0.01   &  3.1 & $   0.05^{+0.01}_{-0.01}$ & $  106^{+4}_{-4}$ &$ 0.14^{+0.04}_{-0.04}$& $ 2.4^{+0.6}_{-0.5}$\\
& 2007C      &  31.61$\,\pm\,$0.40  &  0.04 &  0.55$\,\pm\,$0.04   &  2.4 & $   0.04^{+0.02}_{-0.02}$ & $  120^{+28}_{-15}$ &$ 0.25^{+0.12}_{-0.14}$& $ 2.6^{+1.5}_{-1.1}$\\
& 2007gr     &  30.13$\,\pm\,$0.35  &  0.06 &  0.03$\,\pm\,$0.02   &  3.1 & $   0.05^{+0.02}_{-0.02}$ & $  140^{+35}_{-17}$ &$ 0.41^{+0.18}_{-0.22}$& $ 2.4^{+1.5}_{-1.0}$\\
& 2009jf     &  32.65$\,\pm\,$0.10  &  0.11 &  0.05$\,\pm\,$0.05   &  3.1 & $   0.15^{+0.03}_{-0.02}$ & $  147^{+19}_{-19}$ &$ 0.81^{+0.44}_{-0.44}$& $ 3.3^{+1.1}_{-1.1}$\\
& 2016coi    &  31.29$\,\pm\,$0.16  &  0.08 &  0.13$\,\pm\,$0.03   &  3.1 & $   0.12^{+0.04}_{-0.02}$ & $  141^{+11}_{-9}$ &$ 0.97^{+0.39}_{-0.49}$& $ 2.8^{+1.3}_{-1.0}$\\\midrule
IIb    & 1993J      &  27.78$\,\pm\,$0.18  &  0.07 &  0.10$\,\pm\,$0.10   &  3.1 & $   0.08^{+0.03}_{-0.03}$ & $  100^{+10}_{-9}$ &$ 0.23^{+0.18}_{-0.17}$& $ 3.0^{+0.8}_{-0.6}$\\
& 2006T      &  32.50$\,\pm\,$0.41  &  0.07 &  0.32$\,\pm\,$0.04   &  1.3 & $   0.07^{+0.04}_{-0.04}$ & $  109^{+29}_{-14}$ &$ 0.23^{+0.13}_{-0.13}$& $ 2.5^{+1.4}_{-1.0}$\\
& 2008aq     &  32.15$\,\pm\,$0.40  &  0.04 &  0.00$\,\pm\,$0.00   &    0 & $   0.04^{+0.01}_{-0.01}$ & $   81^{+6}_{-4}$ &$ 0.00^{+0.00}_{-0.00}$& $ 2.8^{+0.8}_{-0.6}$\\
& 2008ax     &  29.91$\,\pm\,$0.29  &  0.02 &  0.38$\,\pm\,$0.10   &  3.1 & $   0.09^{+0.03}_{-0.03}$ & $   96^{+7}_{-7}$ &$ 0.17^{+0.14}_{-0.14}$& $ 2.9^{+0.8}_{-0.6}$\\
& 2010as     &  32.16$\,\pm\,$0.36  &  0.15 &  0.42$\,\pm\,$0.10   &  1.5 & $   0.12^{+0.05}_{-0.05}$ & $   94^{+32}_{-18}$ &$ 0.00^{+0.00}_{-0.00}$& $ 1.7^{+1.1}_{-0.5}$\\
& 2011dh     &  29.46$\,\pm\,$0.28  &  0.03 &  0.04$\,\pm\,$0.07   &  3.1 & $   0.06^{+0.02}_{-0.02}$ & $  103^{+8}_{-7}$ &$ 0.14^{+0.10}_{-0.10}$& $ 3.4^{+1.0}_{-0.9}$\\\midrule
IIP    & 2004et     &  29.18$\,\pm\,$0.20  &  0.34 &  0.07$\,\pm\,$0.07   &  3.1 & $  0.078^{+0.026}_{-0.017}$ & $  419^{+153}_{-48}$ &$ 8.66^{+2.26}_{-2.34}$& $ 2.8^{+1.5}_{-1.3}$\\
& 2005cs     &  29.26$\,\pm\,$0.33  &  0.03 &  0.02$\,\pm\,$0.01   &  3.1 & $  0.004^{+0.001}_{-0.001}$ & $  >628$ &$ 1.05^{+0.32}_{-0.32}$& $ 3.0^{+1.3}_{-1.4}$\\
& 2009N      &  31.67$\,\pm\,$0.11  &  0.02 &  0.11$\,\pm\,$0.02   &  3.1 & $  0.022^{+0.002}_{-0.002}$ & $  >349$ &$ 1.38^{+0.18}_{-0.18}$& -\\
& 2009md     &  31.64$\,\pm\,$0.21  &  0.03 &  0.07$\,\pm\,$0.10   &  3.1 & $  0.005^{+0.001}_{-0.001}$ & $  >169$ &$ 0.65^{+0.20}_{-0.20}$& -\\
& 2012A      &  29.96$\,\pm\,$0.15  &  0.03 &  0.01$\,\pm\,$0.01   &  3.1 & $  0.010^{+0.001}_{-0.001}$ & $  >388$ &$ 1.19^{+0.19}_{-0.19}$& -\\
& 2013ej     &  29.93$\,\pm\,$0.11  &  0.06 &  0.00$\,\pm\,$0.01   &  3.1 & $  0.022^{+0.004}_{-0.003}$ & $  >338$ &$ 4.05^{+0.48}_{-0.47}$& -\\
& 2017eaw    &  29.18$\,\pm\,$0.20  &  0.30 &  0.11$\,\pm\,$0.10   &  3.1 & $  0.066^{+0.017}_{-0.016}$ & $  495^{+258}_{-71}$ &$ 5.93^{+1.83}_{-1.84}$& $ 2.9^{+1.3}_{-1.2}$\\\midrule
II-pec & 1987A      &  18.56$\,\pm\,$0.05  &  - & -   &  - & $  0.069^{+0.004}_{-0.004}$ & $  533^{+78}_{-51}$ &$ 0.28^{+0.13}_{-0.15}$& $ 2.9^{+0.8}_{-0.6}$\\
	\end{tabular}
\begin{tablenotes}
	\item [a] Distance modulus
	\item [b] Galactic extinction towards the SN
	\item [c] Host extinction
	\item [d] Host $R_V=A_V/E(B-V)_\text{host}$
	\item [$ \dagger $] For SN 1987A, the bolometric light curve was taken from the literature, so the extinction estimates are not used.
\end{tablenotes}
\label{tab:results}
\end{threeparttable}
\end{table*}

\begin{figure}
	\includegraphics[width=\columnwidth]{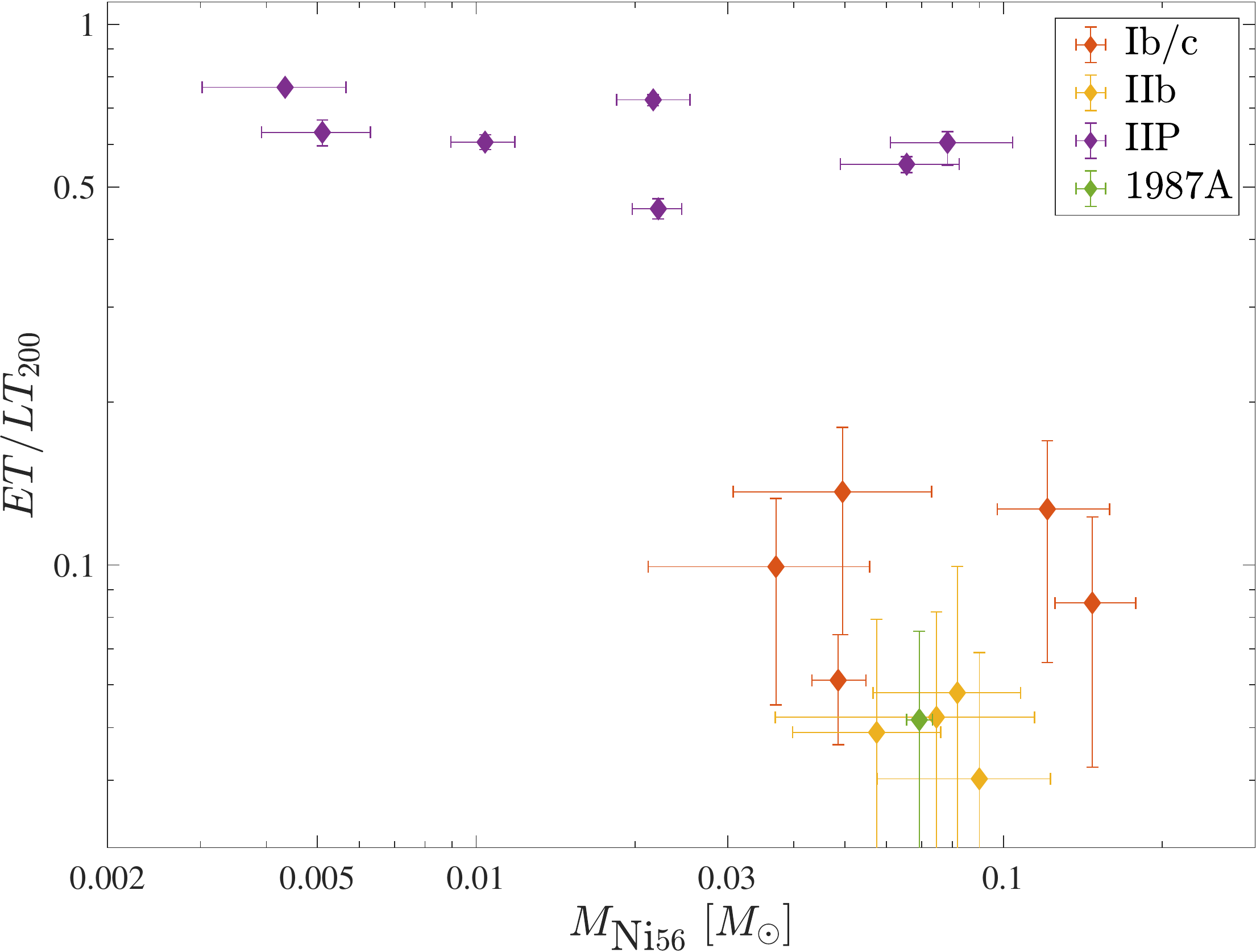}
	\caption{The fraction $ET/LT_{200}$ as a function of $M_{\rm{Ni}56}$. The values of $ET/LT_{200}$ are similar for all Type IIP SNe (purple), despite large variations in $M_{\rm{Ni}56}$ and in the overall luminosity. The derived $ET$ values of SE SNe (red and orange) and of SN 1987A (green) are smaller and more difficult to evaluate. We believe that our systematic errors are of the order of $10$ per cent, so we treat the $ET$ measurement for SE SNe as tentative.}
	\label{fig:ETNi}
\end{figure}

The distribution of $ET/LT_{200}$, where $LT_{200}=LT(200\,\textrm{d})$, as a function of $M_{\rm{Ni}56}$, is shown in Figure~\ref{fig:ETNi}. We choose to compare at $200\,\textrm{d}$ since usually luminosity observations are available at this time and the condition $L=Q_{\text{dep}}$ applies for almost all SNe. However, since a few SNe lack observations at this time, we approximate for all SNe, $LT_{200}\approx QT(200\,\textrm{d})+ET$, using the best-fitting values, which introduces a small error. Note that the derivation of $ET/LT_{200}$ is independent of $M_{\rm{Ni}56}$ and of the distance to the SN. As can be seen in the figure, the fraction $ET/LT_{200}$ is between 0.45 to 0.75 for all Type IIP SNe, despite large variations in $M_{\rm{Ni}56}$ and overall luminosity. The derived $ET$ values of SE SNe are smaller and more difficult to evaluate. We find for all Type Ib/c SNe in our sample $ET/LT_{200}$ values of about $6-14$ per cent. The $ET/LT_{200}$ values of all Type IIb SNe are below $\sim6$ per cent, and for SN 1987A we find an $ET/LT_{200}\sim5$ per cent. As discussed above, Type Ia SNe do not have a detectable $ET$ value. We believe that our systematic errors are of the order of $10$ per cent, so we treat the $ET$ measurement for SE SNe as tentative.

\begin{figure}
	\includegraphics[width=\columnwidth]{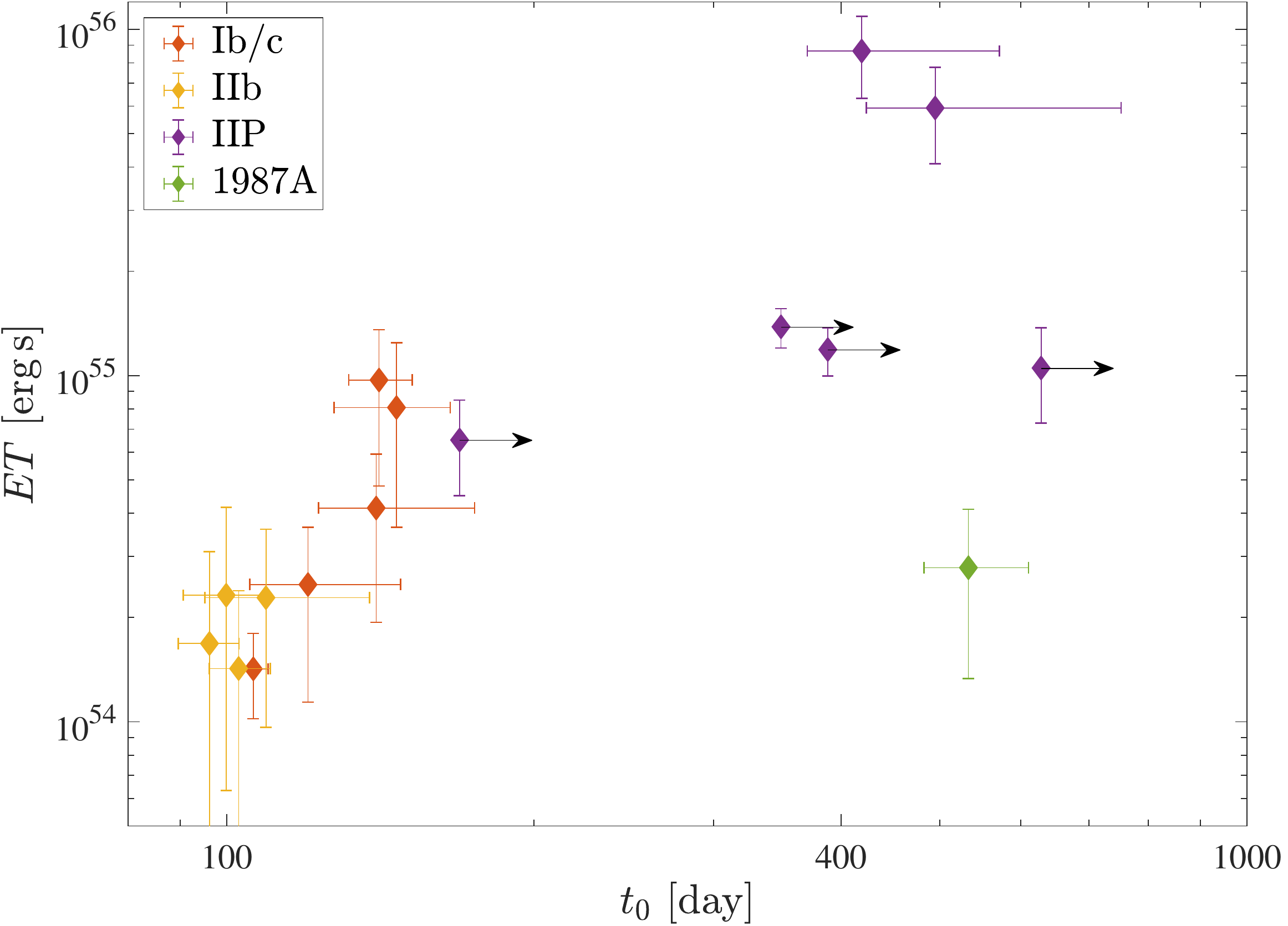}
	\caption{The $ET-t_0$ distribution of our sample. The derived $ET$ values of Type IIP SNe agree to within $20$ per cent with the values given by \citet{Nakar2016}. SE SNe show a positive correlation between $ET$ and $t_0$, though we treat the $ET$ measurement for SE SNe as tentative. Type IIP SNe with a measurable $t_0$ have larger $ET$ values.}
	\label{fig:ETt0}
\end{figure}

We can further calculate $ET$ by using the distances to the SNe. Figure \ref{fig:ETt0} shows the $ET-t_0$ distribution of our sample. As can be seen in the figure, Type IIb SNe have $ET$ values of up to $ \approx2.5\times10^{54}\,\text{erg}\,\text{s}$, with two of the SNe having negligible values and are not shown. Type Ib/c SNe have larger values, ranging from $ \approx1.5\times10^{54}\,\text{erg}\,\text{s} $ to $\approx10^{55}\,\text{erg}\,\text{s}$. There is a possible positive correlation between $ET$ and $t_0$ for SE SNe. We note again that we treat the $ET$ measurement for SE SNe as tentative. Type IIP SNe have the largest $ET$ values, from $ \approx6.5\times10^{54}\,\text{erg}\,\text{s} $ to $ \approx8.5\times10^{55}\,\text{erg}\,\text{s}$.  The derived $ET$ values agree to within $20$ per cent with the values given by \cite{Nakar2016}. Inconsistencies are mainly due to differences between the adopted bolometric light curves. We find that Type IIP SNe with a measurable $t_0$ have larger $ET$ values, but the sample is too small to draw strong conclusions.

\subsection{Comparison to previous works}
\label{sec:previous}

The derived $M_{\rm{Ni}56}$ for the Type Ia SNe and SE SNe sample are almost always lower than the values predicted by Arnett's rule. For example, the $M_{\rm{Ni}56}$ derived for the CSP SNe in our sample are lower than the Arnett's rule estimates of \cite{carAnalysis} for the same SNe by roughly $10$ per cent (SN 2006T), $25$ per cent (SN 2008aq) and $45$ per cent (SN 2007C). \cite{dessart2016inferring} also reported an overestimate of Arnett's rule for their models. This is probably due to the simplifying assumptions used in deriving Arnett's rule. \citet{meza2020strippedenvelope} used pseudo bolometric light curves to estimate the $M_{\rm{Ni}56}$ of SE SN using Arnett's rule. Their light curves were constructed from the $B$ band to the $H$ band, so their luminosity for the same SN was almost always lower than ours, typically by tens of per cents. This underestimate of the bolometric light curve compensated for the higher $M_{\rm{Ni}56}$ derived by Arnett's rule, such that their derived $M_{\rm{Ni}56}$ are similar to our values, with our sample having slightly lower values on average. \citet{meza2020strippedenvelope} used the method of \citet{khatami2019physics}, which is a variant of the Katz integral that uses the peak time and includes a free parameter that cannot be calibrated from the observations \citep[see also][]{Kushnir2019analytical}. The uncertainty of using this method is substantial and hard to estimate. Finally,  \citet{meza2020strippedenvelope} used the direct method, but with the assumption of full $\gamma$-ray deposition (that is unrealistic according to our results), which provides a poor description of the observations (tens of per cent discrepancy).

\begin{figure}
	\includegraphics[width=\columnwidth]{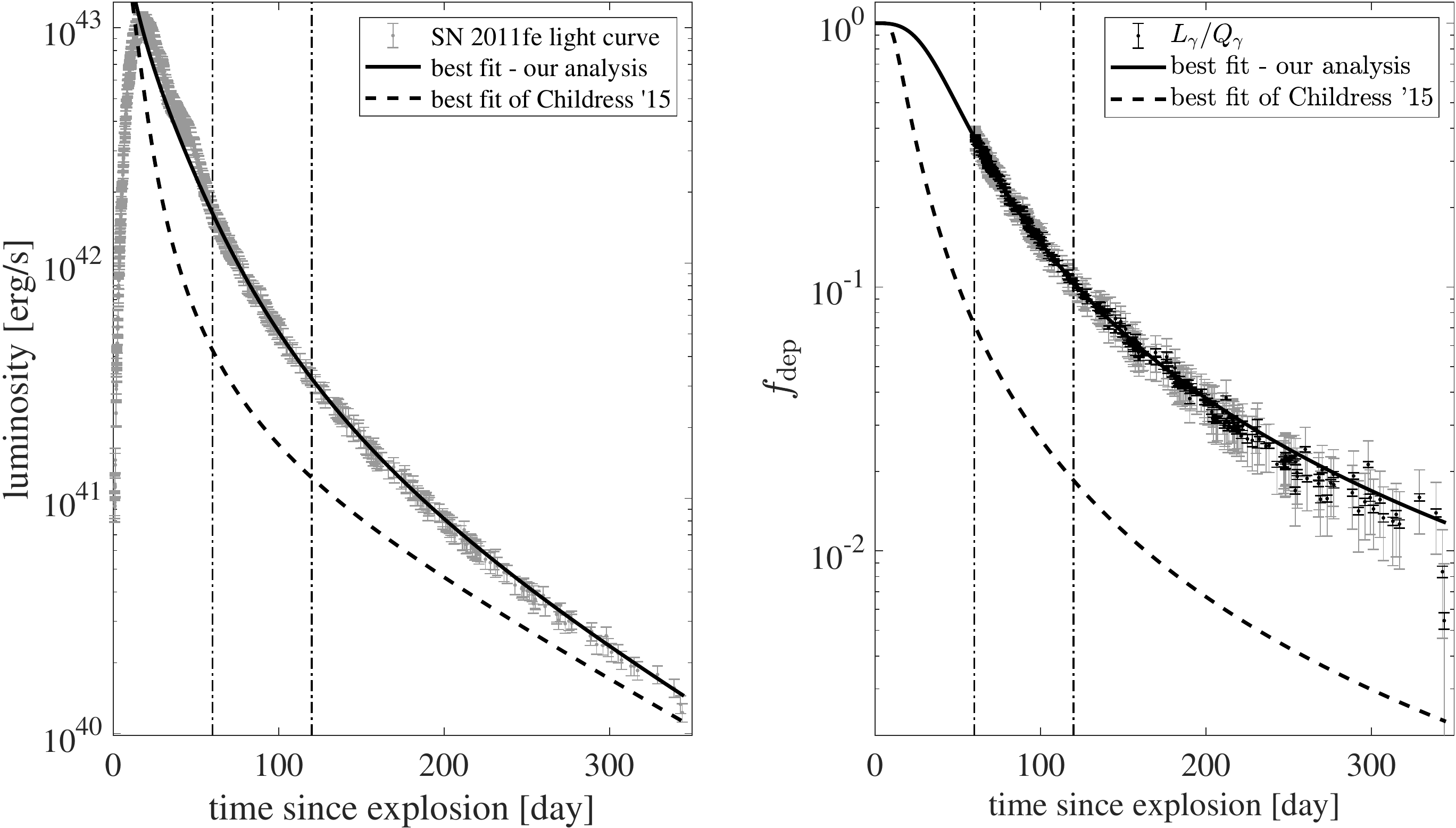}
	\caption{Same as \ref{fig:fits} for SN 2011fe. left-hand side panel: The observed bolometric light curve (grey symbols), our best-fitting model (solid line) and best fit of \citet[][dashed line]{childress2015} with the exponential interpolating function (Equation~\eqref{eq:dep_exp}). right-hand side panel: the $ \gamma $-ray deposition fraction $L_{\gamma}/Q_{\gamma}$, the best-fitting deposition model (solid line), and the deposition model of \citet[][dashed line]{childress2015}. It is clear that the $t_0\approx16\,\textrm{d}$ derived by \citet{childress2015} is too low to explain the light curve evolution.}
	\label{fig:2011fe_comp}
\end{figure}

We further compare our results to those previously derived for Type Ia SNe by different methods. \citet{Scalzo2019} estimated $M_{\rm{Ni}56}$ and $t_0$ for their Type Ia SNe sample using hydrodynamic modelling, which includes various simplifying assumptions, and found, for the same SNe, values that are generally lower than ours. Using $^{56}$Co emission lines in nebular phase spectra, \cite{childress2015} derived $t_c$ (related to $t_0$ by $t_0\approx0.18t_c$) for six Type Ia SNe. They found a span of $13.2-48.2\,\textrm{d}$ in $t_0$ values, much larger than our results and from those of \citet{WygodaI2019}. Four of these SNe were also analysed by us. For SN 2003du, \cite{childress2015} derived a value of $35.1\pm4.1\,\textrm{d}$, consistent with our result of $35.8\pm4.2\,\textrm{d}$. The value of SN 2012fr was somewhat lower than ours, $ 34.7\pm4.0\,\textrm{d} $ compared to $ 43.3\pm0.9\,\textrm{d} $, respectively. For SN 2007af and SN 2011fe, the $t_0$ values derived by \cite{childress2015} are $13.3\pm3.8\,\textrm{d}$ and $16.4\pm2.1\,\textrm{d}$, which are much lower than our values of $36.7\pm1.2\,\textrm{d}$ and $39.4\pm0.4\,\textrm{d}$, respectively. In Figure~\ref{fig:2011fe_comp}, we compare the light curve and the $\gamma$-ray deposition fraction of SN 2011fe to the best fit values of \cite{childress2015}. It is clear that the $t_0$ derived by \citet{childress2015} is too low to explain the light curve evolution. We suggest that the reason for this discrepancy is their assumption that the ejecta is optically thin already at $t\sim100\,\textrm{d}$. 

We find that the $t_0$ values we derived for the SE SNe sample are lower than the results of \cite{wheeler2015analysis}, which were derived by fitting the late time light-curve luminosity to the radioactive energy deposition rate. In their analysis, they used the exponential interpolating deposition function, Equation~\eqref{eq:dep_exp}, and incomplete positron trapping was assumed \citep[which is not realistic, see][]{kushnir2020constraints}. Their values are $\approx30$ per cent (SN 1993J) to $\approx140$ per cent (SN 2009jf) higher than our values. We try to reproduce their results by using the same interpolating function and assuming incomplete positron trapping, and we find that $t_0$ increase by $5-10$ per cent. This increment brings some $t_0$ values closer to the ones derived by \cite{wheeler2015analysis}, like SN 1993J, but most SNe still differ by tens of per cents.

\section{Constraining ejecta parameters with $\gamma$-ray transfer simulations}
\label{sec:simulations}

In this section, we demonstrate that the derived $\gamma$-ray deposition histories from Section~\ref{sec:results} offer a powerful tool for constraining models. The application for Type Ia SNe was already discussed by \cite{WygodaI2019}, so we focus on SE SNe (Section~\ref{sec:SE SNe}) and on Type II SNe (Section~\ref{sec:type II SNe}). We apply a Monte Carlo (MC) $\gamma$-ray transfer simulations to models from the literature, which allows us to measure the values of $t_0$ and $n$ \citep[note that $t_0$ can be obtained directly from the models without performing transfer simulation, by averaging over the $^{56}$Ni optical depth, see][]{WygodaI2019}. The MC $\gamma$-ray transfer code is similar to the one described by \cite{WygodaI2019,WygodaII2019}. In Section~\ref{sec:Compare ET} we compare the observed $ET$ values of our core-collapse sample to the models predictions and we discuss the implications of this comparison. The model parameters and the results of the simulations are presented in Appendix \ref{tab:models}. The parameters consist of the ejecta mass, $M_{\rm{ej}}$, the kinetic energy, the amount of $ ^{56} $Ni mixing, and the ratio of ejecta mass to the square root of kinetic energy. The amount of $ ^{56} $Ni mixing is defined as
\begin{equation}\label{eq:mixing}
	\frac{\int X_\text{Ni56}(m)mdm}{M_\text{ej}M_\text{Ni56}}
\end{equation}
where $ X_\text{Ni56} $ is the $^{56}$Ni mass fraction and $ m $ is the enclosed mass. This is a measure of the $^{56}$Ni normalized 'centre of mass'. Profiles with centered $^{56}$Ni distribution will have mixing values close to 0, whereas a fully and uniformly mixed distribution will have a mixing value of 0.5. 

\subsection{SE SNe comparison with models -- $n$ and $t_0$}
\label{sec:SE SNe}

\begin{figure}
	\includegraphics[width=\columnwidth]{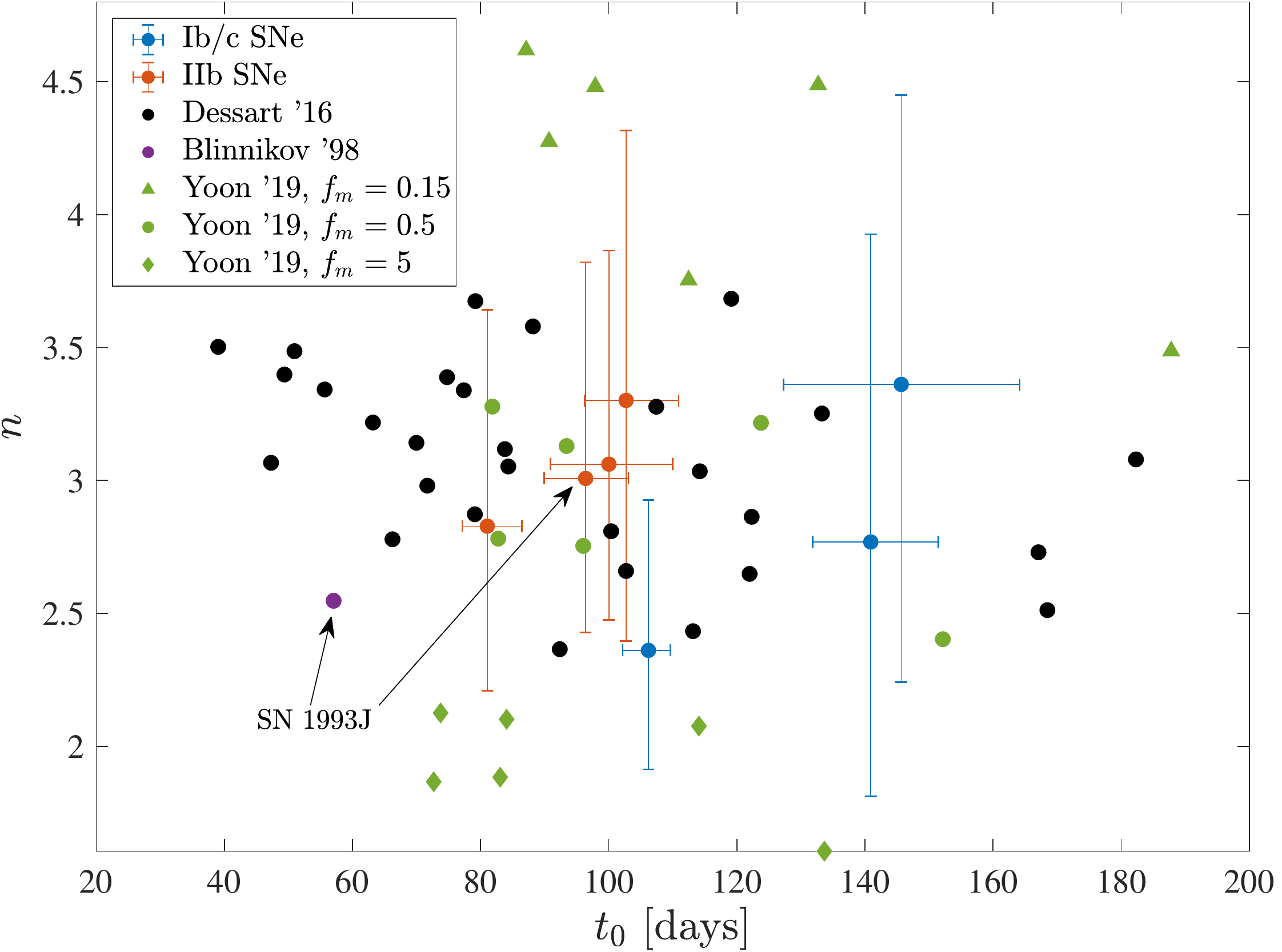}
	\caption{The $n-t_0$ distribution of SE SNe with relatively low $t_0$ error (blue and red), together with the MC $\gamma$-ray transfer results for the ejecta of \citet[][black]{dessart2016inferring}, \citet[][purple]{blinnikov1998comparative}, and \citet[][green symbols, the shape marks the degree of mixing]{Yoon2019}.}
	\label{fig:SEmodels}
\end{figure}

The $n-t_0$ distribution of SE SNe with relatively a low $t_0$ error (three Type Ib/c, four Type IIb), is shown in Figure~\ref{fig:SEmodels}. Also shown are these parameters for several models from the literature \citep[kindly provided to us by the authors of][]{blinnikov1998comparative,dessart2016inferring,Yoon2019}, calculated with the MC $\gamma$-ray transfer code. 

\cite{Yoon2019} simulated Type Ib/c SN progenitors with different explosion energies and $^{56}$Ni mixing. We examine two models, shown as green symbols in Figure~\ref{fig:SEmodels}. The first, HE3.87, simulated a Type Ib SN, with an ejecta mass of $M_{\text{ej}}=2.4\,M_\odot $, and the second, CO3.93, simulated a Type Ic SN, with an ejecta mass of $M_{\text{ej}}=2.49\,M_\odot$. Each model has several kinetic energy values and different levels of $^{56}$Ni mixing. \cite{Yoon2019} compared the light curves and the early-time color evolutions of observed Type Ib/c SNe to the models. They found that models with moderate mixing, $ f_m=0.15-0.5 $, agree well with Type Ib SNe, while Type Ic SNe are better described by models with high mixing, $ f_m=0.5-5 $. We find that the values of $t_0$ are mainly determined by the kinetic energy (see discussion below) and, to some extent, by the amount of mixing. The values of $n$ are almost solely determined by the level of mixing. As can be seen in the figure, profiles with low mixing, shown as triangles \citep[$f_m =0.15$, see][for details]{Yoon2019} have large values of $n$ and are in disagreement with our analysis, while profiles with moderate mixing ($f_m =0.5$, shown as circles), or high mixing ($f_m=5$, diamonds) have lower values of $ n $ and are in better agreement with our analysis. Models with relatively high kinetic energies ($E_{\text{kin}}>1.5\times10^{51}\,\text{erg}$) agree well with Type IIb SNe, while moderate kinetic energies ($E_{\text{kin}}<1.5\times10^{51}\,\text{erg}$) agree well with Type Ib/c SNe. This result is in contrast with the expected kinetic energies of each type \citep[see]{Kushnir2015progenitors}. However, both progenitor models of \cite{Yoon2019} are a star stripped from its envelope through a binary companion, and the ejecta masses of these models are relatively low (see Appendix \ref{tab:models}). Type Ib/c SNe progenitors might be massive Wolf\--Rayet (WR) stars \citep{Kushnir2015progenitors}, which can lead to a different range of ejecta masses and energies, so the lower $ t0 $ values of the Type IIb SNe might be the result of lower ejecta mass and not larger kinetic energies.

\cite{dessart2016inferring} considered 29 profiles of five different models that simulated the explosion of the mass donor in a close-binary system. The profiles vary in $M_{\text{ej}}$, $E_{\text{kin}}$, and have two levels of $^{56}$Ni mixing. The derived values of $t_0$ and $n$ are shown as black symbols in Figure~\ref{fig:SEmodels}. These models span a larger range of $t_0$ values than the ones that have been observed. This can be understood by inspecting the $^{56}$Ni mass-averaged column density of each model, which determines the value of $t_0$ \citep{WygodaI2019}. For the models considered by \cite{dessart2016inferring}, the two levels of $^{56}$Ni mixing are very similar, such that $t_0$ is mainly determined by the column density of the ejecta
\begin{eqnarray}\label{eq:alpha}
 t_0\propto\sqrt{\int \rho dv}\propto\frac{M_{ej}}{\sqrt{E_{\text{kin}}}}\propto\alpha,\nonumber\\
 \textrm{where}\;\alpha\equiv \frac{M_\text{ej}/M_\odot}{\sqrt{E_\text{kin}/10^{51}\,\text{erg}}}.
 \end{eqnarray}
The $t_0$ values of our sample, using the density profile of \cite{dessart2016inferring}, correspond to ejecta with $2.1\lesssim\alpha\lesssim3.5$, while the full set of models have $\alpha$-values within $1\lesssim\alpha\lesssim4$.

SN 1993J was modelled by \cite{blinnikov1998comparative}, using a profile with $M_{\text{ej}}=2.26\,M_\odot$ and $E_{\text{kin}}=1.32\times10^{51}\,\text{erg}$, which corresponds to $\alpha\approx2$. Using radiation transfer codes, \cite{blinnikov1998comparative} compared their models to the observed $U$, $B$ and $V$ magnitudes, and to the bolometric luminosity, from shock breakout to $ 120\,\textrm{d} $ since explosion. They found good agreement with the bolometric light curve, and decent agreement with the individual magnitudes. We derived $t_0\approx57\,\textrm{d}$ (purple symbol in the figure), which is much lower than the observed value of SN 1993J ($\approx100\,\textrm{d}$). 

We find that $ \gamma $-ray transfer simulations can be used to constrain models in a number of ways. The observed range of $ t_0 $ is incompatible with models having too small or too large $\alpha$ values. This constrains the allowed range of masses and kinetic energies of the profiles, and indicates that the models of \cite{dessart2016inferring} with very small $ t_0 $ values of $<80\,\textrm{d}$ are inconsistent with the observations. As can be seen in Appendix \ref{tab:models}, the level of $ ^{56} $Ni mixing is portrayed in the value of $ n $, where higher amount of mixing leads to lower $ n $ values. Although harder to determine, the $n$ parameter of the observed SNe can be used to identify models with insufficient amount of $ ^{56} $Ni mixing, like the profiles with low $ ^{56} $Ni mixing of \cite{Yoon2019}. Some of the models that are inconsistent with the observed $\gamma$-ray deposition were found to agree with optical observations. Since the optical radiation transfer calculations are more uncertain compared with the $\gamma$-ray transfer calculations, which are accurate and easy to implement, we believe that the observed $\gamma$-ray deposition provides a more robust constrain. 

\subsection{Type II SNe comparison with models -- $n$ and $t_0$}
\label{sec:type II SNe}

\begin{figure}
	\includegraphics[width=\columnwidth]{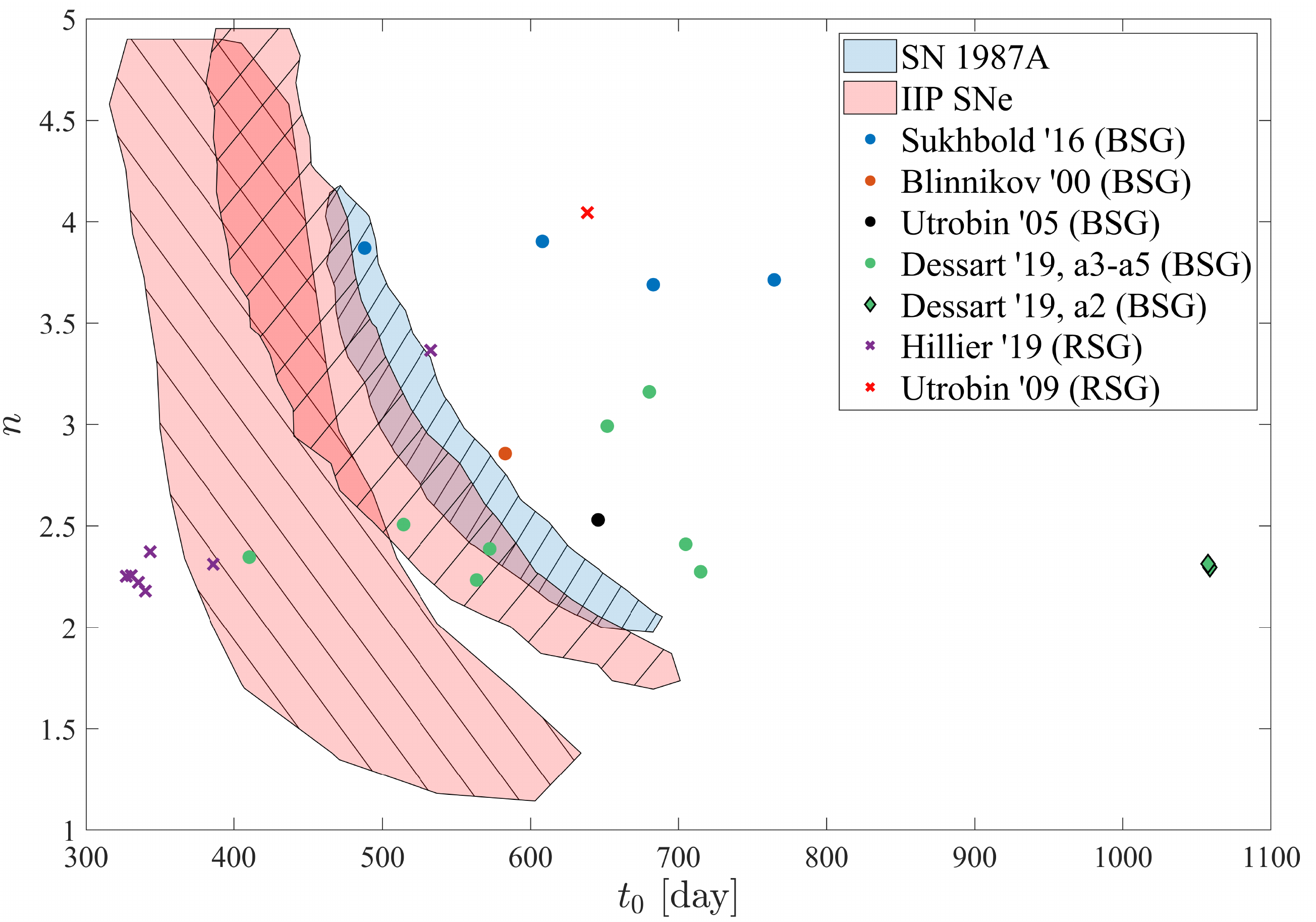}
	\caption{The $n-t_0 $ 68\% confidence region for the sample of Type II SNe with measured $t_0$: SN 1987A (blue shaded region), SN 2004et and SN 2017eaw (both in red shaded region), together with the MC $\gamma$-ray transfer results for the ejecta of BSG (circles and diamonds) and RSG progenitors (crosses) from the literature. The BSG progenitors are of \citet[][blue circles]{sukhbold2016core}, \citet[][red circle]{blinnikov2000radiation}, \citet[][black circle]{utrobin2005supernova}, and \citet[][green circles for the a3,a4,a5 profiles and green diamonds for the a2 profiles]{dessart2019supernovae}. The RSG progenitors are of \citet[][purple crosses]{hillier2019photometric} and \citet[][red cross]{utrobin2009high}.}
	\label{fig:IImodels}
\end{figure}

We apply the analysis of the previous section for Type II SN. The sample of Type II SNe with measured $t_0$ consists of SN 1987A, SN 2004et and SN 2017eaw. Since the optical depth for $\gamma$-rays does not become very small during the phases used for parameter fitting, there is a large degeneracy between $t_0$ and $n$ (and an especially large uncertainty in $n$). The $n-t_0$ $68$ per cent confidence region derived for these SNe are shown in Figure~\ref{fig:IImodels}. Also shown are these parameters for several models from the literature, which include the blue super-giant (BSG) progenitor explosion models of \cite{sukhbold2016core}, \cite{blinnikov2000radiation} and \cite{dessart2019supernovae}, and the red super-giants (RSG) progenitor models of \cite{utrobin2009high} and \cite{hillier2019photometric}. The model parameters are calculated with the MC $\gamma$-ray transfer code for the ejecta profiles, kindly given to us by the authors of the papers.

The ejecta of \cite{sukhbold2016core}, marked with blue circles, include four profiles with an initial progenitor zero-age-main-sequence (ZAMS) of $ (16.2,18.2,19.2,20.2)\times M_\odot$, evolved to pre-SN, and exploded with the calibration model W18 \citep[see][for details]{sukhbold2016core}, which was designed to match the explosion parameters of SN 1987A. The $ t_0 $ values derived for the models with ZAMS solar masses of $ (16.2,18.2,20.2) M_\odot$ match the observed $ t_0 $ of SN 1987A, while the derived $t_0$ of the $19.2\, M_\odot$ ZAMS model, which has a low explosion energy, is larger than the observed value. The relatively unmixed $ ^{56} $Ni distribution of these profiles (see Appendix \ref{tab:models}) results in larger $n$ values than the observed value.

\cite{blinnikov2000radiation} modelled the light curve of SN 1987A using $M_{\text{ej}}=14\,M_\odot$ ejecta with various kinetic energies and isotope distributions. Using a radiation transfer code, \cite{blinnikov2000radiation} compared the monochromatic and bolometric light curves up to $ 180\,\textrm{d} $ since explosion, and found a good match between the observations and the simulation. The derived value of $t_0$ and $n$ (red symbol in the figure) for the profile that was provided (14E1, with $E_{\text{kin}}\approx1.0\times10^{51}\,\text{erg}$ and a relatively mixed isotope distribution) is consistent with the observed values. 

\cite{utrobin2005supernova} constructed SN ejecta profiles to match the light curve of SN 1987A. We were given the $18\,M_\odot$ ejecta profile, which provided the best match to the light curve in the original work. This agreement with the observations also appears for the derived $t_0$ and $n$ values (black symbol in the figure). \cite{dessart2019supernovae} simulated the explosion of a $15\,M_\odot$ ZAMS star that collapsed as a BSG. The resulting ejecta (green symbols) have a range of $M_{\text{ej}}$, $E_{\text{kin}}$, and $M_{\rm{Ni}56}$. The a3,a4,a5 profiles, shown in circles, have derived $ t_0 $ values in the range $ 400-700\,\textrm{d} $. The derived values of $t_0$ and $n$ for the a4 profiles are consistent with the observed SN 1987A values, and are also the ones with the closest match to the early bolometric light curve in the original work. The a2 profiles, shown as diamonds (notice that there are two overlapping symbols), have small kinetic energies, and result in substantially large $t_0$ values of $\approx1050\,\textrm{d}$, which is inconsistent with SN 1987A. They also substantially deviate from the early bolometric light curve in the original work.

\cite{hillier2019photometric} studied the explosions of an RSG evaluated from ZAMS mass of $15\,M_\odot$. Different values for the final mass of the star and different masses for possible circumstellar matter were considered. The range of ejecta mass was $\approx11-13\,M_\odot$. \cite{hillier2019photometric} compared the magnitude of individual bands and the spectra of their models, calculated with a radiation transfer code, to Type IIP SN observations. They found a good match between SN 2004et and the x1p5ext3 profile. The derived $t_0$ values (purple crosses) are relatively low compared with the observed Type IIP SNe, although some of them are consistent with $t_0=419^{+135}_{-47}$ of SN 2004et (and especially, the $ t_0 $ value of the x1p5ext3 profile is $\approx380\,\textrm{d}$). \cite{utrobin2009high} modelled the light curve of SN 2004et with high $M_{\text{ej}}=22.9\, M_\odot$, which explains the large derived value $t_0\approx640\,\textrm{d}$ (red cross), compared with the observed $t_0$ of this SN. In the original work, the bolometric light curve of the model was calculated using a radiation transfer code and compared to the observed light curve from the explosion to $200\,\textrm{d}$ since explosion, and resulted in a very good match.

Similar to other types of SNe, $\gamma$-ray transfer simulations of Type II SNe can discriminate models with $ t_0 $ values that are inconsistent with observations. In the case of Type II SNe, $t_0$ is hard to determine, often being larger than the last observed epoch. Nevertheless, models with low $t_0$ values, such as some of the RSG profiles of \cite{hillier2019photometric}, are below all observed values or the lower limits of Type II SNe with observations exceeding $300\,\textrm{d}$ since explosion. Additionally, the observed values of some specific SNe, such as SN 1987A and SN2004et, is inconsistent with some proposed ejecta, such as the high mass ejecta of \cite{utrobin2009high} that deviates from the observed value (despite the good match to the observed light curve in the original work). The parameter $n$ and the amount of mixing is also much harder to determine than for SE SNe, but it is quite constraining for the case of SN 1987A. We find that the unmixed models of \citet{sukhbold2016core} do not fit SN 1987A and some mixing is required. This is supported by comparison to other observations as well \citep{blinnikov2000radiation,utrobin2005supernova}.

\subsection{Comparison with models -- $ET$}
\label{sec:Compare ET}

Using the derived $ET$ values and the analytical relations of \cite{Shussman2016}, we can attempt to estimate the radius of the progenitor stars, $R_*$. In their work, the explosions of RSG progenitors were simulated, and the resulting $ ET $ values were compared with:
\begin{equation}\label{ET_R}
ET\approx \beta v_\text{ej} M_\text{ej}R_*\approx 1.5\beta(E_\text{exp}M_\text{ej})^{1/2}R_*,
\end{equation}
where $E_\text{exp}$ is the explosion energy, $ M_\text{ej} $ is the ejected mass, {$ v_\text{ej} = \sqrt{2E_\text{exp}/M_\text{exp}} $ is the root mean square velocity of the ejecta, and $\beta$ is a scaling parameter. They found that a constant value of $ \beta=0.1 $ can describe the results of the simulations to $30$ per cent accuracy for progenitors with large envelope masses, but for progenitors with small envelope masses the scaling parameter decreases as the envelope mass decreases (if envelope mass is considered instead of total mass, then low envelope mass progenitors can also be well approximated). Although \cite{Shussman2016} were focused on Type IIP SNe, we can use Equation~\eqref{ET_R} to estimate the characteristic progenitor radius for other types of SNe in our sample. For Type IIb SNe, we assume $v_{\text{ej}}\approx4-7\times10^3\,\text{km}\,\text{s}^{-1}$, $M_{\text{ej}}\approx3-6\,M_\odot $ (corresponding to $E_\text{exp}\approx1-1.5\times 10^{51}\,\text{erg} $) and $ \beta=0.01$. The $\beta$ value roughly corresponds to the scaling factor of the low envelope mass progenitors of \cite{Shussman2016}. We find that the Type IIb SNe of our sample with observed $ET$ have progenitor star radii of $R_*\approx250-450\,R_\odot$. Previous studies found that the progenitor of SN 1993J was a yellow supergiant with a radius of $\approx600\,R_\odot$ \citep{Woosley1994ApJ...429..300W,piro2015using}. This is in a rough agreement with our results given the large uncertainties. For SN 1987A, we keep the original value of $\beta=0.1$, and assume $v_{\text{ej}}\approx2.5-3.5\times10^3\,\text{km}\,\text{s}^{-1}$ and $M_{\text{ej}}\approx12-18\,M_\odot$ (corresponding to $E_\text{exp}\approx1-1.5\times 10^{51}\,\text{erg}$). We find that the radius of the progenitor is $R_*\approx 38-60\,R_\odot$. This is in agreement with the BSG progenitor of SN 1987A, with an estimated radius of $R\approx43\pm14\,R_*$ \citep{Arnett1989ARA&A..27..629A}.
The procedure described above is not applicable for Type Ib/c SNe, for which the progenitors are expected to have very little hydrogen, and predicts much larger progenitor radii than expected. 

We next compare the observed $ET$ to the values predicted from the models in Sections~\ref{sec:SE SNe} and~\ref{sec:type II SNe}. The $ET$ values of the models are calculated by summing the internal energy of the ejecta profiles and multiplying with the time since explosion. This is only possible if the internal energies (or temperatures) and radii are provided at early enough times, such that the energy deposited from $^{56}$Ni decay is small compared to the initial energy. Only the ejecta of the BSG progenitors of \cite{dessart2019supernovae}, given in $ \sim1\,\textrm{d} $ since explosion, satisfy this condition (the other profiles were provided at very late times, so even the derived upper limits on $ET$, see below, are not useful). The $ET$ values of these ejecta are $0.5-1.7\times10^{54}\,\text{erg}\,\text{s}$, which are somewhat lower than the observed value of SN 1987A, $1.8-4.2\times10^{54}\,\text{erg}\,\text{s}$. 

For all SE SNe ejecta profiles that we obtained, the energy deposited from $ ^{56} $Ni decay is larger than the thermal energy at the times of the profiles, so we can only provide an upper limit for $ET$. For the SN 1993J profile of \cite{blinnikov1998comparative} the upper limit is $ ET<9.9\times10^{53}\,\text{erg}\,\text{s} $, which is consistent with the observed value of this SN, $0.1-2.2\times10^{54}\,\text{erg}\,\text{s} $. The upper limits for the SE SNe profiles of \cite{dessart2016inferring} are in the range of $ ET<0.8-5\times10^{53}\,\text{erg}\,\text{s} $. These values are lower than the observed $ET$ of our SE SNe sample. Confirmation of our measured values would thus place a stringent constrain on the models. 


\section{Discussion and Conclusions}
\label{sec:discussion}

We have constructed (Section~\ref{sec:bolometric}) and analysed (Sections~\ref{sec:fit}, \ref{sec:results}) bolometric light curves to constrain the $\gamma$-ray deposition history of several types of SNe (Figure~\ref{fig:t0Ni}). We have recovered the tight range of $\gamma$-ray escape time, $t_0\approx30-45\,\textrm{d}$, for Type Ia SNe \citep{WygodaI2019}, and have been able to correct a small $\approx10$ per cent systematic error in the $t_0$ values of these SNe. We have found a new tight range $t_0\approx80-140\,\textrm{d}$, where different subtypes of this class have quite similar $t_0$ values despite their different spectral characterization. Type Ib/c have slightly higher $t_0$ values compared with Type IIb, and a non-negligible $ET$ value (Figure~\ref{fig:ETNi}). We treat these $ET$ measurements as tentative, since they could be an artifact of systematic errors. Type IIP SNe are clearly separated from other SNe types with $t_0\gtrsim400\,\textrm{d}$, and there is a possible negative correlation between $t_0$ and the synthesized $^{56}$Ni mass. We have found that the typical masses of the synthesized $^{56}$Ni in SE SNe are larger than those in Type IIP SNe, in agreement with \citet{Kushnir2015progenitors}, a fact that disfavours progenitors with the same initial mass range for these explosions \citep[see detailed discussion in][]{Kushnir2015progenitors}. Instead, the progenitors of SE SNe explosions could be massive Wolf-Rayet stars, which are predicted to yield strong explosions with low ejecta masses according to the collapse-induced thermonuclear explosions mechanism for core-collapse SNe \citep{KushnirKatz2015failure,Kushnir2015thermonuclear,Kushnir2015progenitors}. 

We have applied a simple $ \gamma$-ray radiation transfer code to calculate the $\gamma$-ray deposition histories of models from the literature (Section~\ref{sec:simulations}), and we have shown that the observed histories are a powerful tool to constrain models. 

The sample of core-collapse SNe is quite limited, although in principle it is straightforward to increase the size of this sample. This would require a collaborated effort to acquire UV to IR coverage from early to late times, with the IR observation being of a special importance, reaching up to $\sim40$ per cent at some phases. Accurate bolometric observations might help to determine the $ET$ values of SE SNe, which would allow stringent constraints on the progenitor system. Additional Type IIP SNe might help determine whether $t_0$ and $M_{\rm{Ni}56}$ are correlated, which would have consequences for the explosion mechanism. 

\section*{Acknowledgements}

We thank Boaz Katz, Nahliel Wygoda, Eran Ofek and Subo Dong for useful discussions. We thank Stan Woosley, Luc Dessart, Sung-Chul Yoon, Tuguldur Sukhbold, Sergei Blinnikov, and Victor P. Utrobin for sharing their ejecta profiles with us. DK is supported by the Israel Atomic Energy Commission --The Council for Higher Education --Pazi Foundation --and by a research grant from The Abramson Family Center for Young Scientists. 
The data underlying this article are available in the article and in its online supplementary material.



\bibliographystyle{mnras}
\bibliography{bibliography} 




\appendix
\section{The case of SN 2011dh}
\label{app:2011dh}

The Type IIb SN 2011dh is one of the best observed SE SN in the last decade. It has been under extensive observations for a period of $ 2\,\text{yr} $, from the UV to the medium IR \citep{ergon2015type}. Although IR observations are available up to $\sim400\,\textrm{d}$ from explosion, our fitting procedure works well only up to $\sim150\,\textrm{d}$ from explosion, with best-fitting parameters $M_{\text{Ni}56}\approx0.059\,M_\odot$, $ t_0\approx100\,\textrm{d}$ and $n\approx2.8$, well within the range of Type IIb typical values (see Table~\ref{tab:results}). At later epochs, the optical flux declines faster than the well-observed Type IIb SN 1993J and SN 2008ax \citep{ergon2015type}, and so the bolometric luminosity declines faster as well. Applying the fitting procedure for the entire time range results in a very poor fit, with an extremely high value of the $ n\approx9 $, far beyond any other SNe (see Figure \ref{fig:2011dh}). \cite{ergon2015type} suggested that dust formation in the ejecta might cause this phenomenon, supported by the fractional increase of the mid-IR luminosity at these epochs. In this scenario, the missing energy is emitted in wavelengths above the \textit{Spitzer} $4.5\,\mathrm{\mu}\text{m}$ band, and are not taken into account.

\begin{figure}
	\includegraphics[width=\columnwidth]{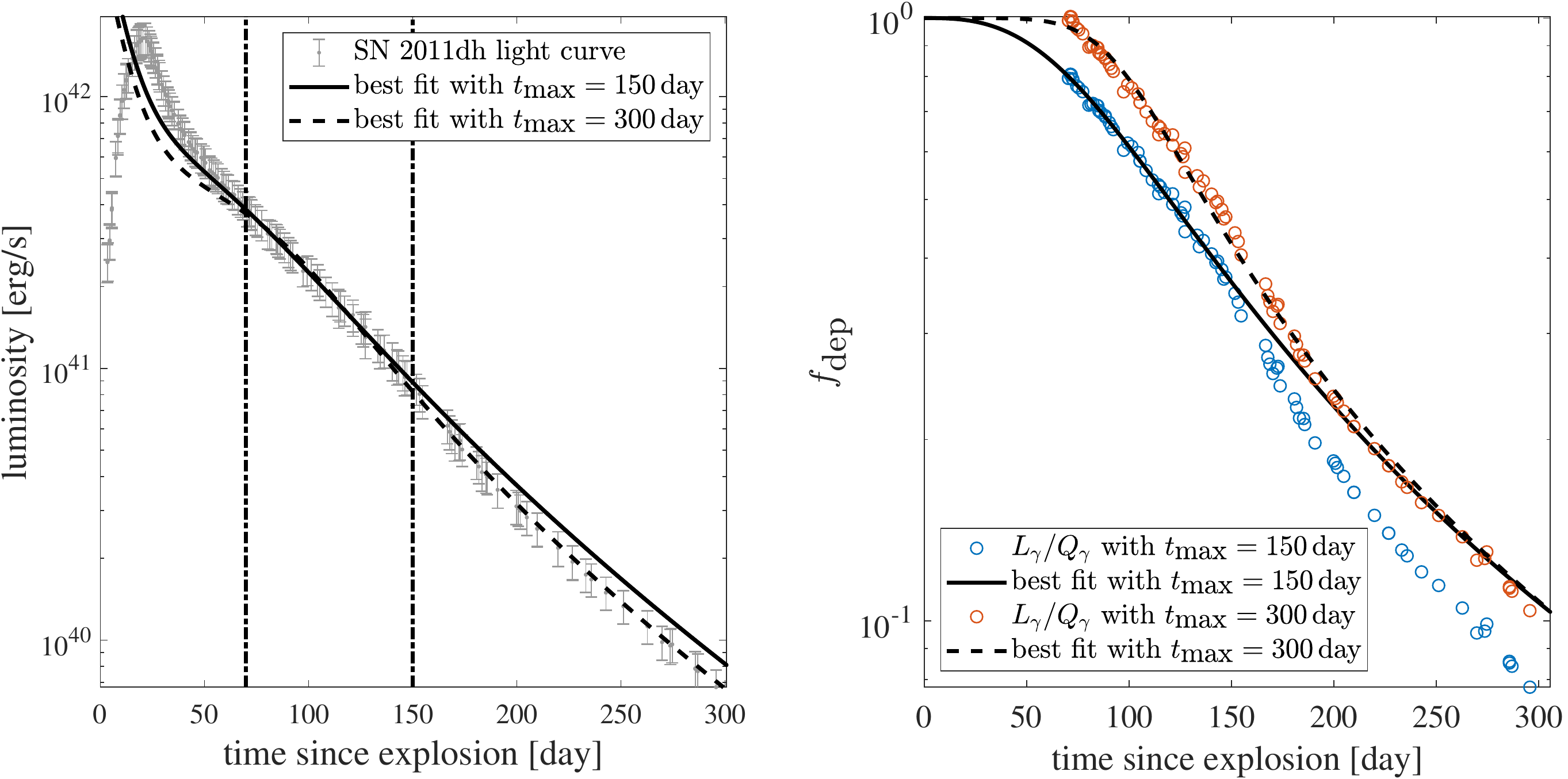}
	\caption{Same as Figure \ref{fig:fits}, for the case of SN 2011dh, comparing the best-fitting results of the shortened time range, $ t_\text{max}=150\,\textrm{d} $, and the usual time range of SE SNe, $ t_\text{max}=300\,\textrm{d} $ (starting from $ t_\text{min}=70\,\textrm{d} $). left-hand side panel: the observed bolometric light curve (grey symbols), the best-fitting model with the shortened time range (solid black line) and the best-fitting model with the usual time range (dashed line). The vertical dotted lines indicate $ t_\text{min}$ of both time ranges and $ t_\text{max} $ of the shortened time range. right-hand side panel: the $ \gamma $-ray deposition fraction $L_{\gamma}/Q_{\gamma}$ for the shortened time range (blue symbols), its best-fitting deposition model (solid line), the deposition fraction for the usual time range (orange symbols) and its best-fitting deposition model (dashed line).} 
	\label{fig:2011dh}
\end{figure}

\section{Bolometric light curves and the results of the integral method for the entire sample}
\label{app:deposition plots}

We present here, in Fig. \ref{fig:allfits}, the best-fitting bolometric light curves and the results of the integral method  for the entire sample, in the same format as in Figure~\ref{fig:fits}. The order is the same as in Table~\ref{tab:results}: Type Ia, Type Ib/c, Type IIb, Type IIP and, finally, SN 1987A.
\begin{figure*}
	\includegraphics[width=0.67\textwidth]{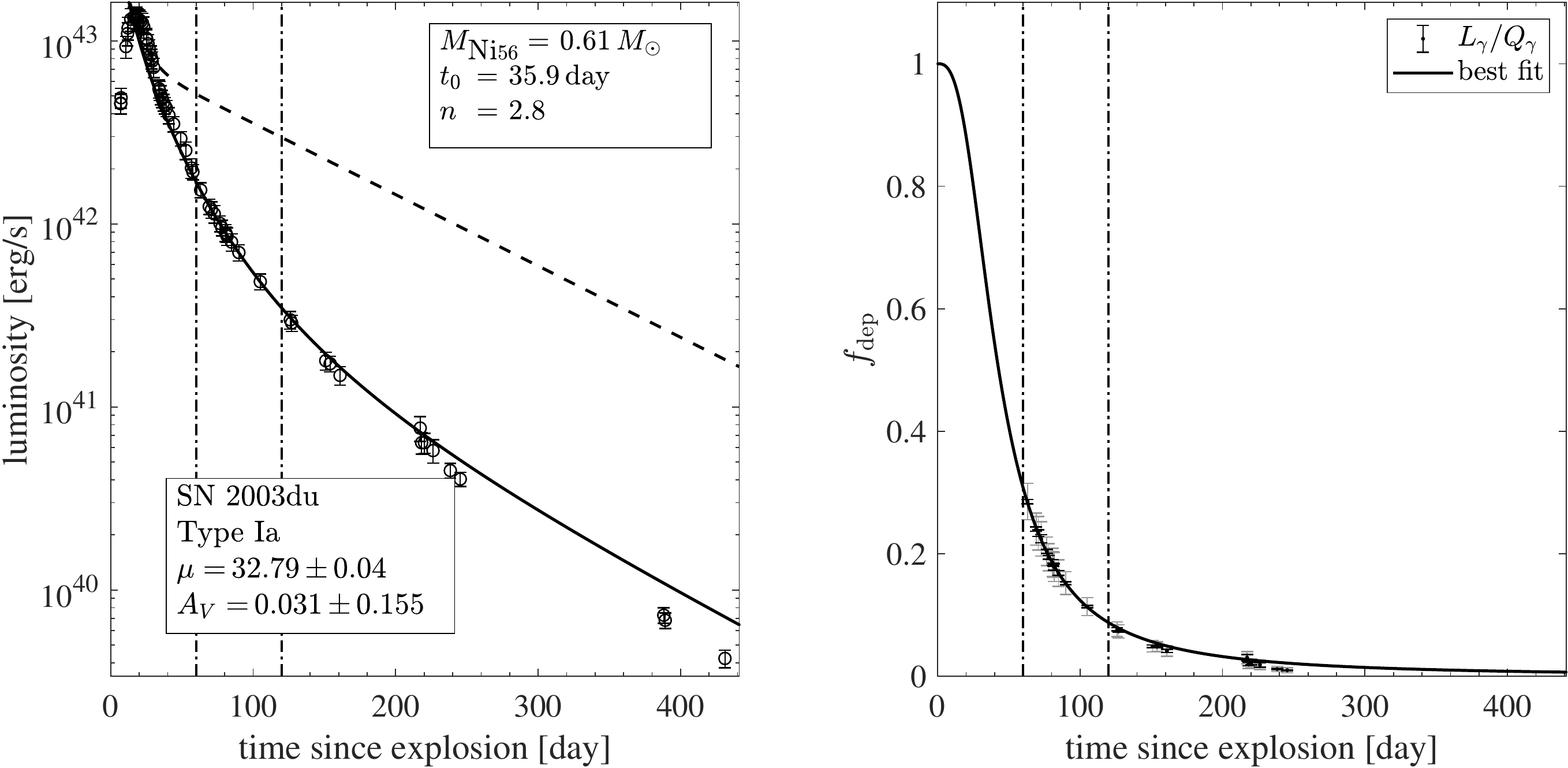}	
		
	\vspace{0.25 cm}
	\includegraphics[width=0.67\textwidth]{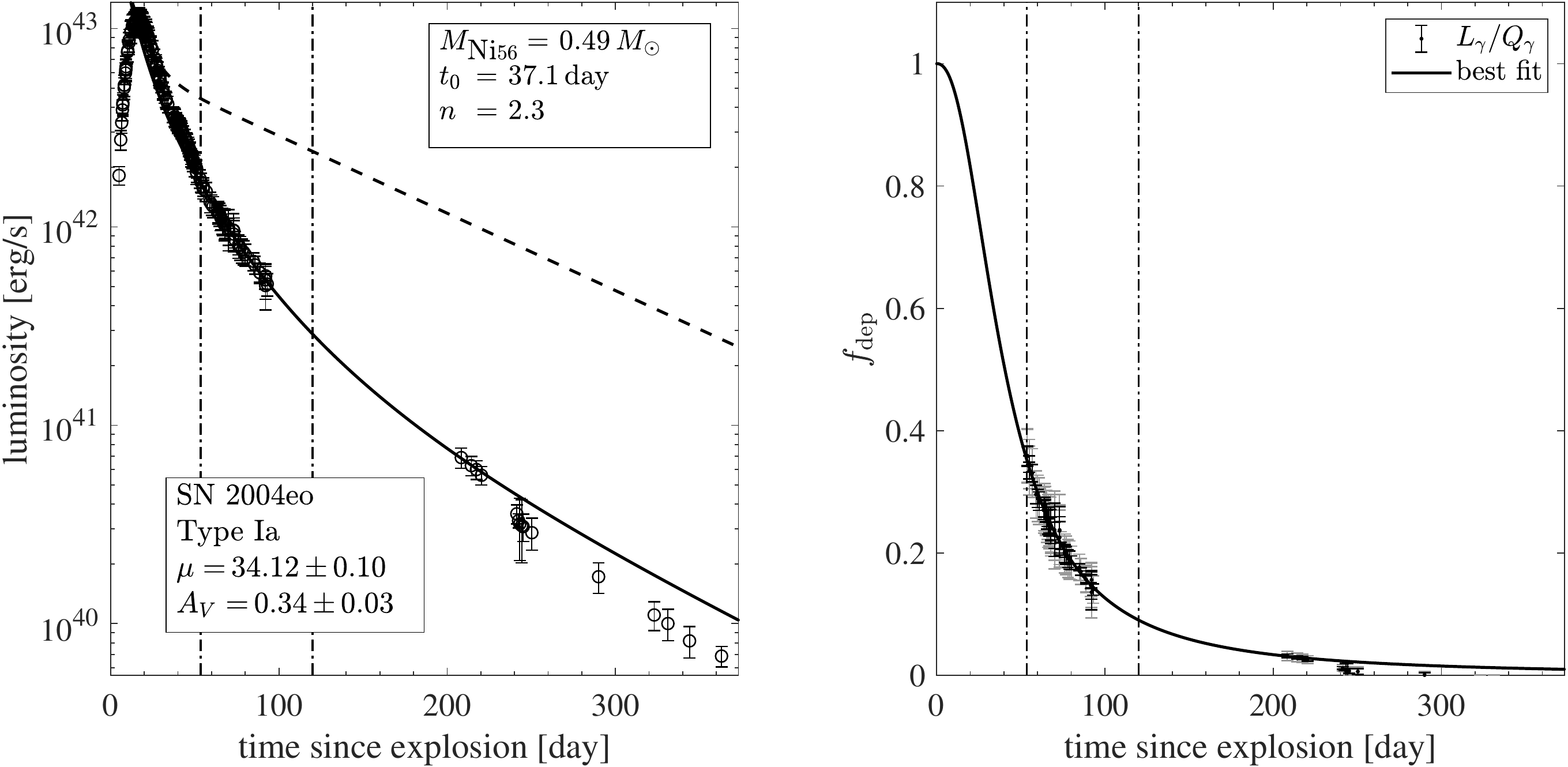}
	
	\vspace{0.25 cm}
	\includegraphics[width=0.67\textwidth]{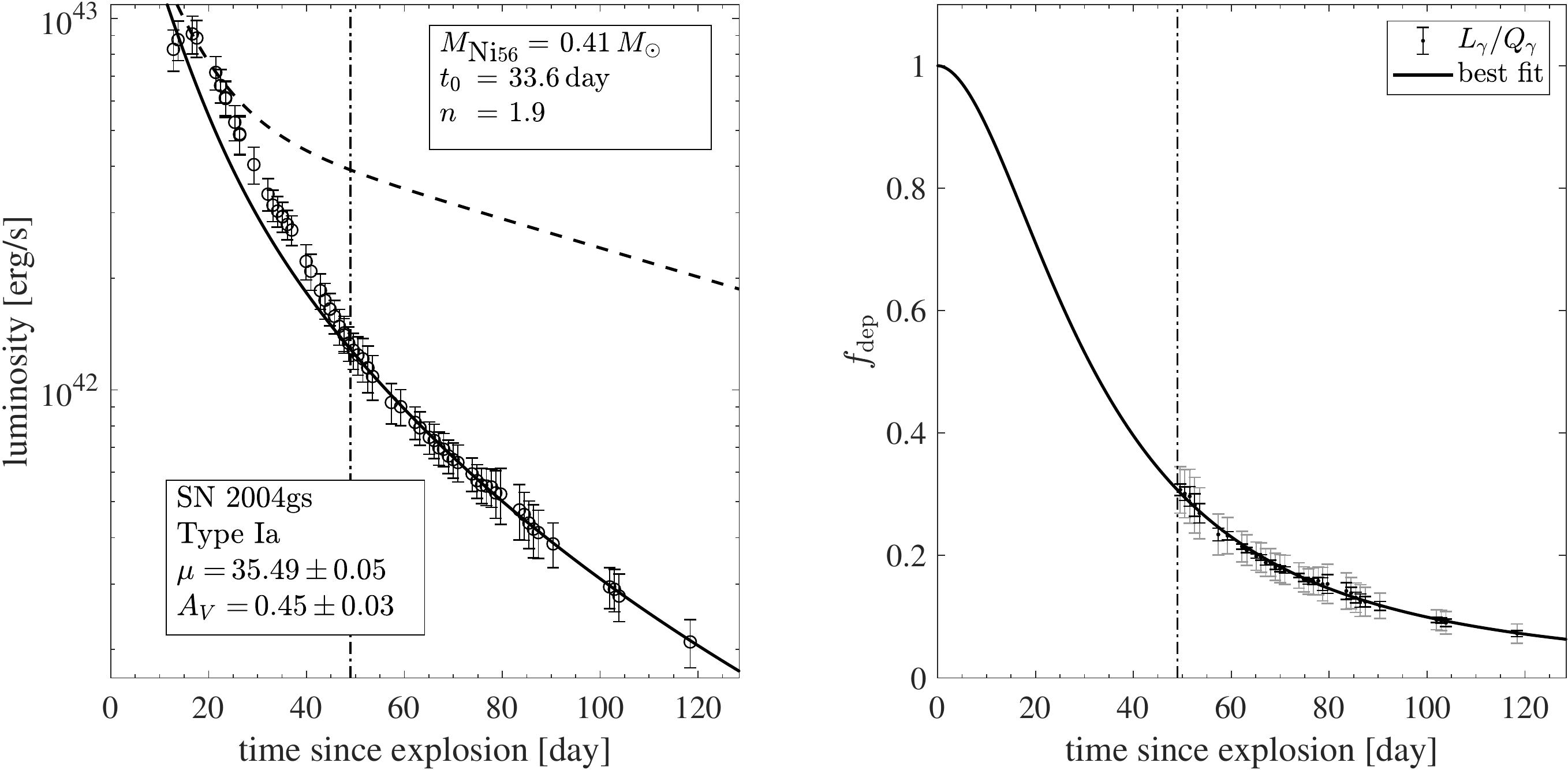}
	
	\vspace{0.25 cm}
	\includegraphics[width=0.67\textwidth]{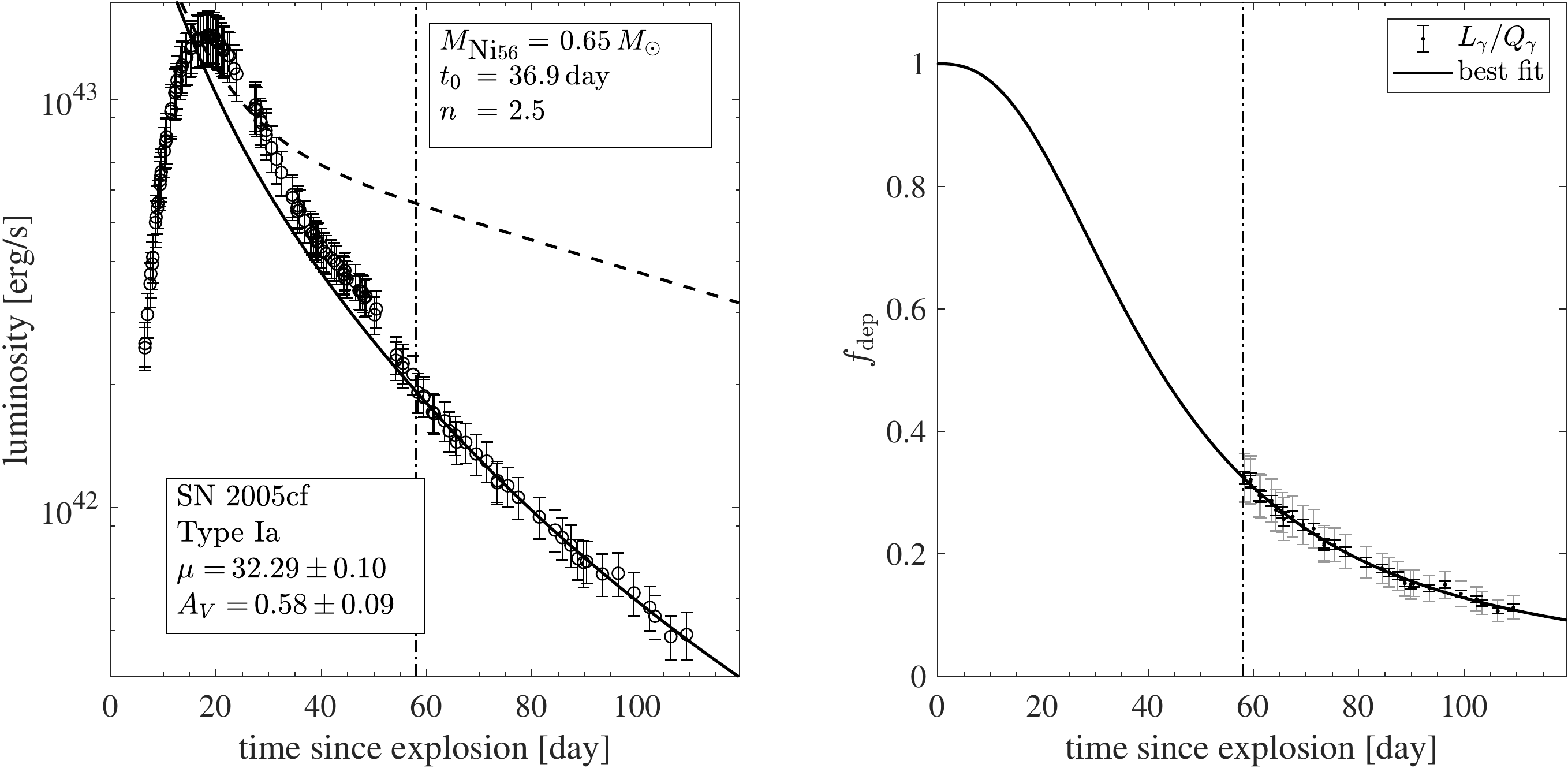}
	\caption{Same as Figure~\ref{fig:fits} for the full SNe sample.}
	 
	\label{fig:allfits}
\end{figure*}
\begin{figure*} 		
\ContinuedFloat	
	\includegraphics[width=0.67\textwidth]{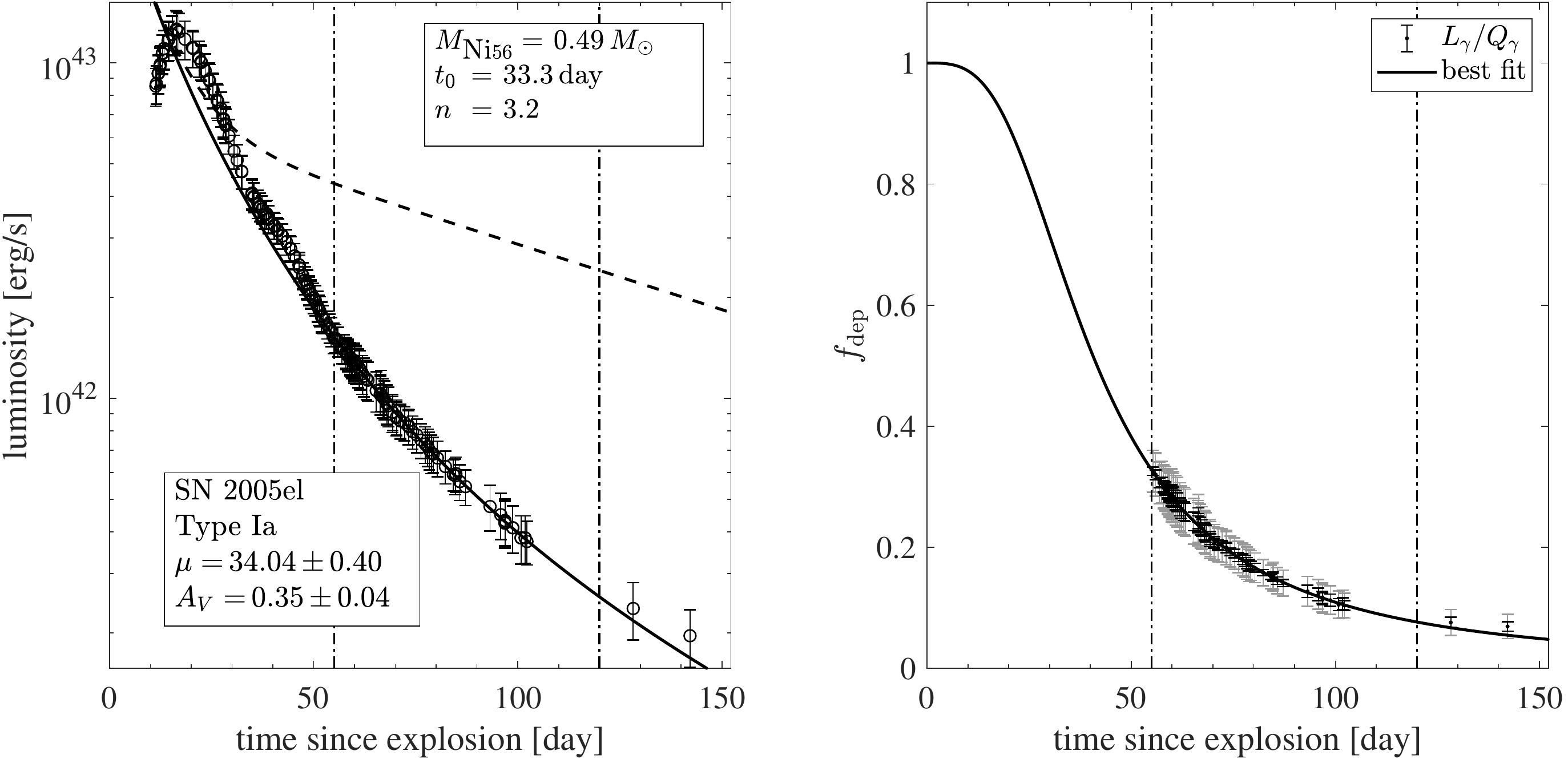}	
				
	\vspace{0.25 cm}
	\includegraphics[width=0.67\textwidth]{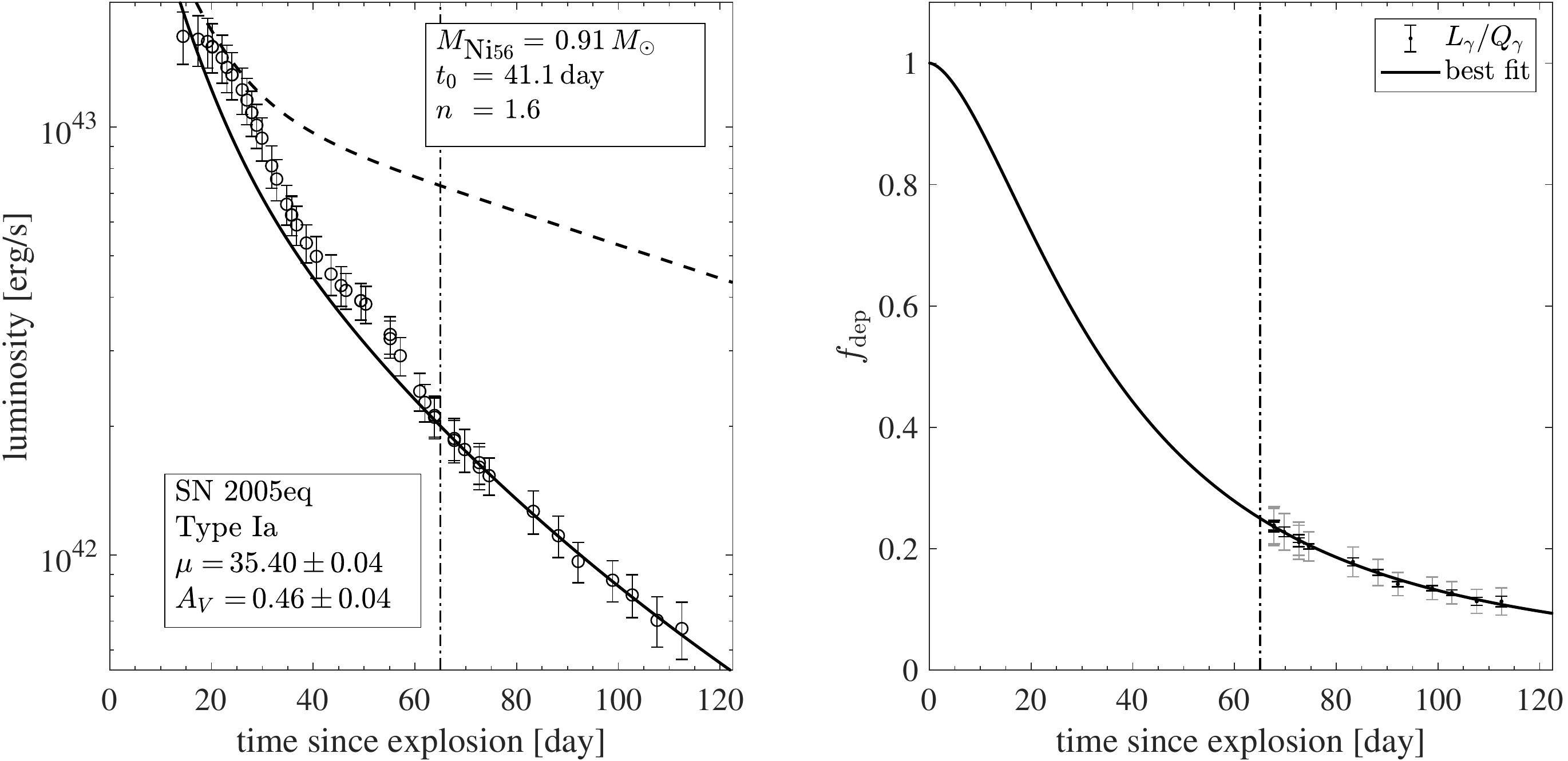}

	\vspace{0.25 cm}
	\includegraphics[width=0.67\textwidth]{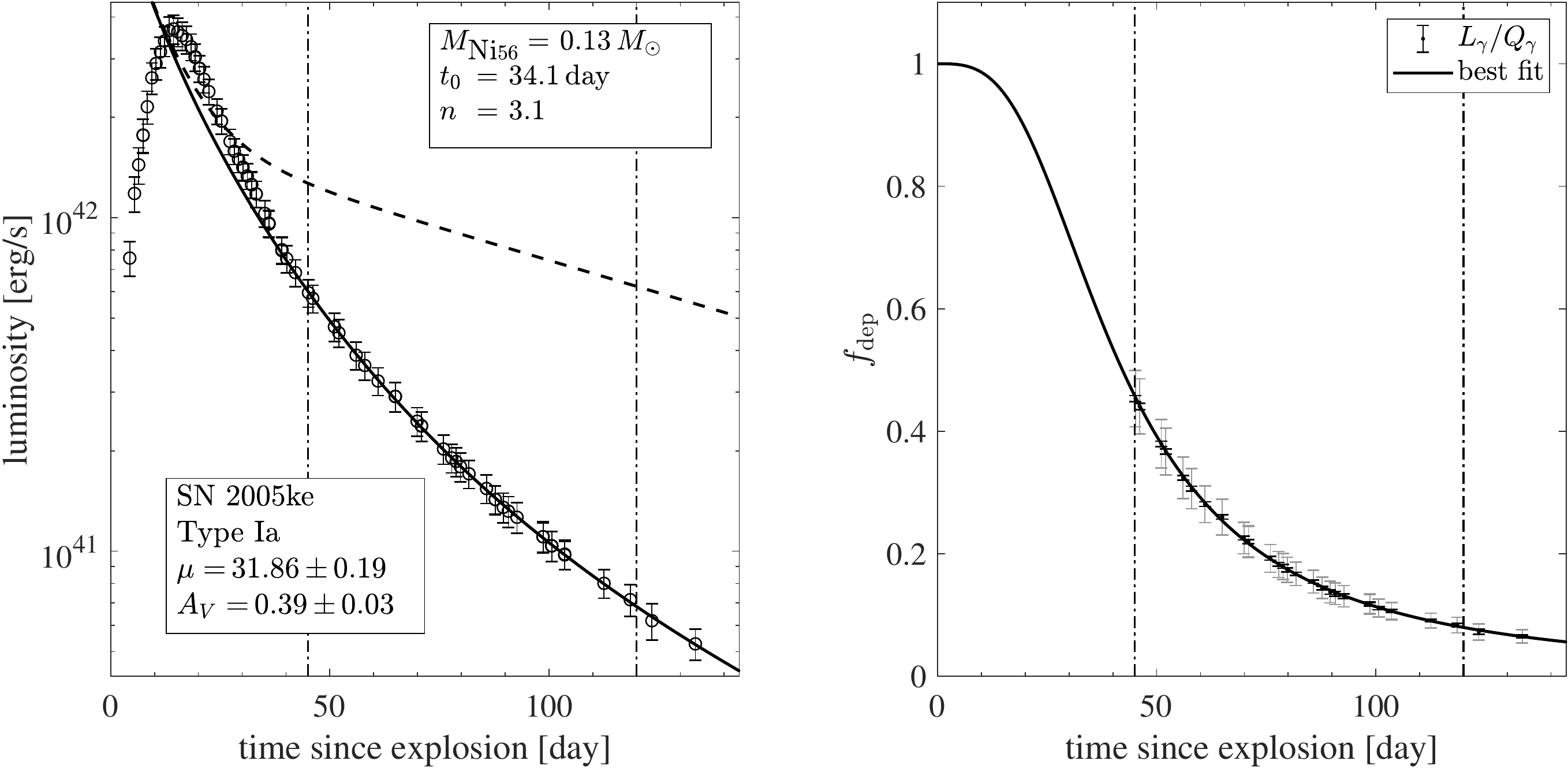}

	\vspace{0.25 cm}
	\includegraphics[width=0.67\textwidth]{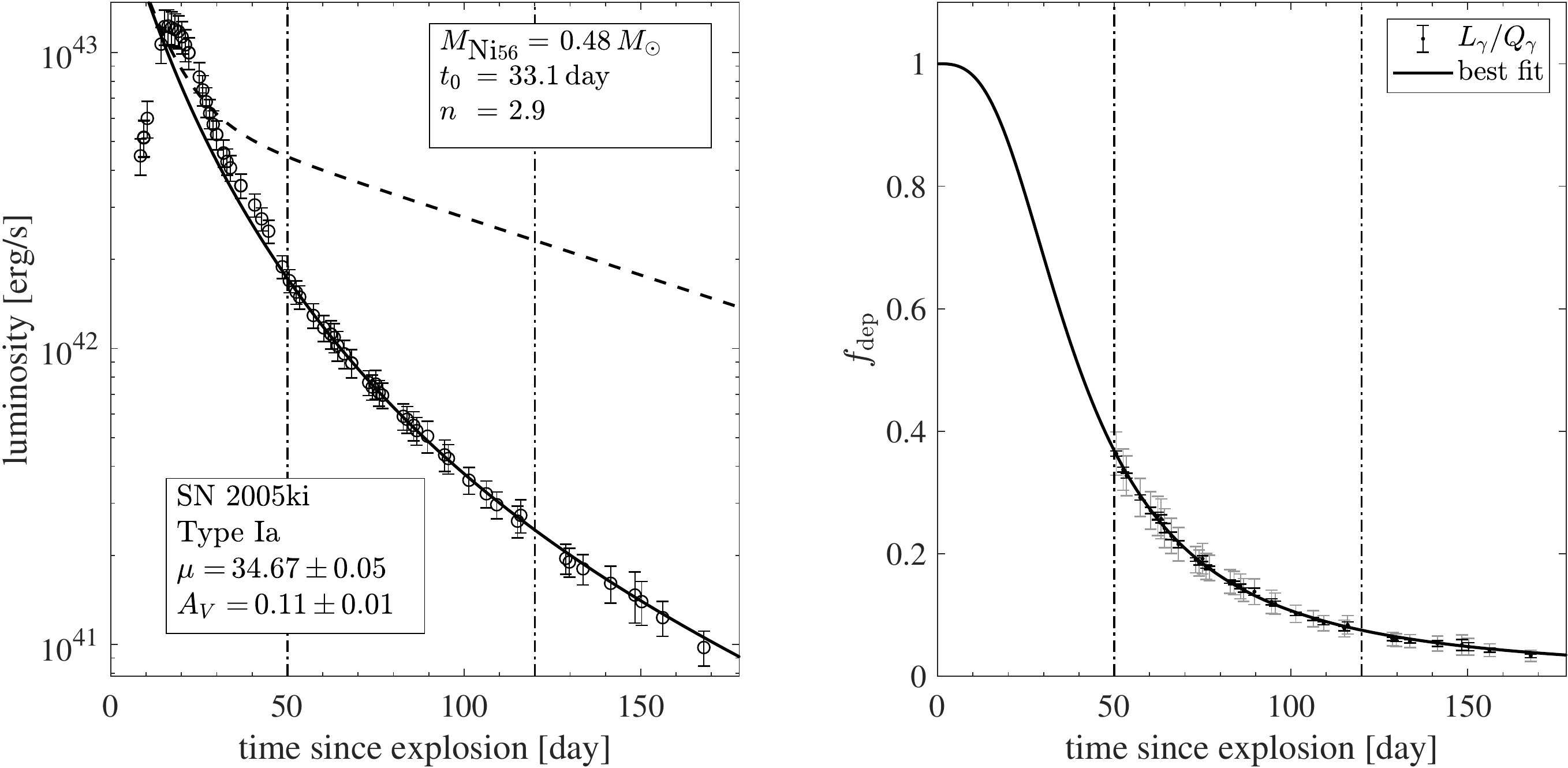}
		\caption{(continued) Same as Figure~\ref{fig:fits} for the full SNe sample.}
\end{figure*}
\begin{figure*} 	
\ContinuedFloat	
	\includegraphics[width=0.67\textwidth]{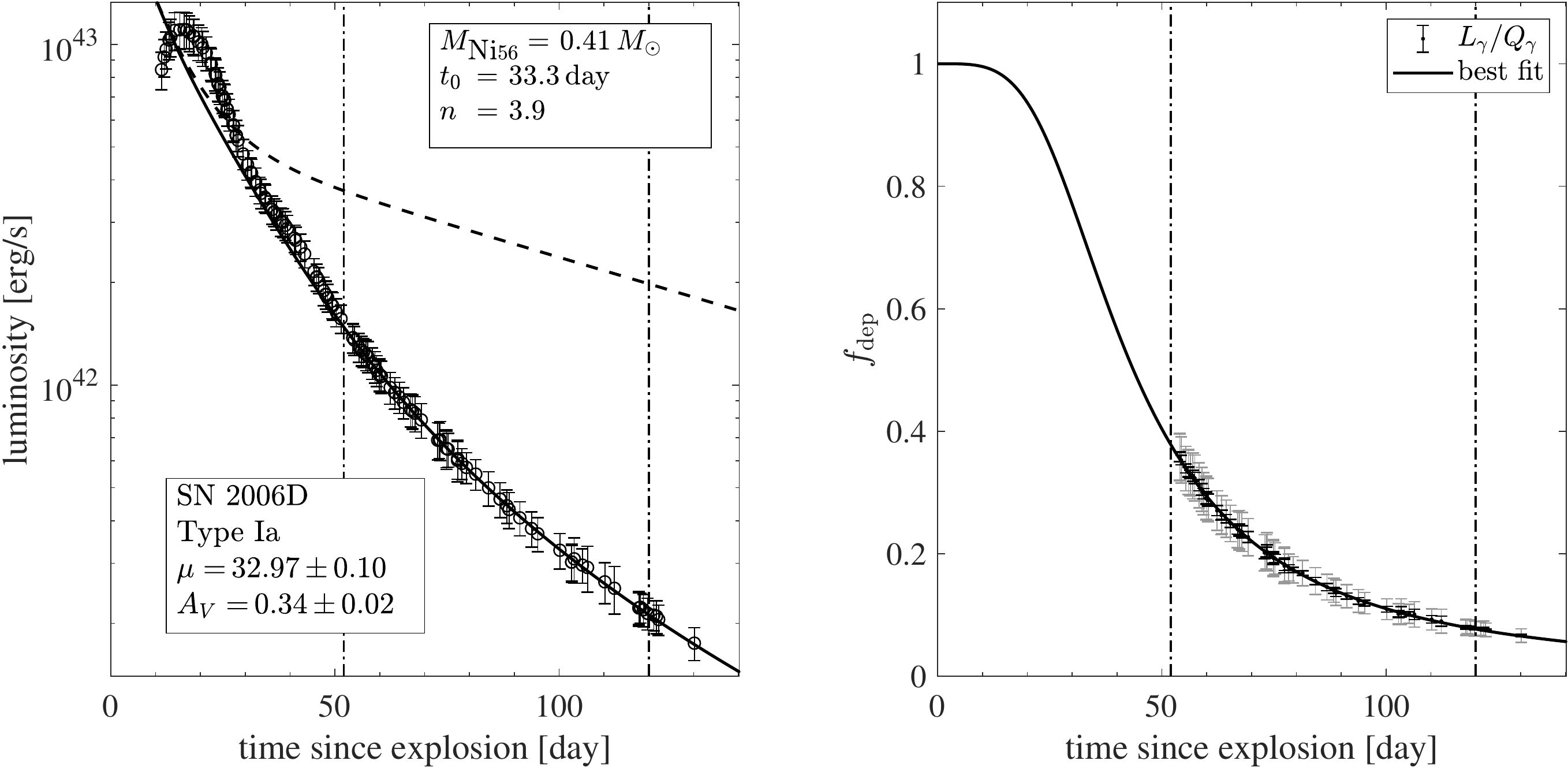}
		 	
	\vspace{0.25 cm}
	\includegraphics[width=0.67\textwidth]{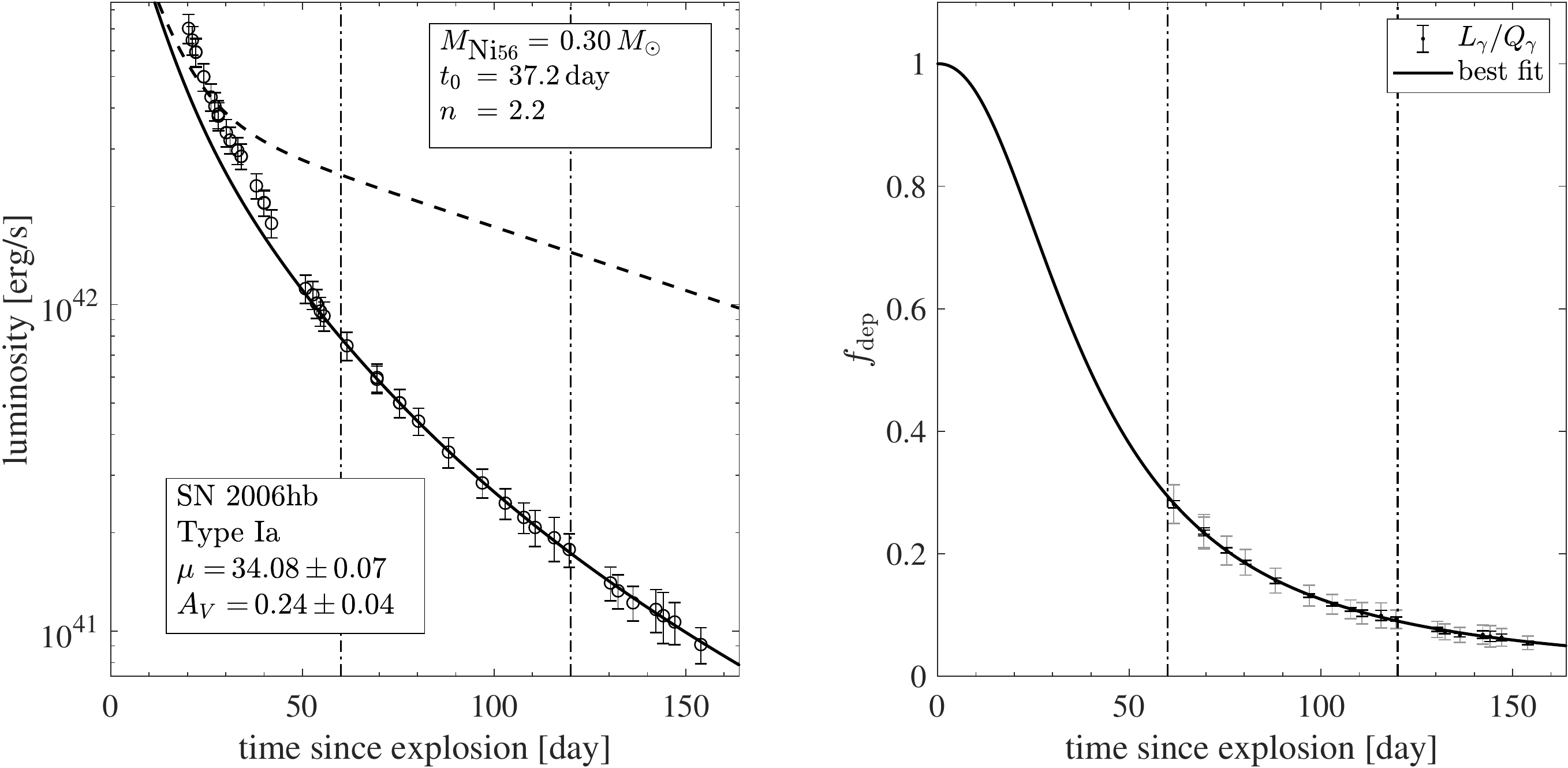}
	
	\vspace{0.25 cm}
	\includegraphics[width=0.67\textwidth]{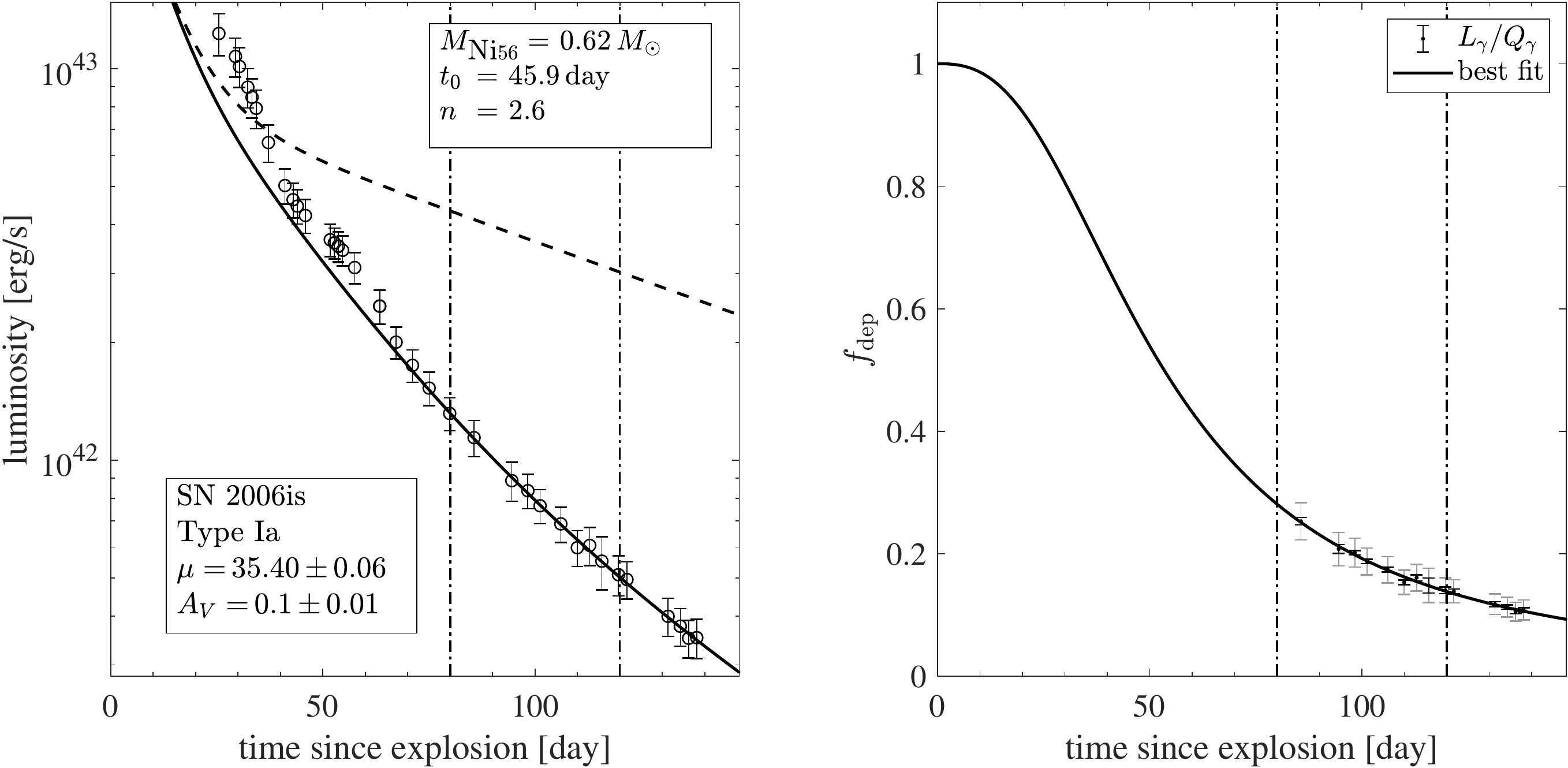}
	
	\vspace{0.25 cm}
	\includegraphics[width=0.67\textwidth]{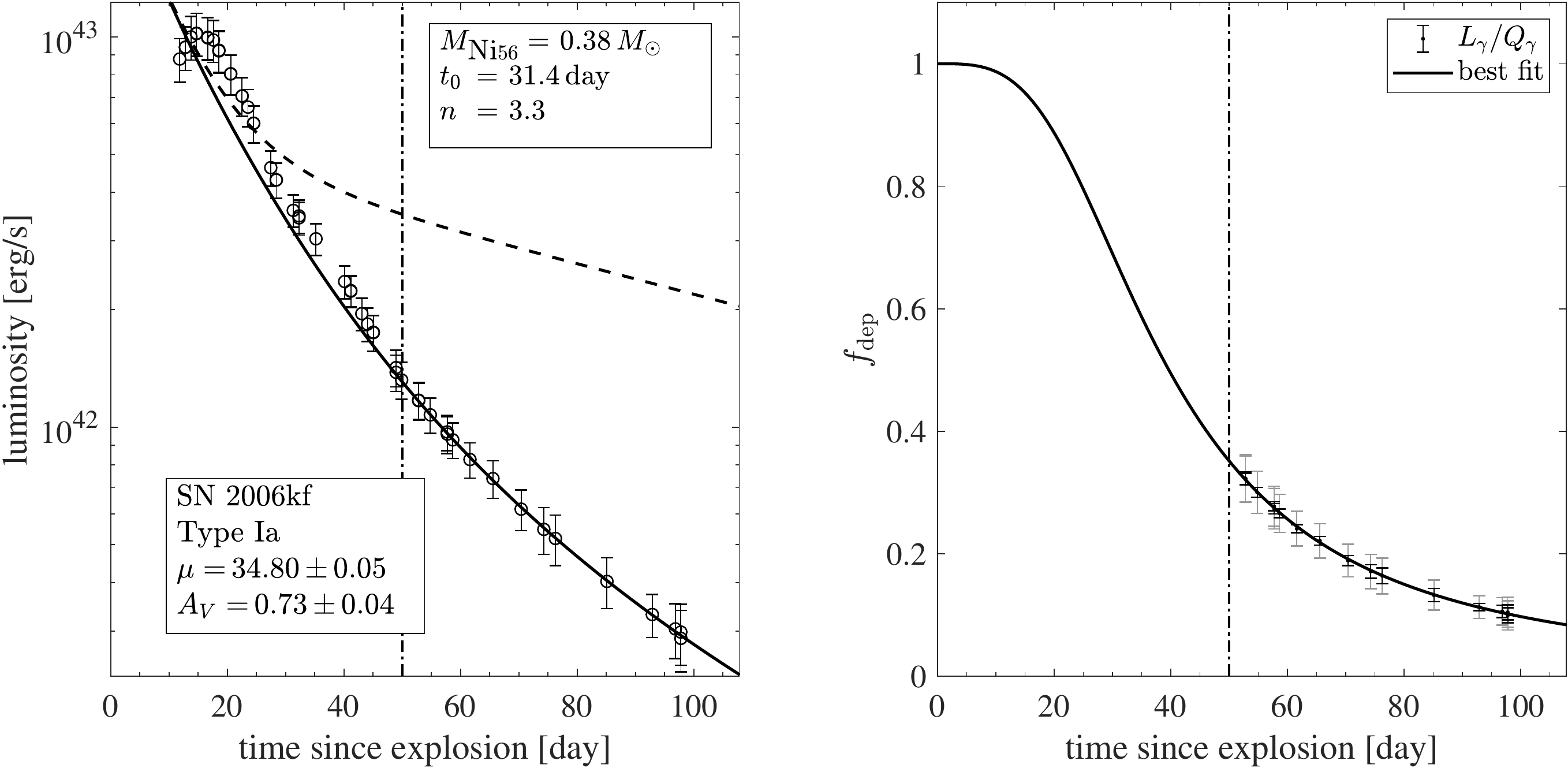}
	\caption{(continued) Same as Figure~\ref{fig:fits} for the full SNe sample.}
\end{figure*}
\begin{figure*} 	
\ContinuedFloat			
	\includegraphics[width=0.67\textwidth]{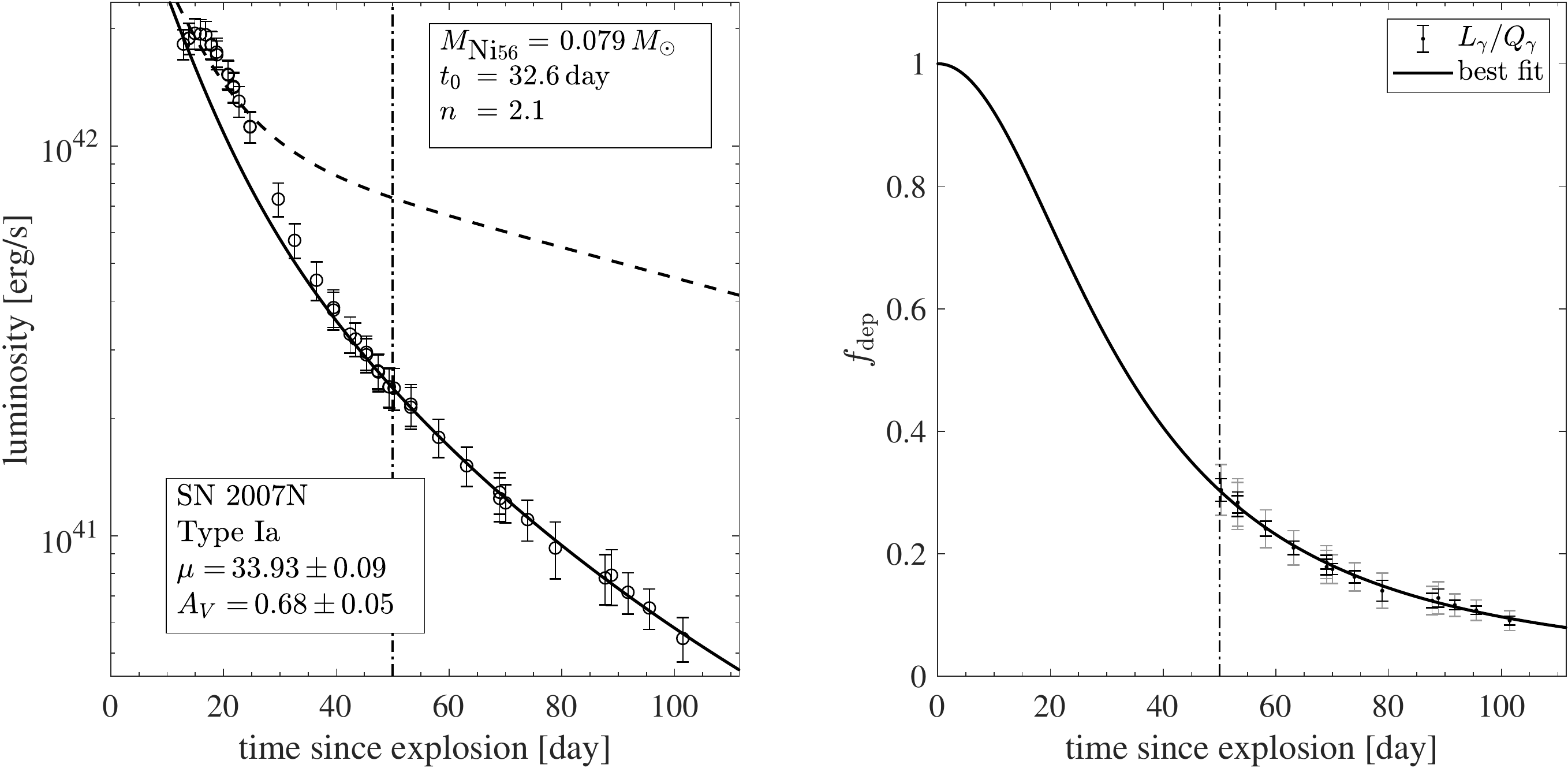}
		 	
	\vspace{0.25 cm}
	\includegraphics[width=0.67\textwidth]{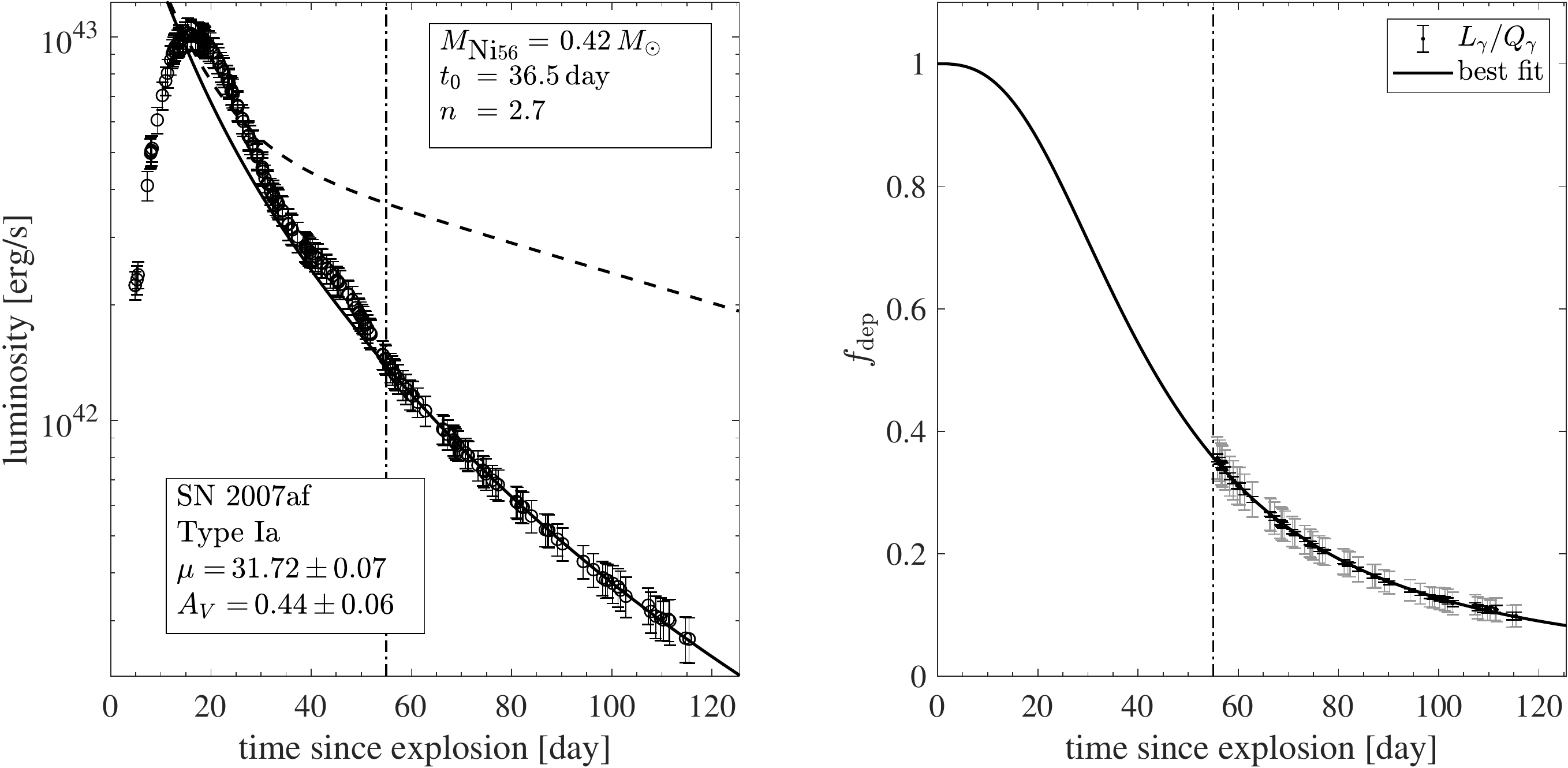}	
			
	\vspace{0.25 cm}		
	\includegraphics[width=0.67\textwidth]{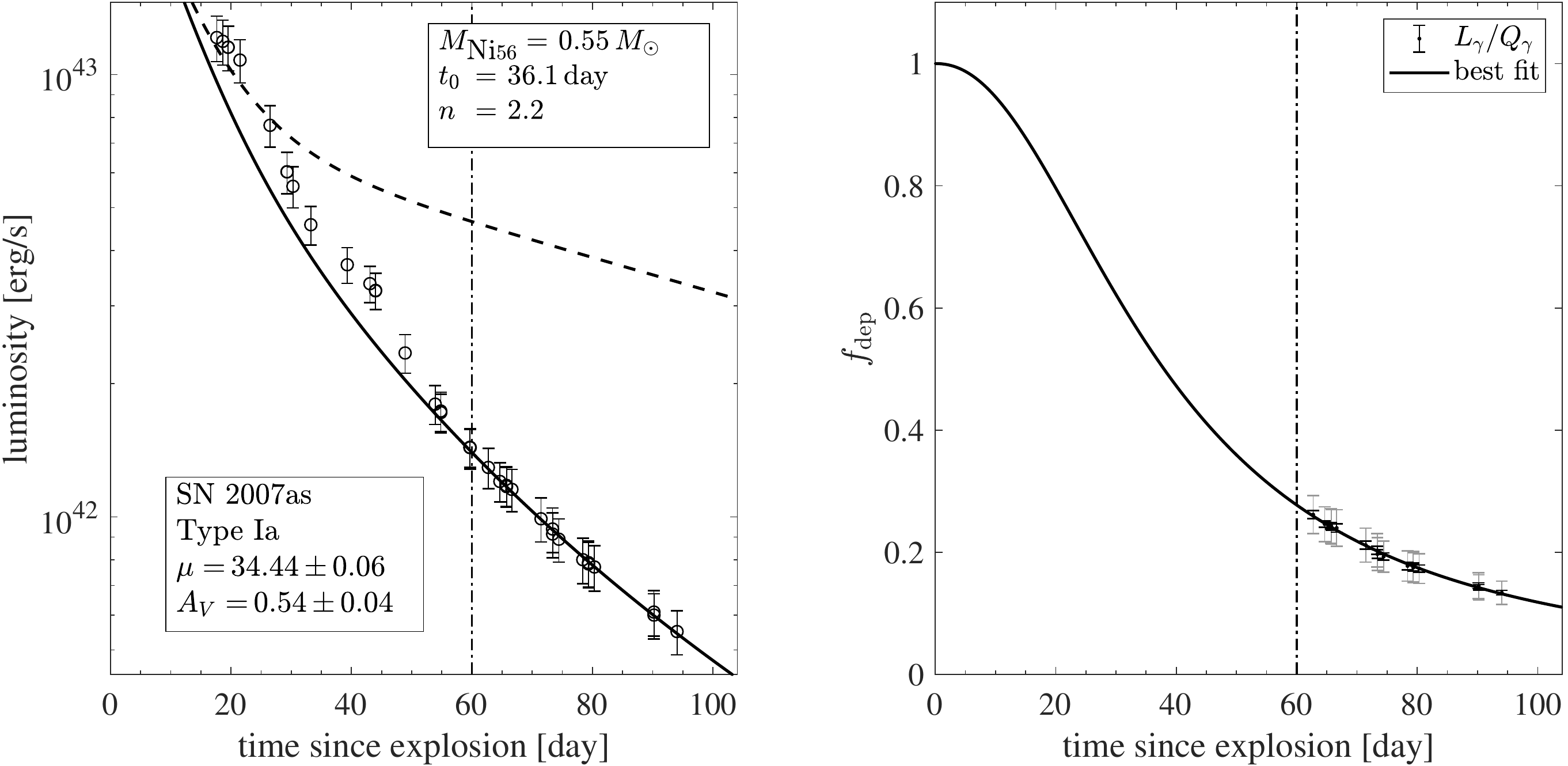}

	\vspace{0.25 cm}
	\includegraphics[width=0.67\textwidth]{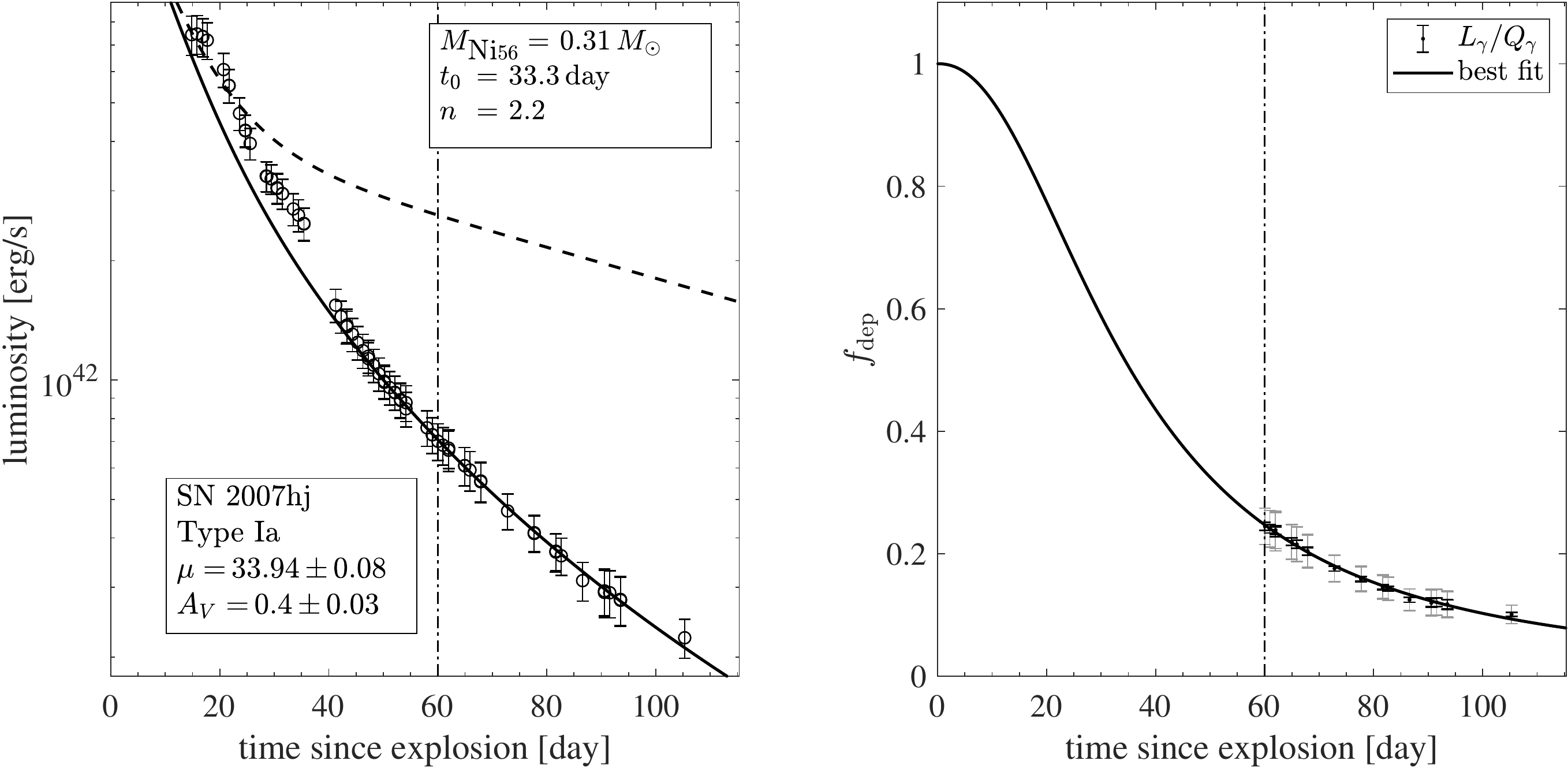}
	\caption{(continued) Same as Figure~\ref{fig:fits} for the full SNe sample.}
\end{figure*}
\begin{figure*} 
\ContinuedFloat		
	\includegraphics[width=0.67\textwidth]{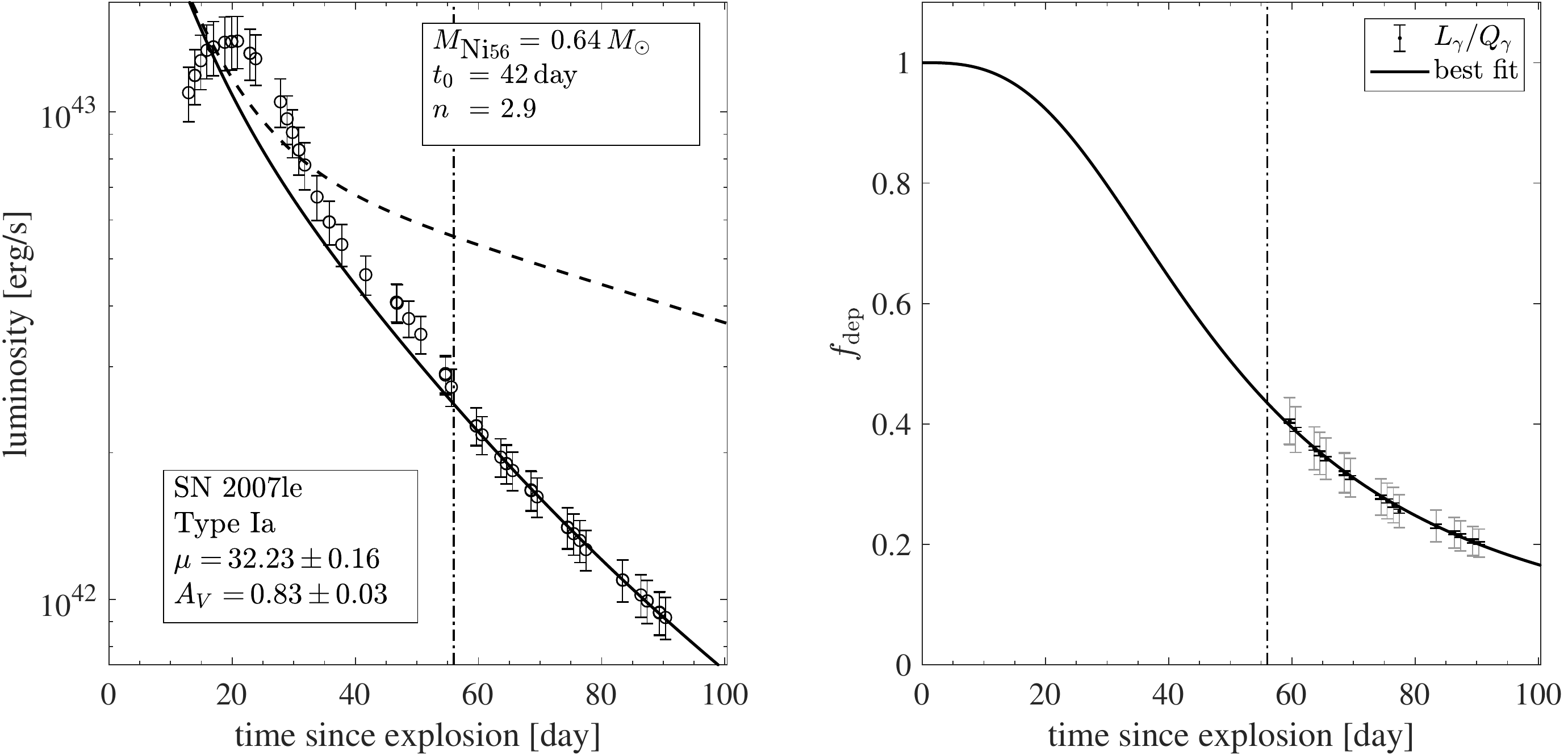}
 				
 	\vspace{0.25 cm}	
	\includegraphics[width=0.67\textwidth]{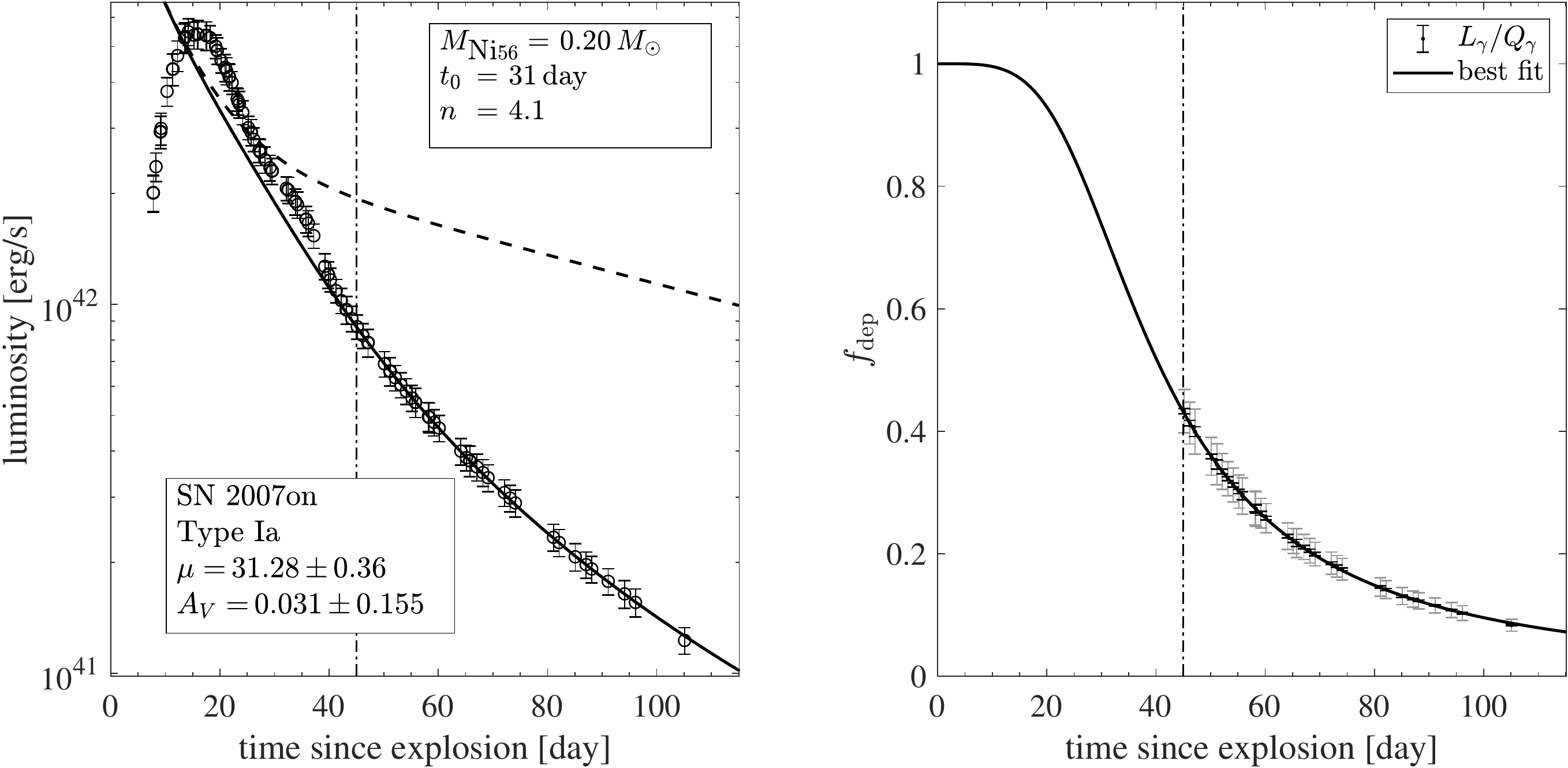}
 	
	\vspace{0.25 cm}
	\includegraphics[width=0.67\textwidth]{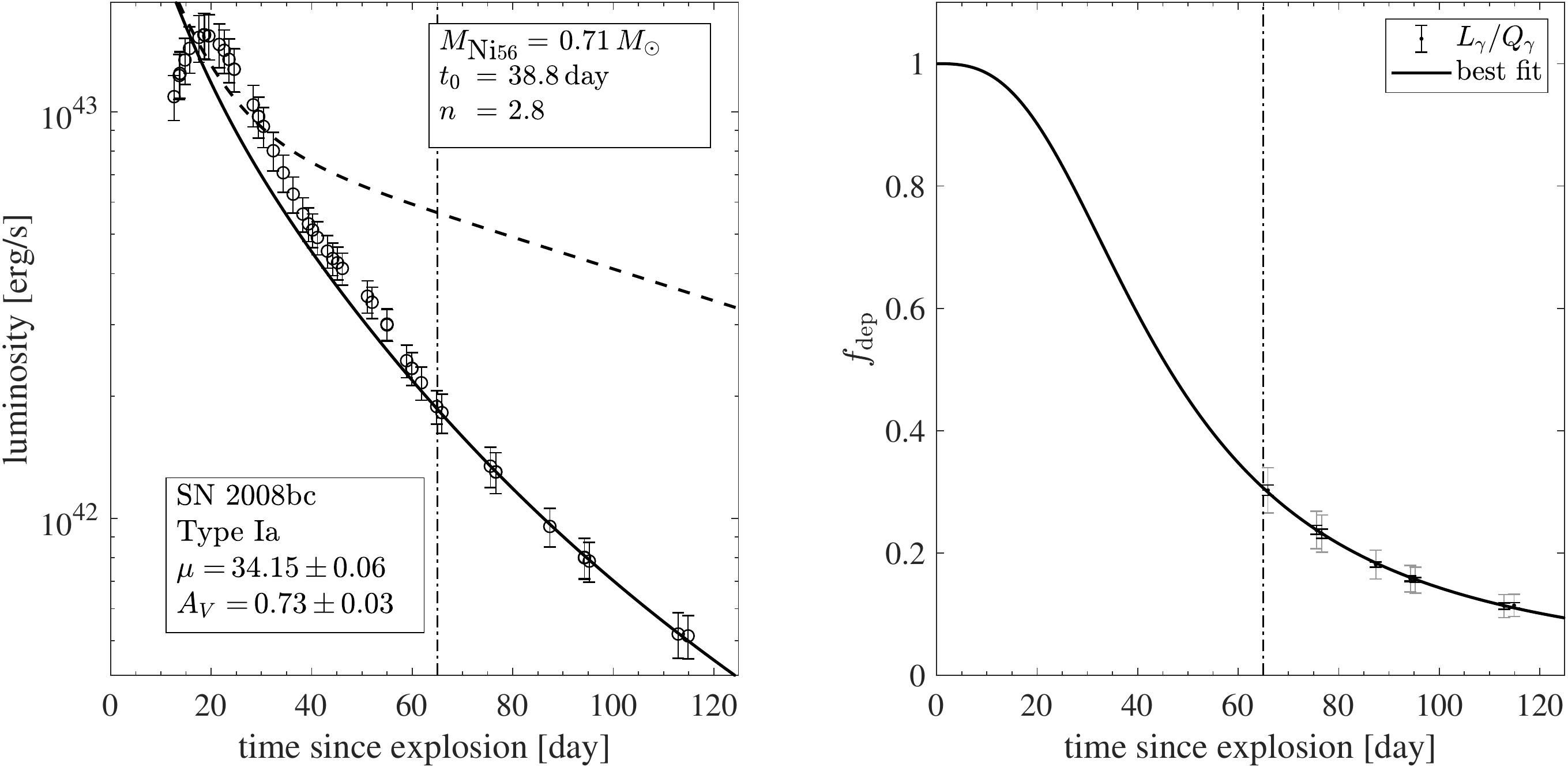}
	
	\vspace{0.25 cm}		
	\includegraphics[width=0.67\textwidth]{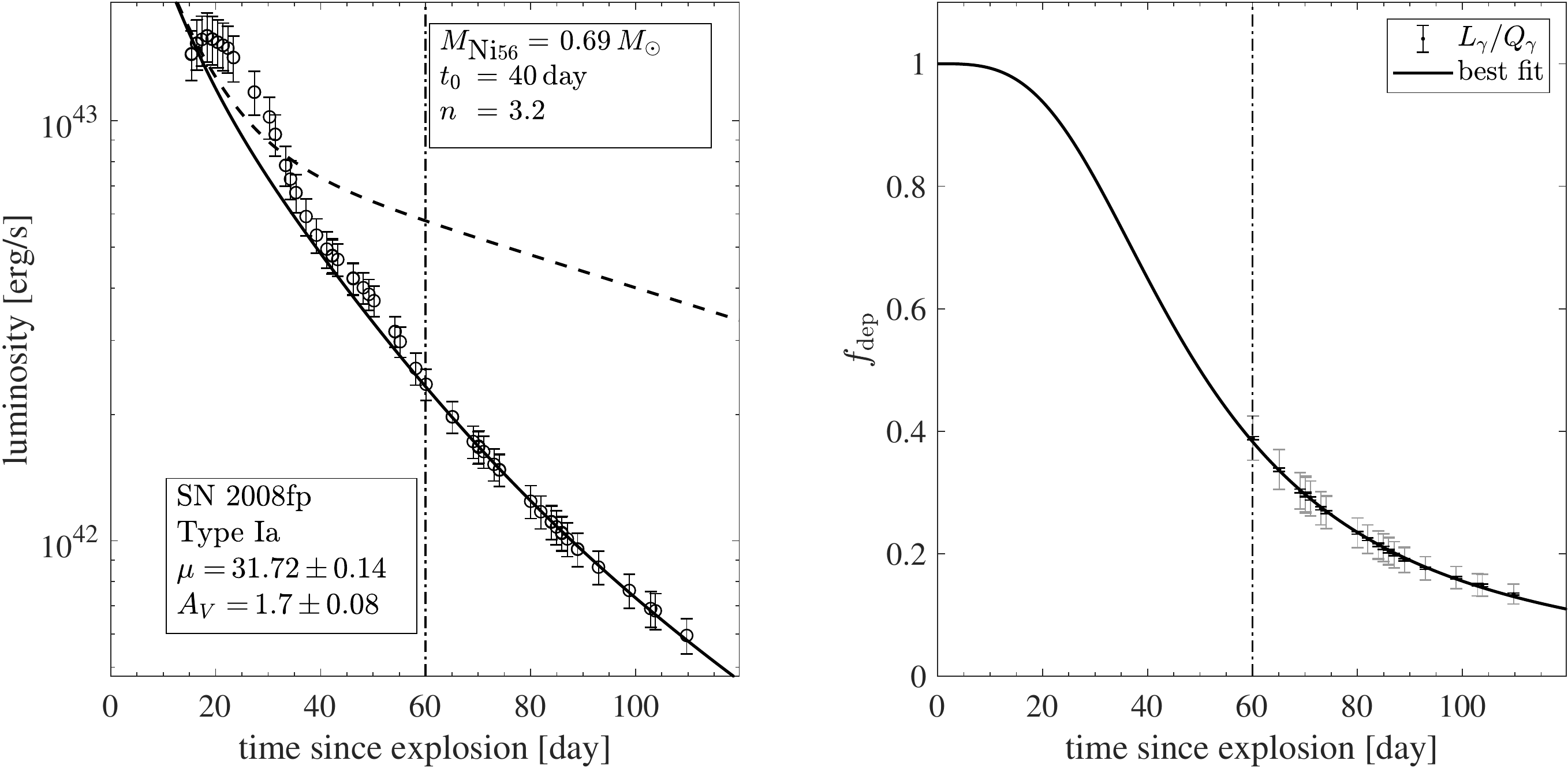}		
	\caption{(continued) Same as Figure~\ref{fig:fits} for the full SNe sample.}		
\end{figure*}
\begin{figure*} 
\ContinuedFloat	
	\includegraphics[width=0.67\textwidth]{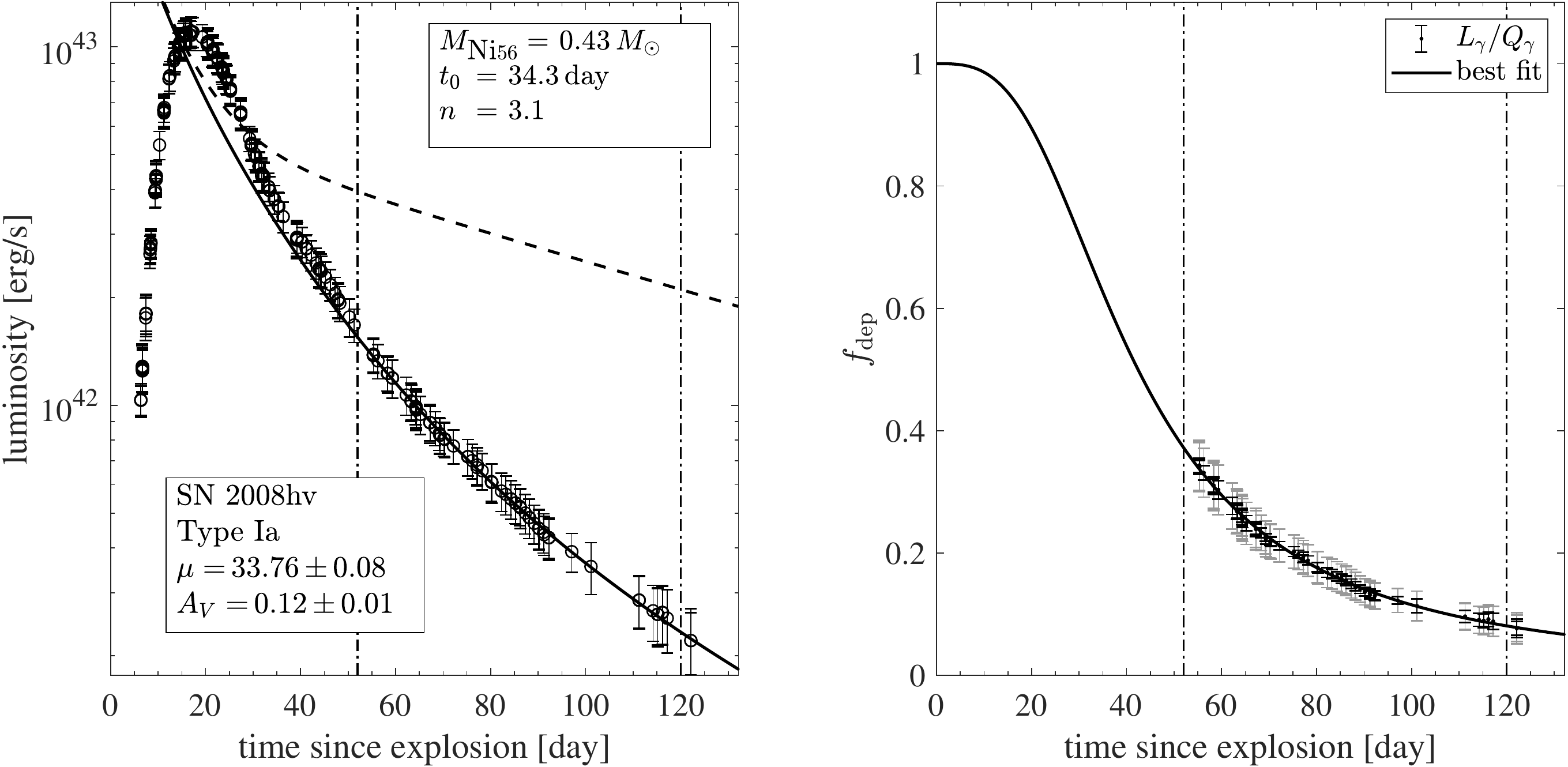}
 	
	\vspace{0.25 cm}
	\includegraphics[width=0.67\textwidth]{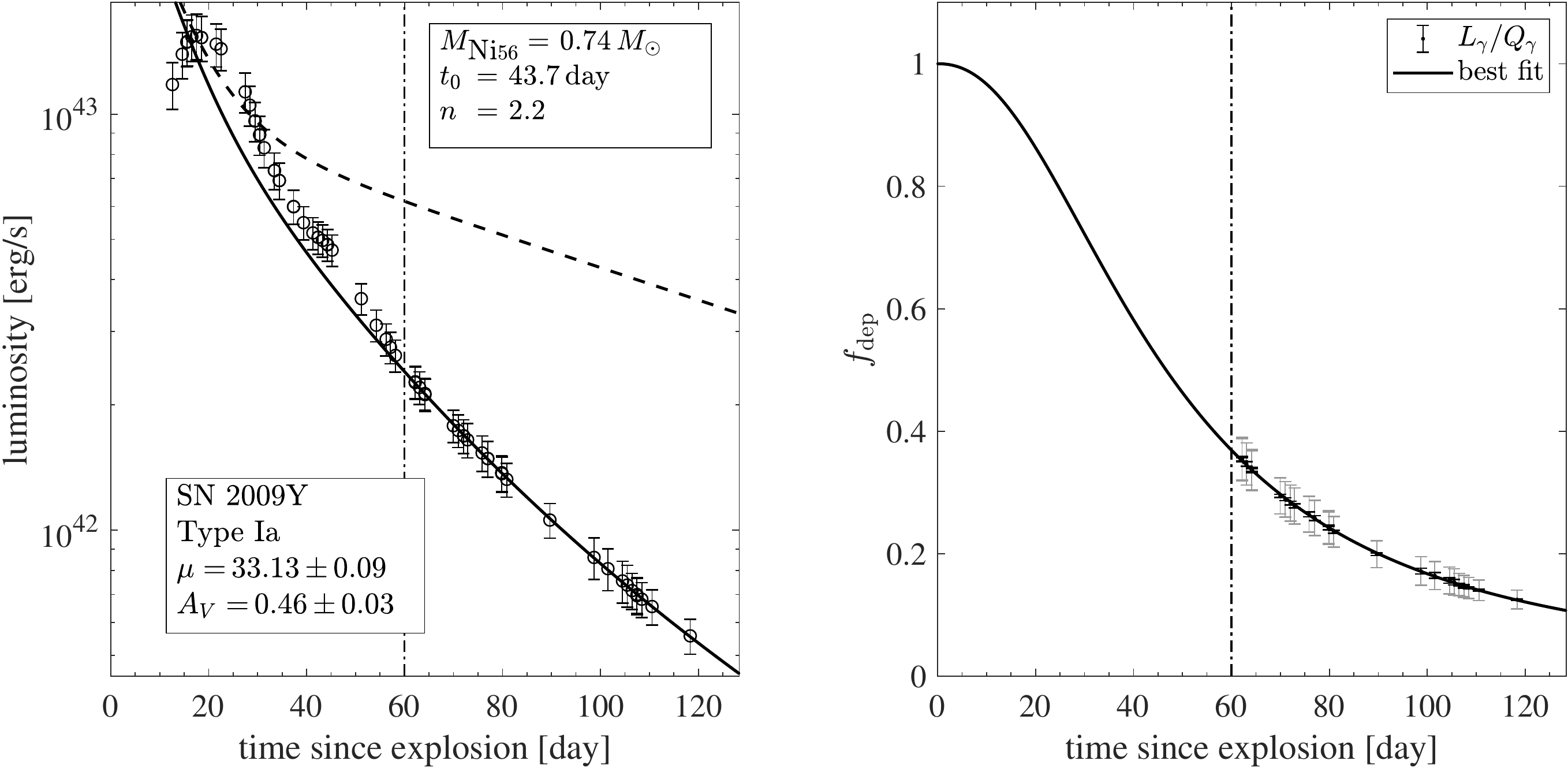}
	
	\vspace{0.25 cm}
	\includegraphics[width=0.67\textwidth]{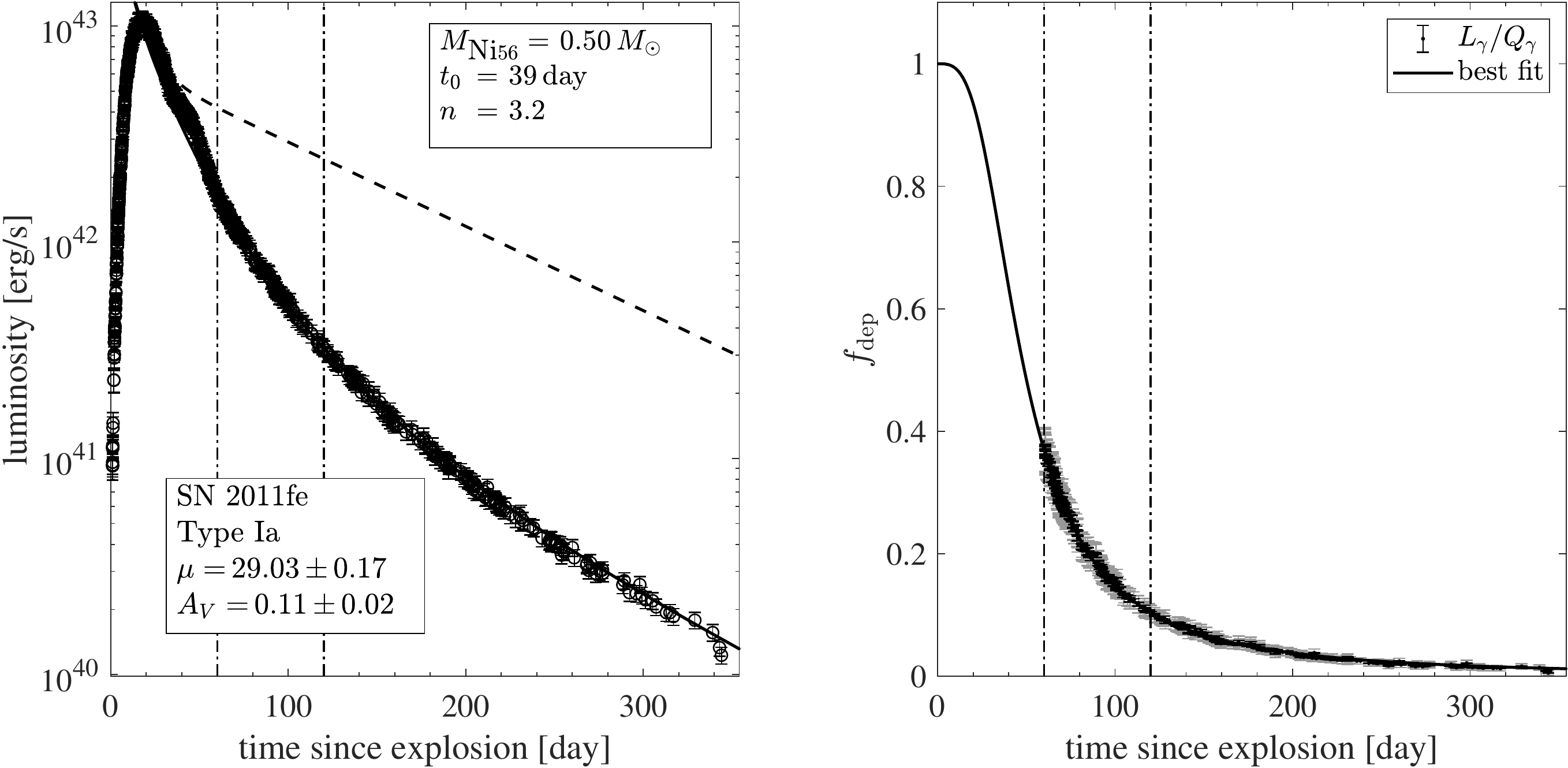}
	
	\vspace{0.25 cm}
	\includegraphics[width=0.67\textwidth]{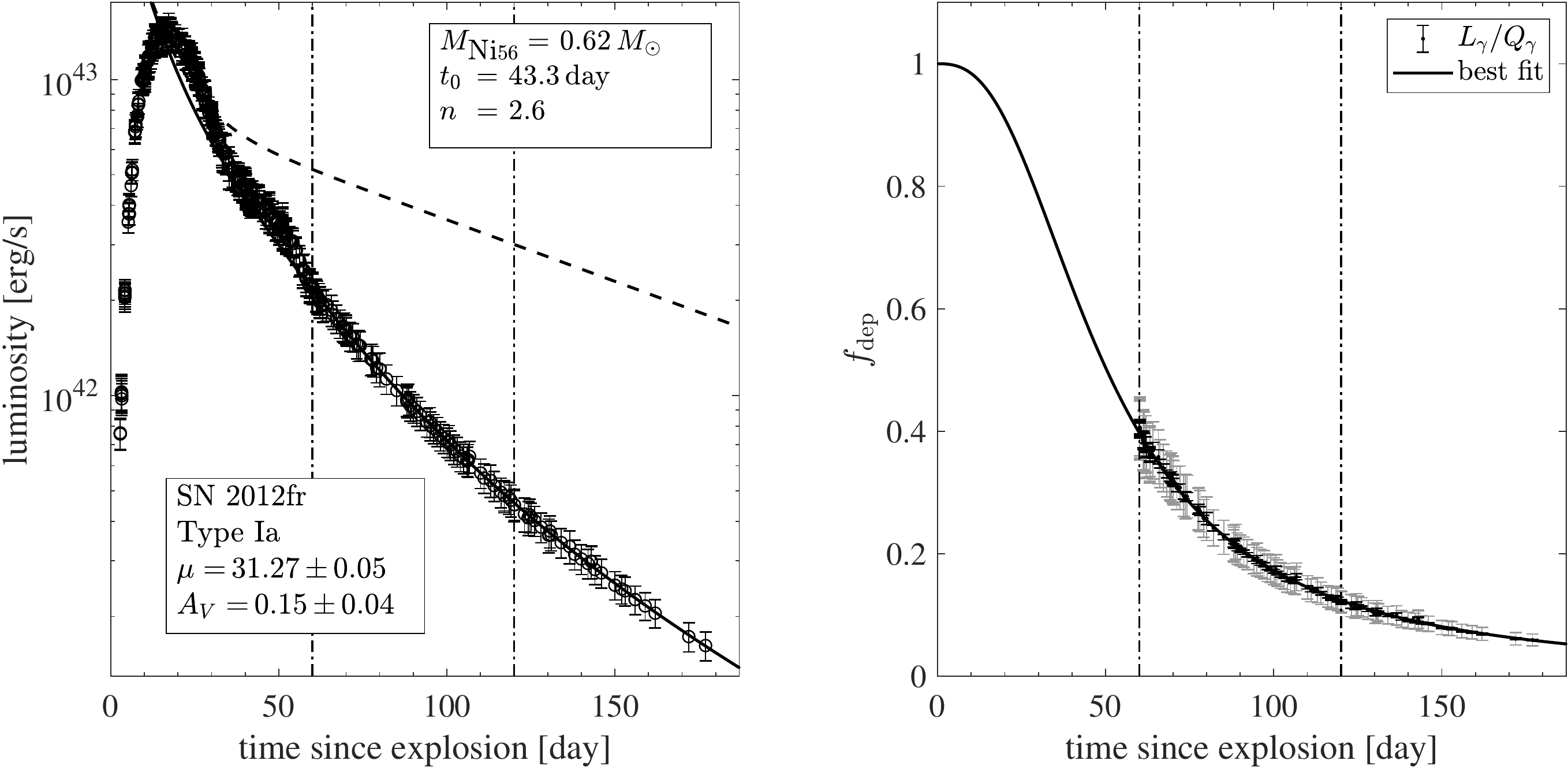}
	\caption{(continued) Same as Figure~\ref{fig:fits} for the full SNe sample.}
\end{figure*}
\begin{figure*} 
\ContinuedFloat	
	\includegraphics[width=0.67\textwidth]{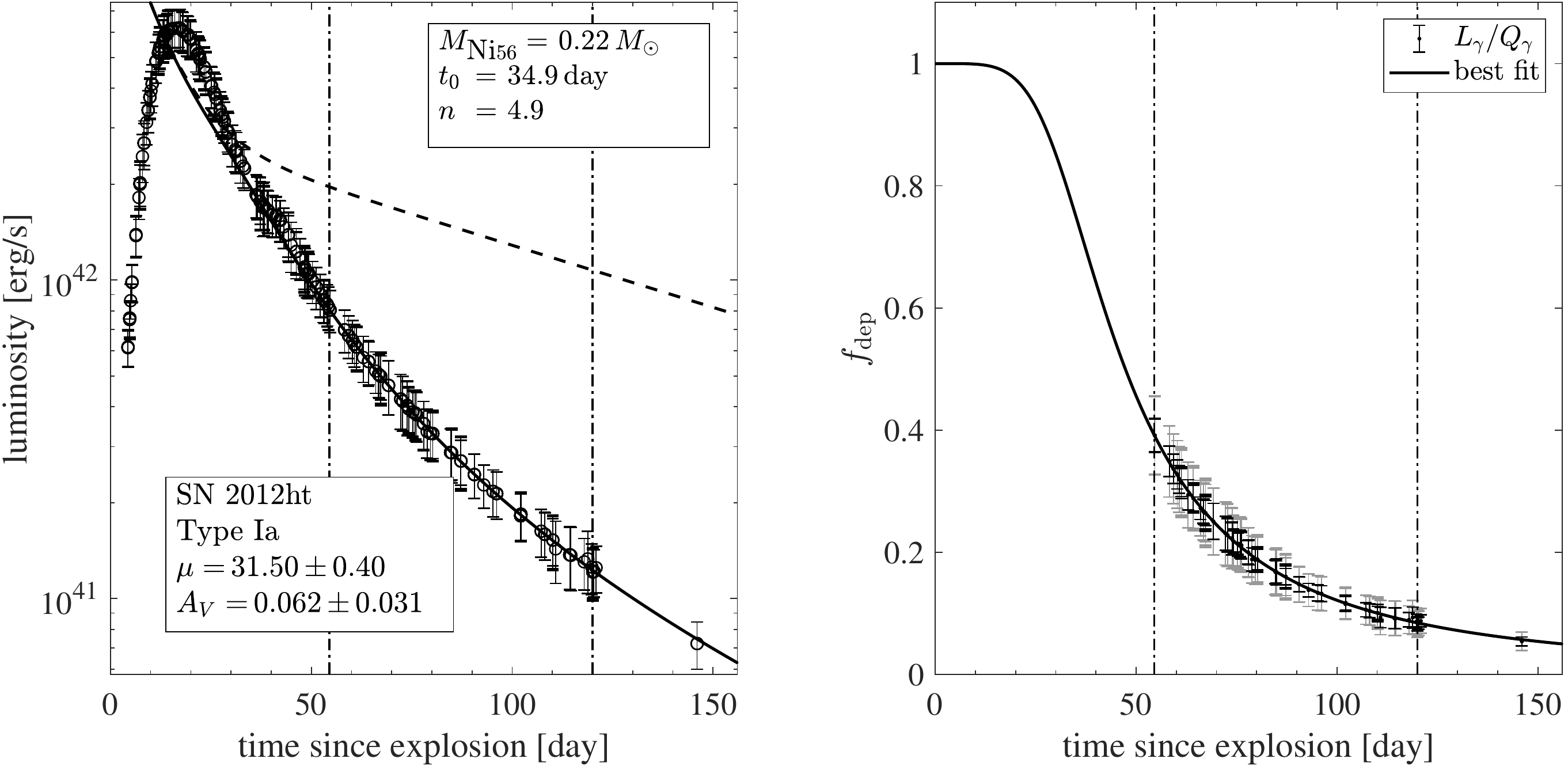}
		
	\vspace{0.25 cm}
	\includegraphics[width=0.67\textwidth]{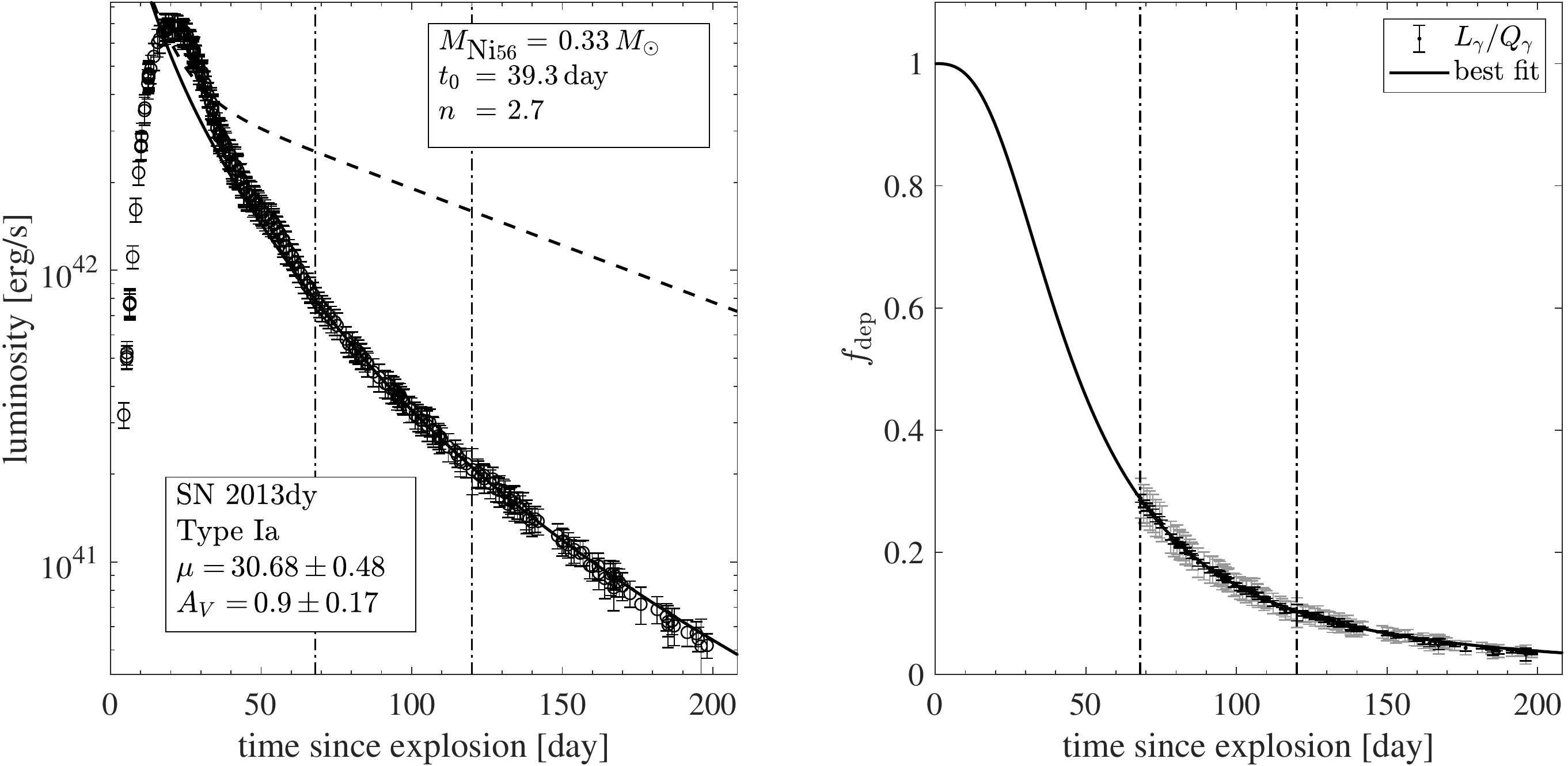}

	\vspace{0.25 cm}
	\includegraphics[width=0.67\textwidth]{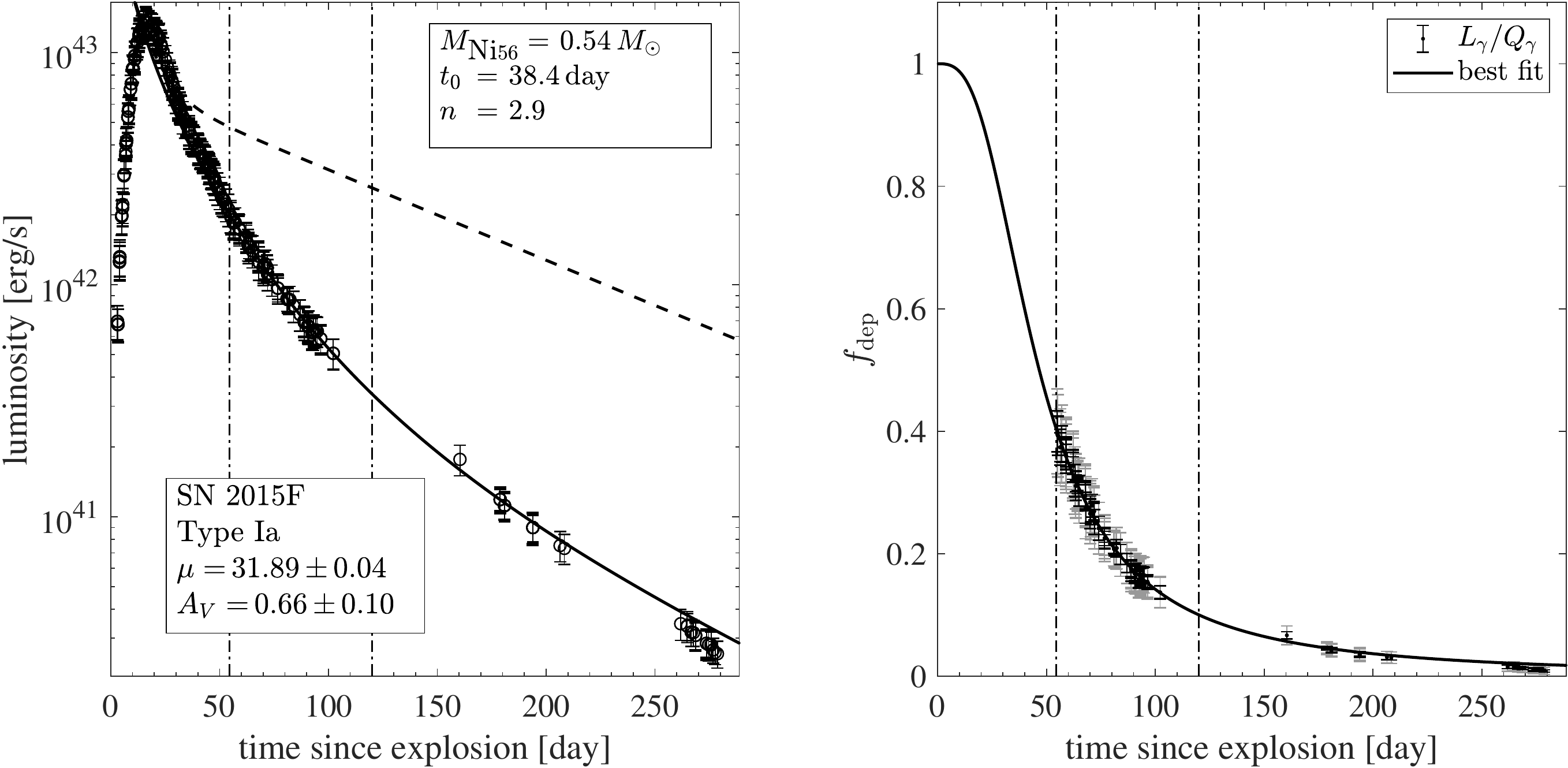}

	\caption{(continued) Same as Figure~\ref{fig:fits} for the full SNe sample.}		
\end{figure*}
\begin{figure*} 
\ContinuedFloat	
	\includegraphics[width=0.67\textwidth]{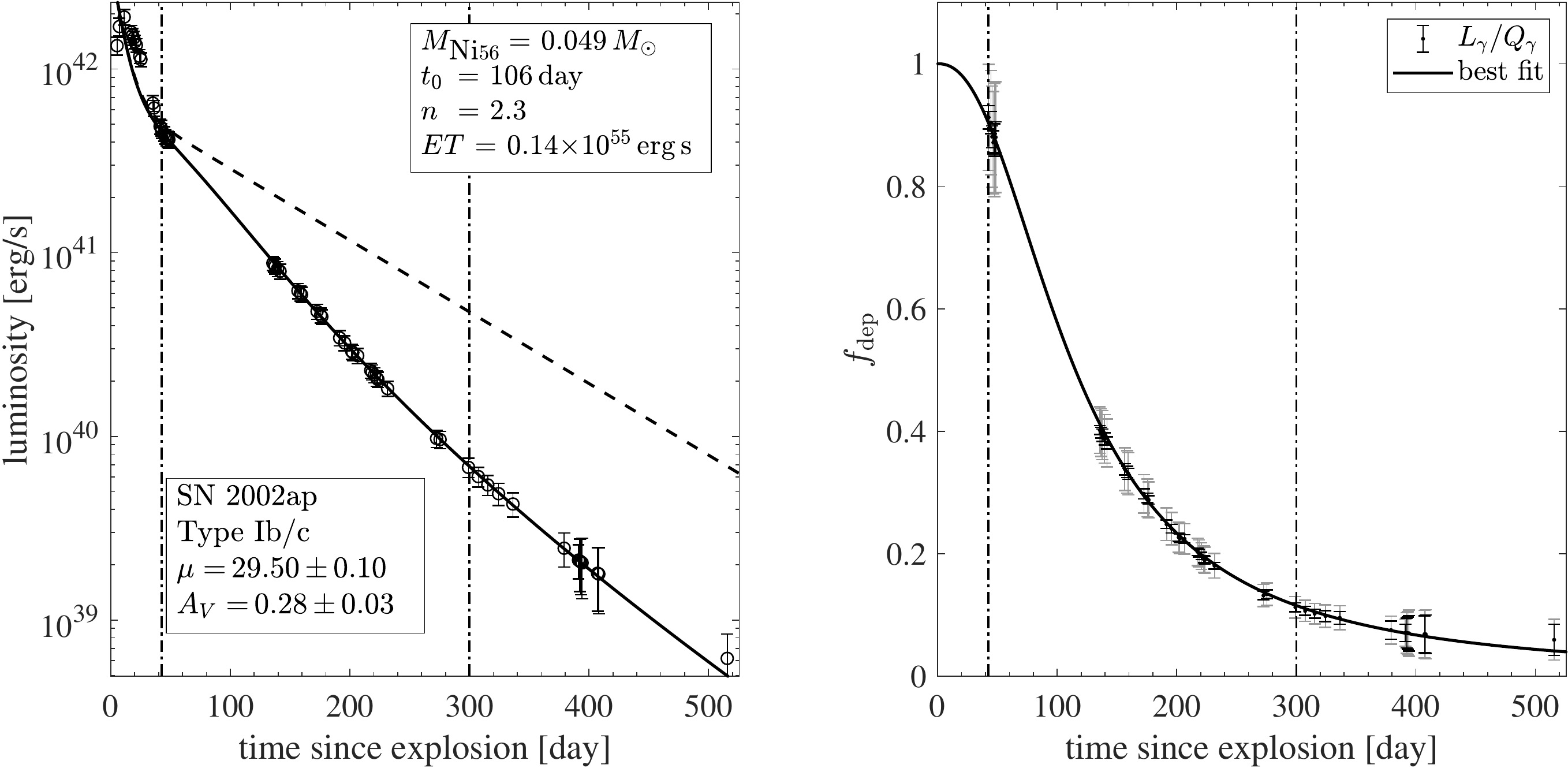}
		
	\vspace{0.25 cm}		
	\includegraphics[width=0.67\textwidth]{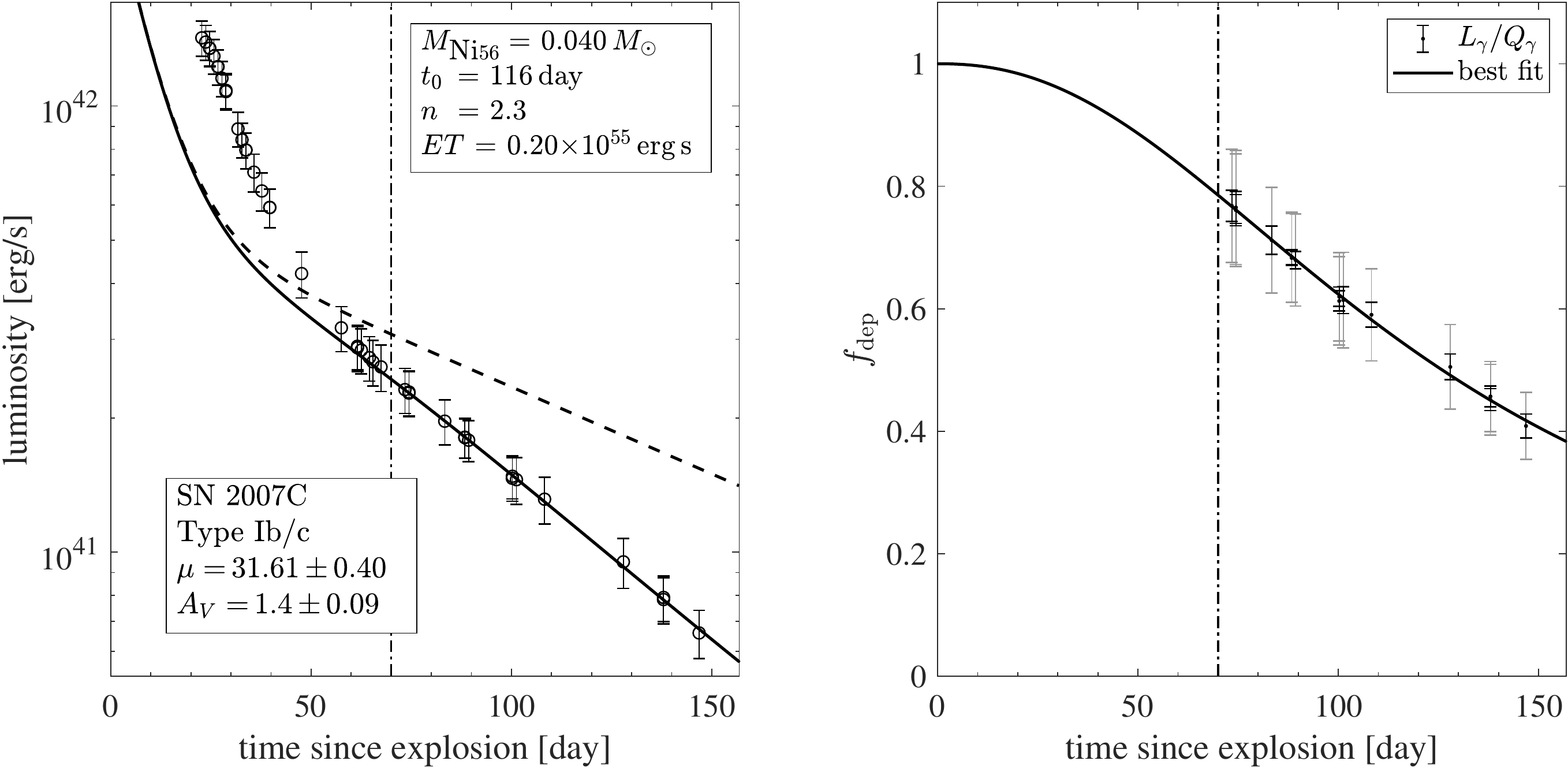}
			
	\vspace{0.25 cm}
	\includegraphics[width=0.67\textwidth]{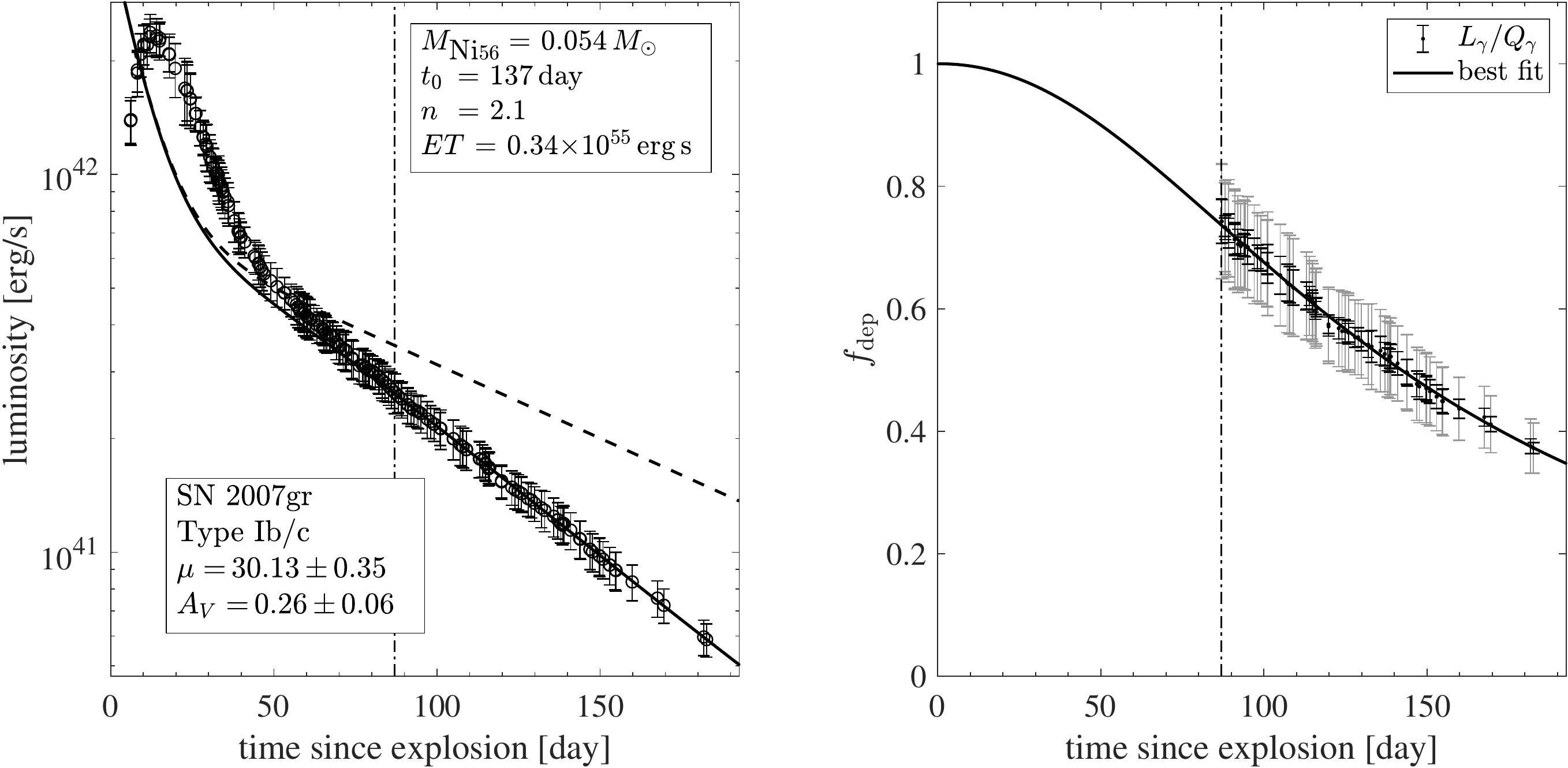}
	 			
	\vspace{0.25 cm}
	\includegraphics[width=0.67\textwidth]{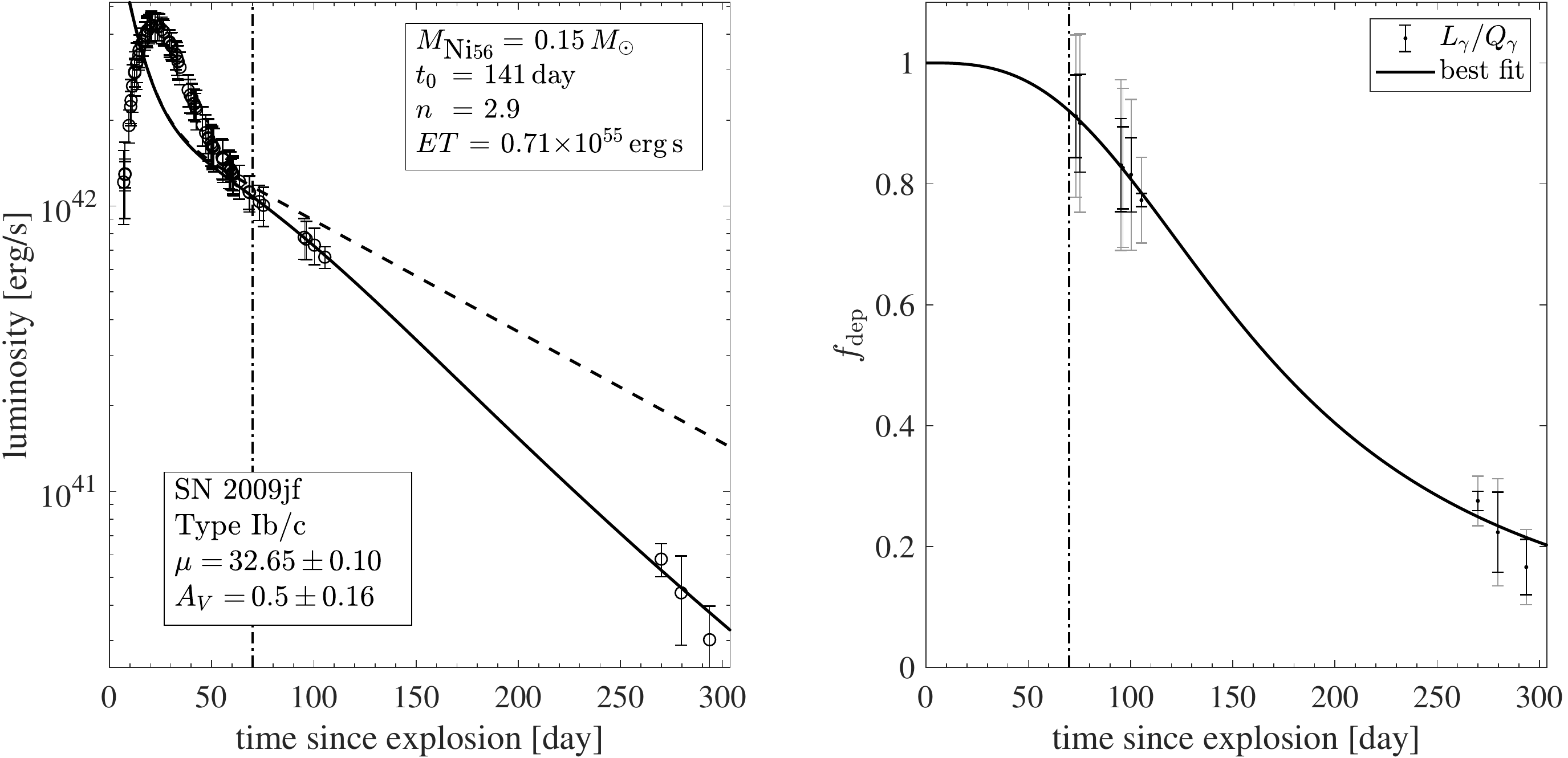}
	\caption{(continued) Same as Figure~\ref{fig:fits} for the full SNe sample.}
\end{figure*}
\begin{figure*} 
\ContinuedFloat
	\includegraphics[width=0.67\textwidth]{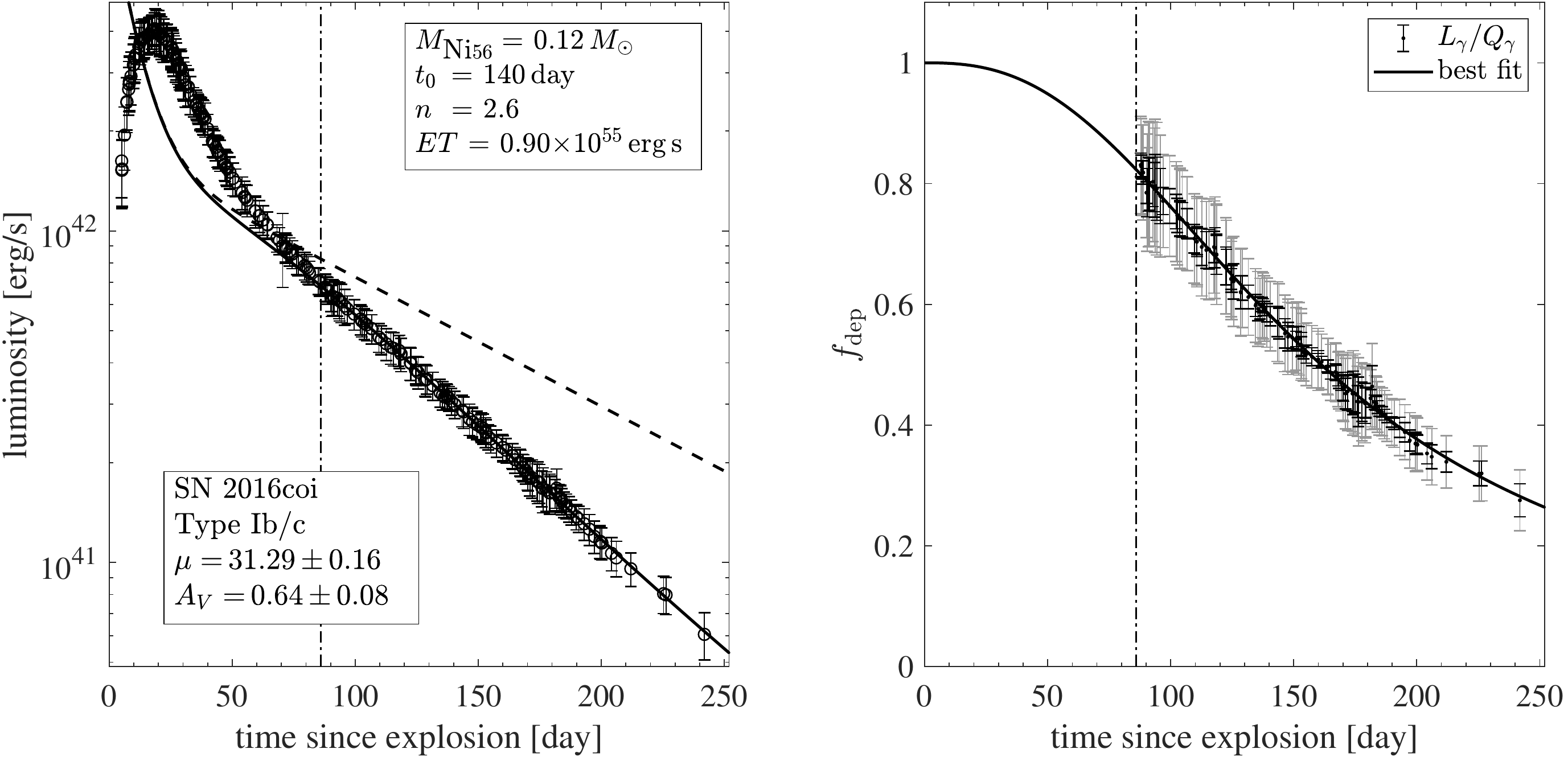}	
			\caption{(continued) Same as Figure~\ref{fig:fits} for the full SNe sample.}
\end{figure*}
\begin{figure*} 
	\ContinuedFloat	
	\includegraphics[width=0.67\textwidth]{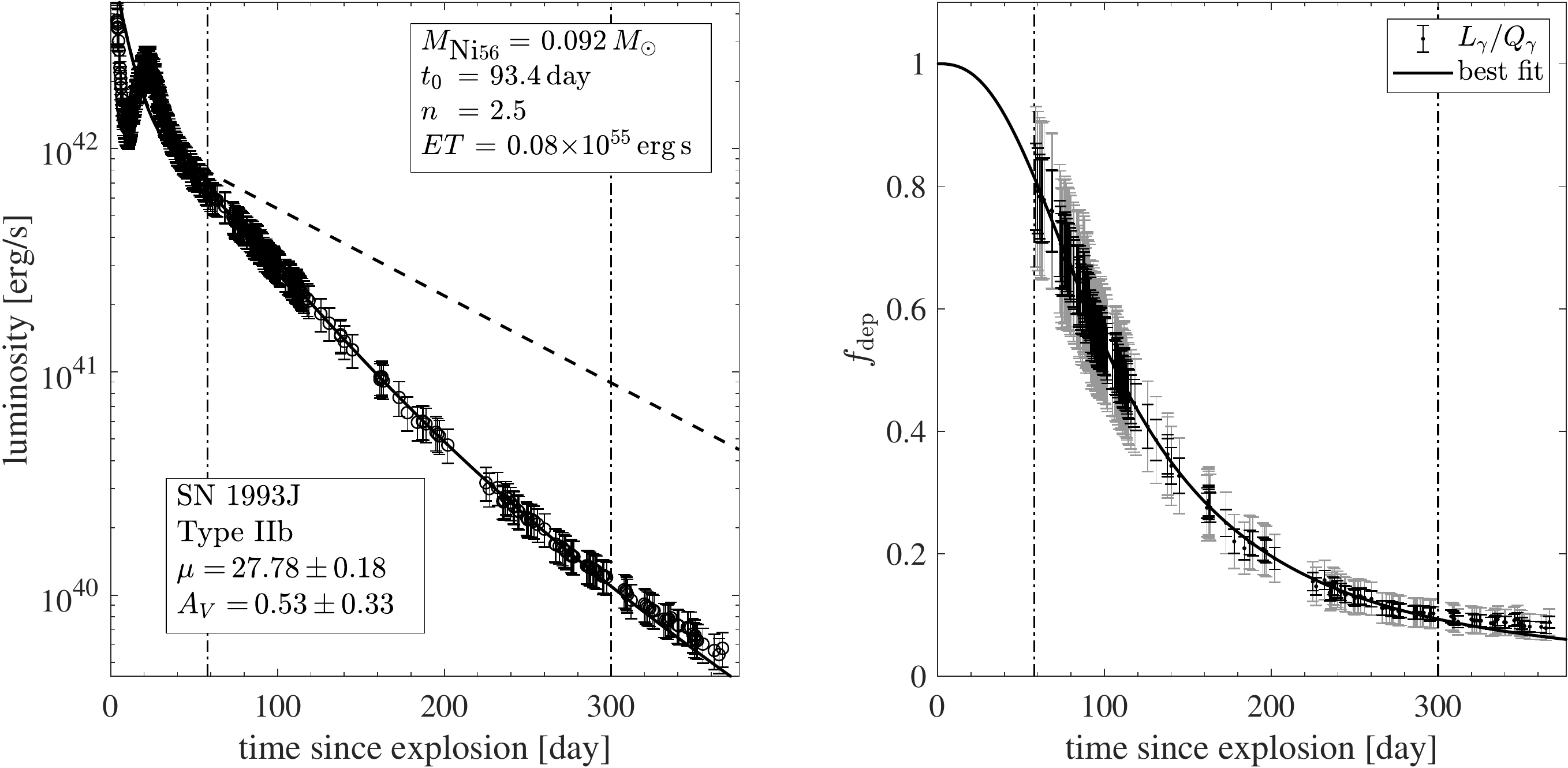}
				
	\vspace{0.25 cm}
	\includegraphics[width=0.67\textwidth]{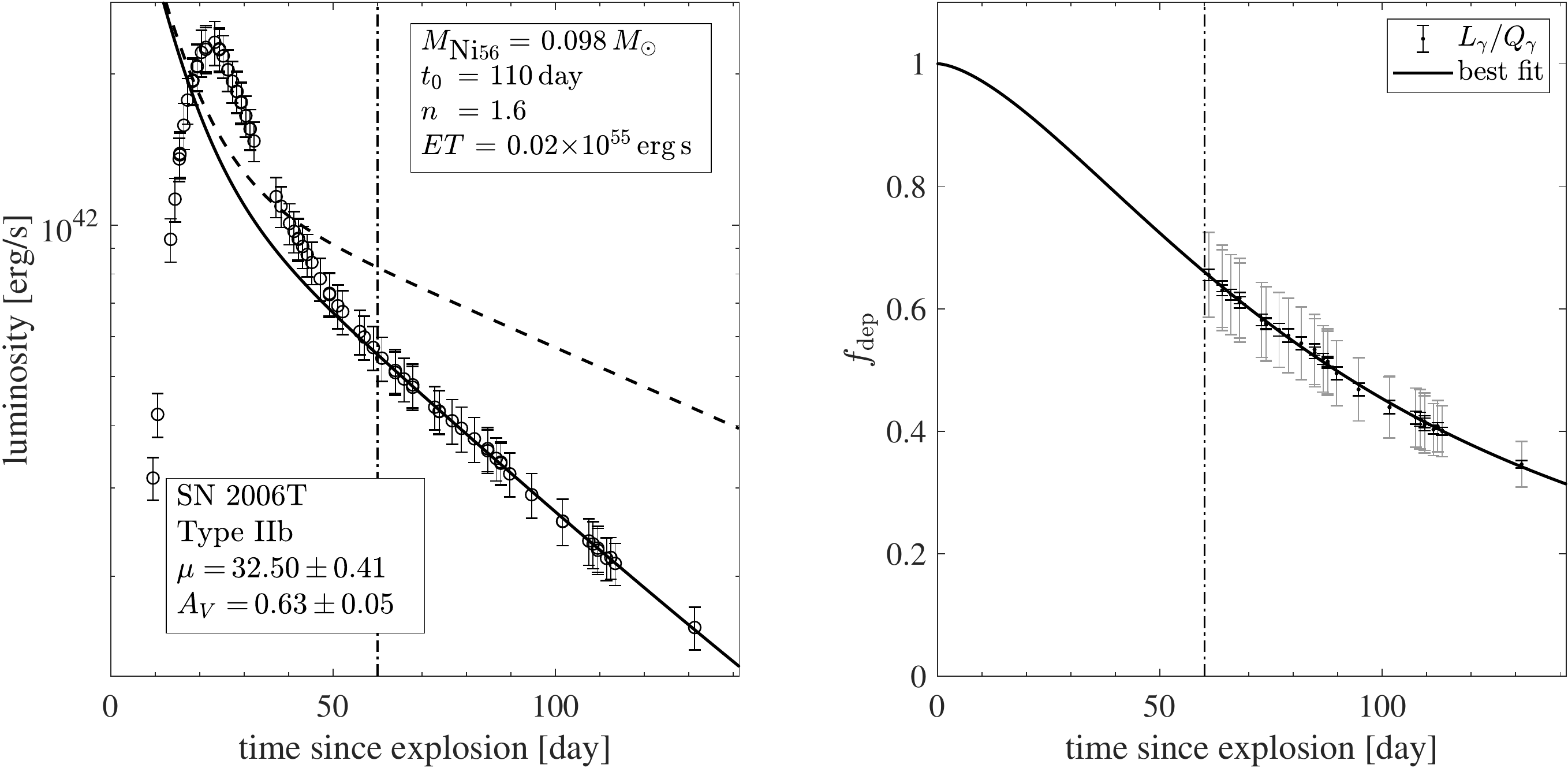}
	
	\vspace{0.25 cm}
	\includegraphics[width=0.67\textwidth]{figures/2008aq-eps-converted-to.pdf}
								
	\vspace{0.25 cm}
	\includegraphics[width=0.67\textwidth]{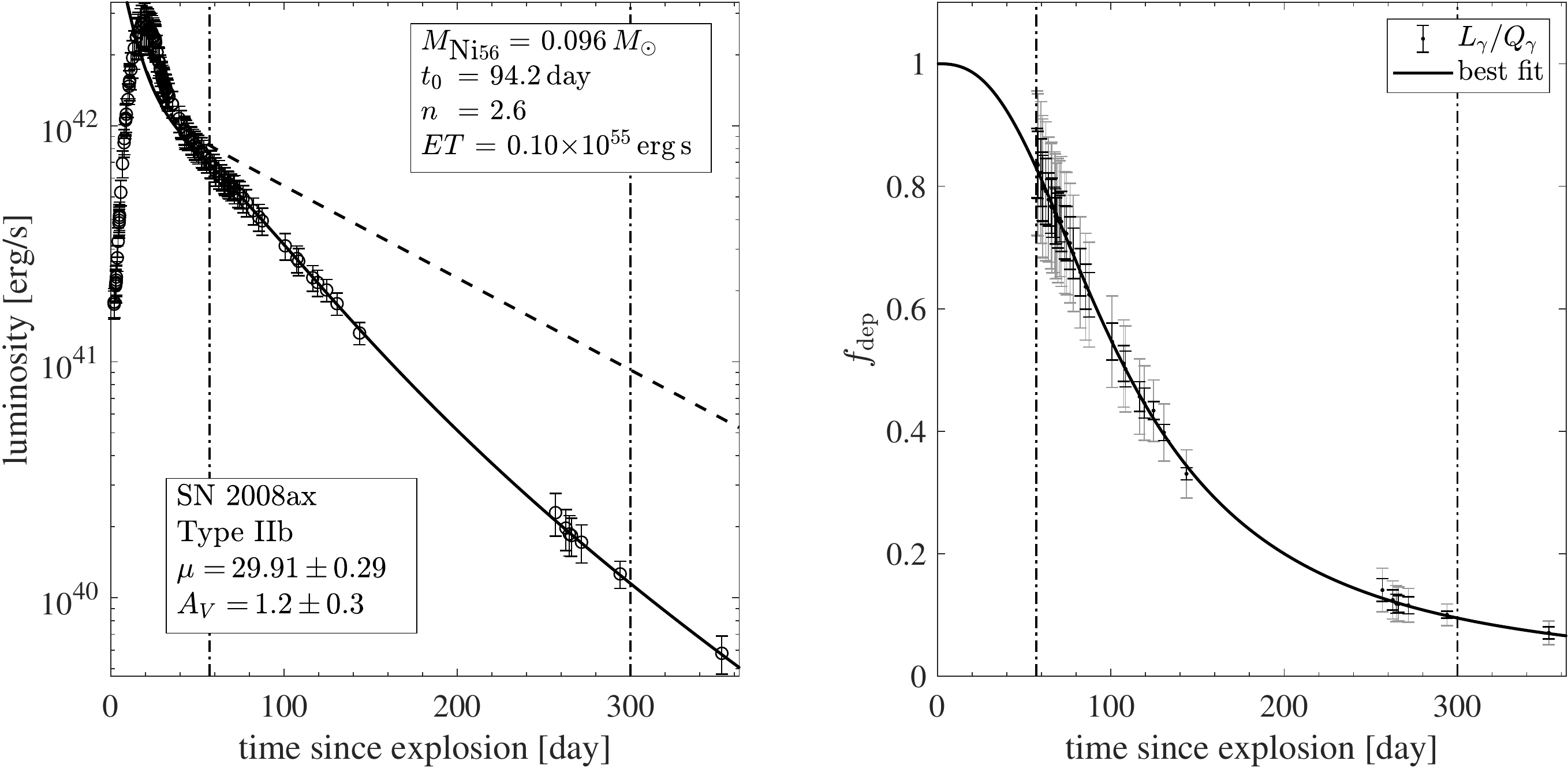}
	 \caption{(continued) Same as Figure~\ref{fig:fits} for the full SNe sample.}
\end{figure*}
\begin{figure*} 
\ContinuedFloat	
	\includegraphics[width=0.67\textwidth]{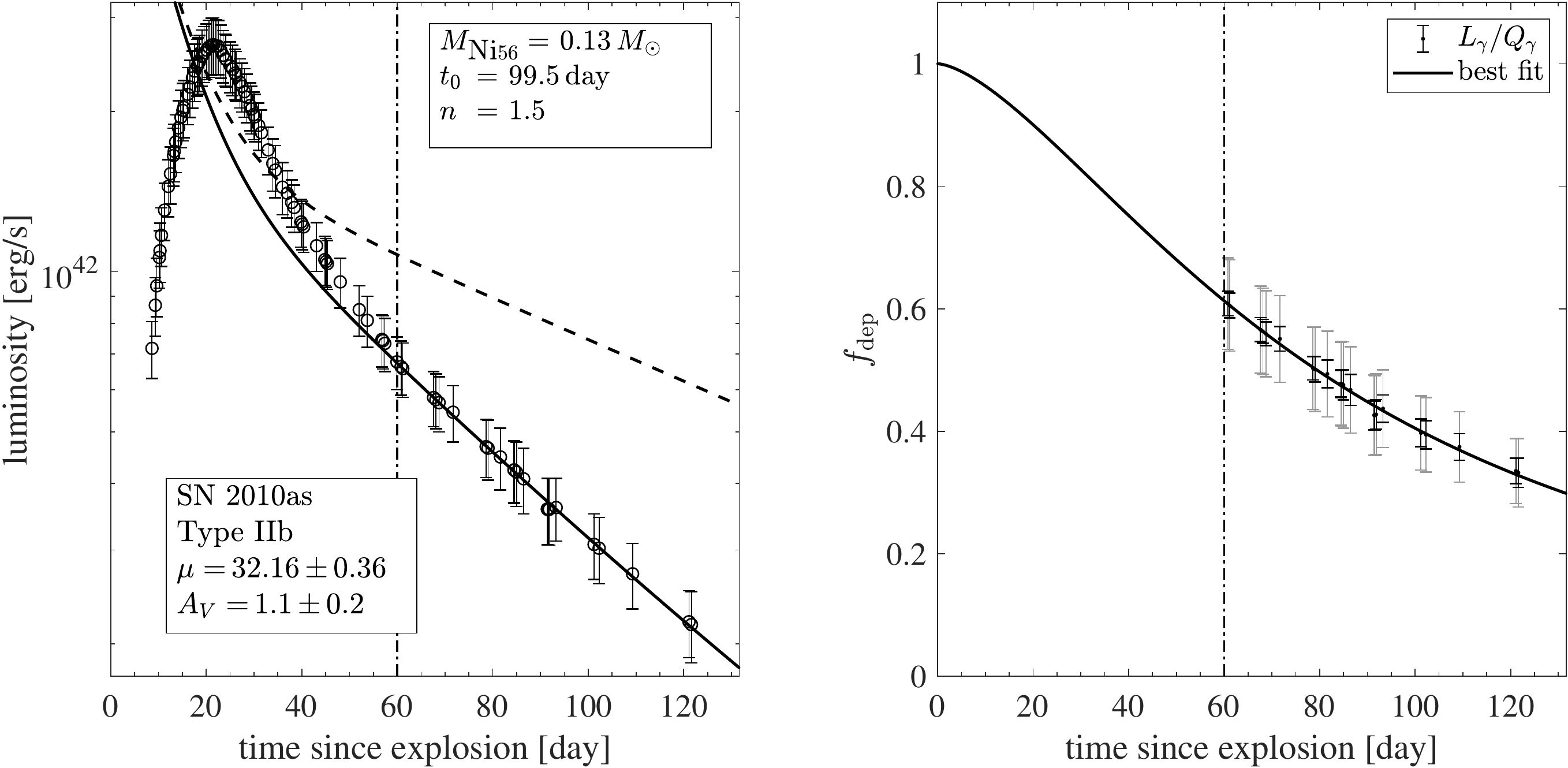}
	
	\vspace{0.25 cm}
	\includegraphics[width=0.67\textwidth]{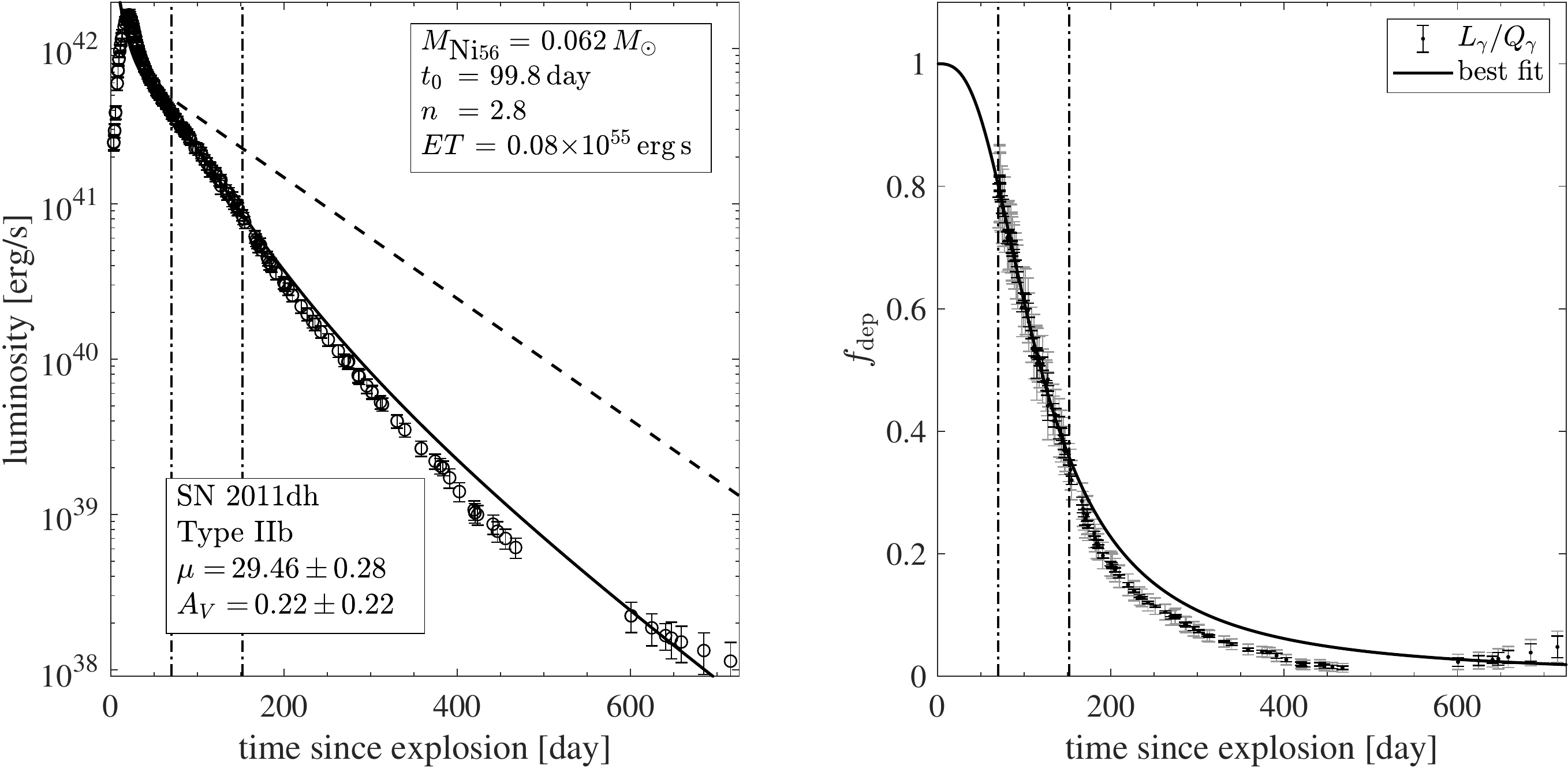}
	
	\caption{(continued) Same as Figure~\ref{fig:fits} for the full SNe sample.}
\end{figure*}
\begin{figure*} 
\ContinuedFloat	
	\includegraphics[width=0.67\textwidth]{figures/2004et-eps-converted-to.pdf}	
								
	\vspace{0.25 cm}					
	\includegraphics[width=0.67\textwidth]{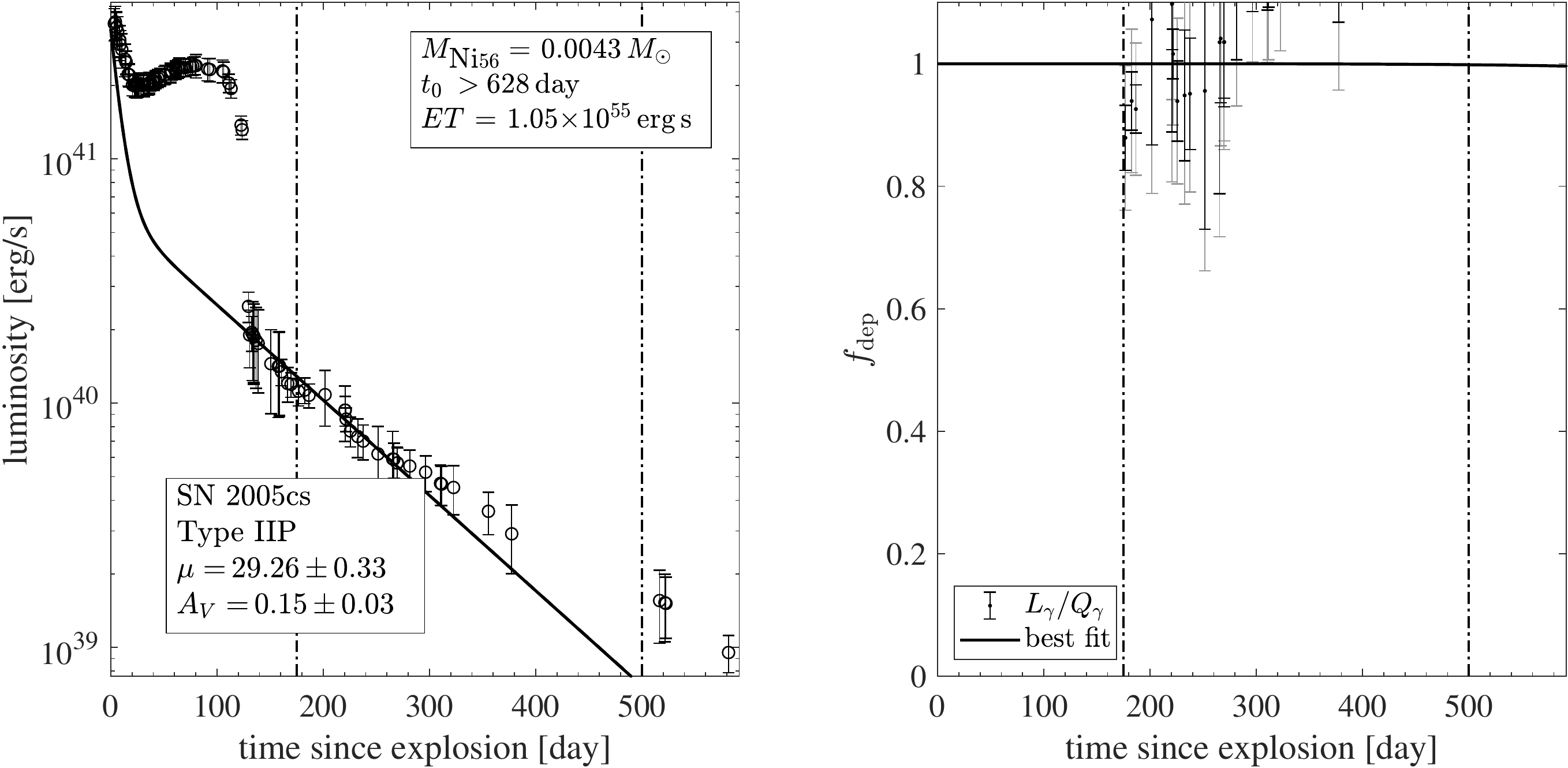}
	
	\vspace{0.25 cm}
	\includegraphics[width=0.67\textwidth]{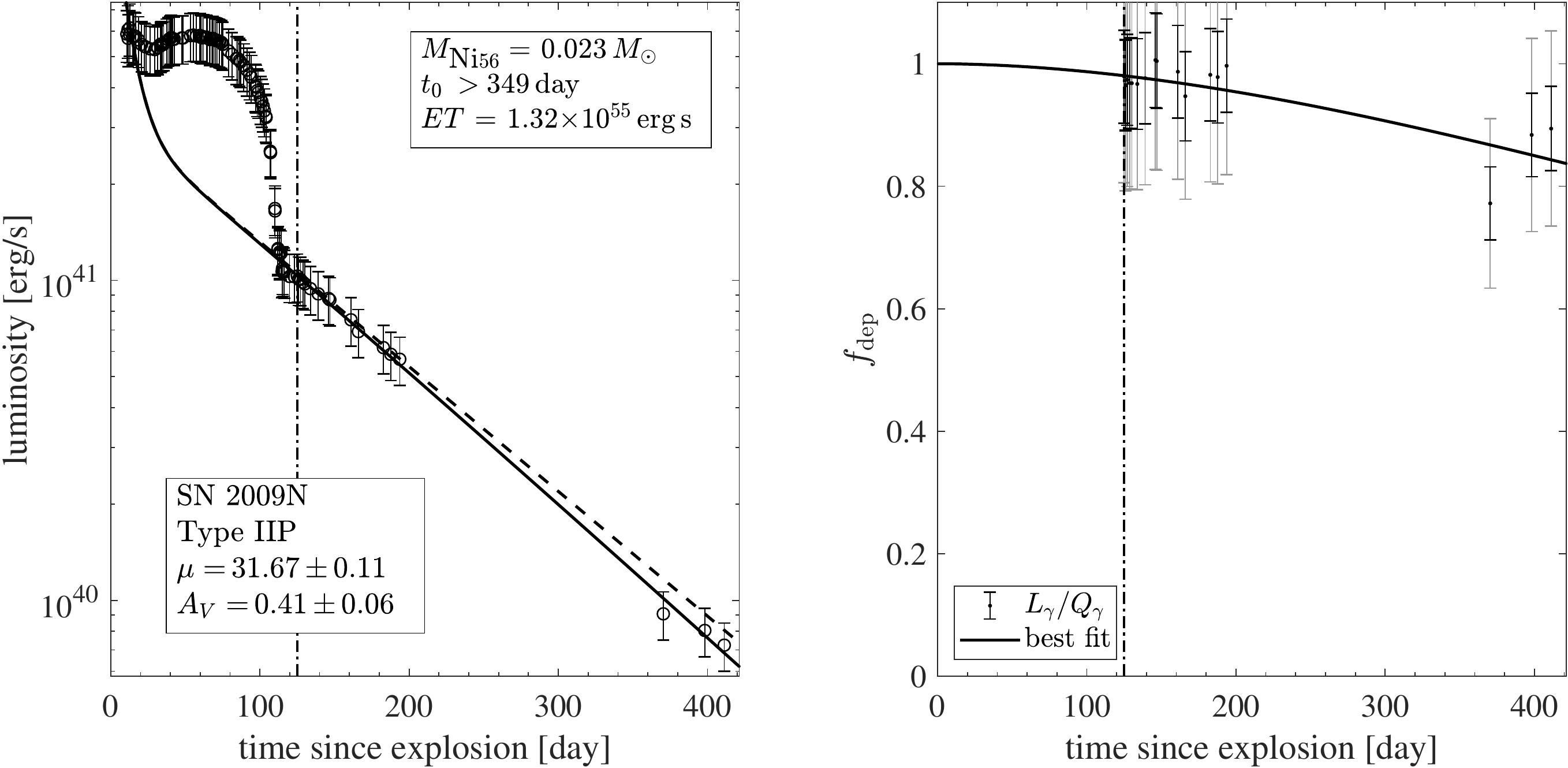}		
									
	\vspace{0.25 cm}	
	\includegraphics[width=0.67\textwidth]{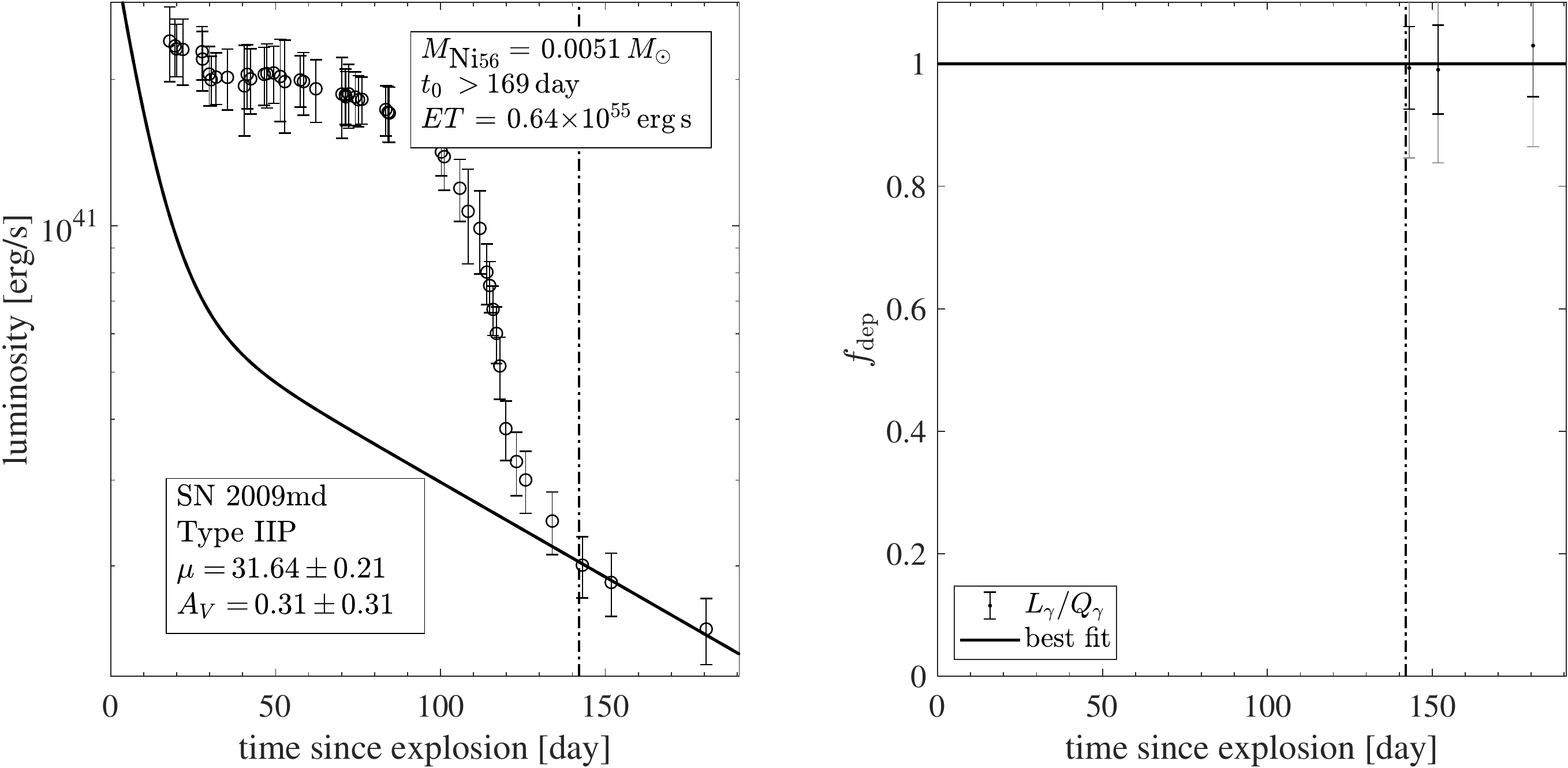}
	
	\caption{(continued) Same as Figure~\ref{fig:fits} for the full SNe sample.}
\end{figure*}
\begin{figure*} 
\ContinuedFloat
	\includegraphics[width=0.67\textwidth]{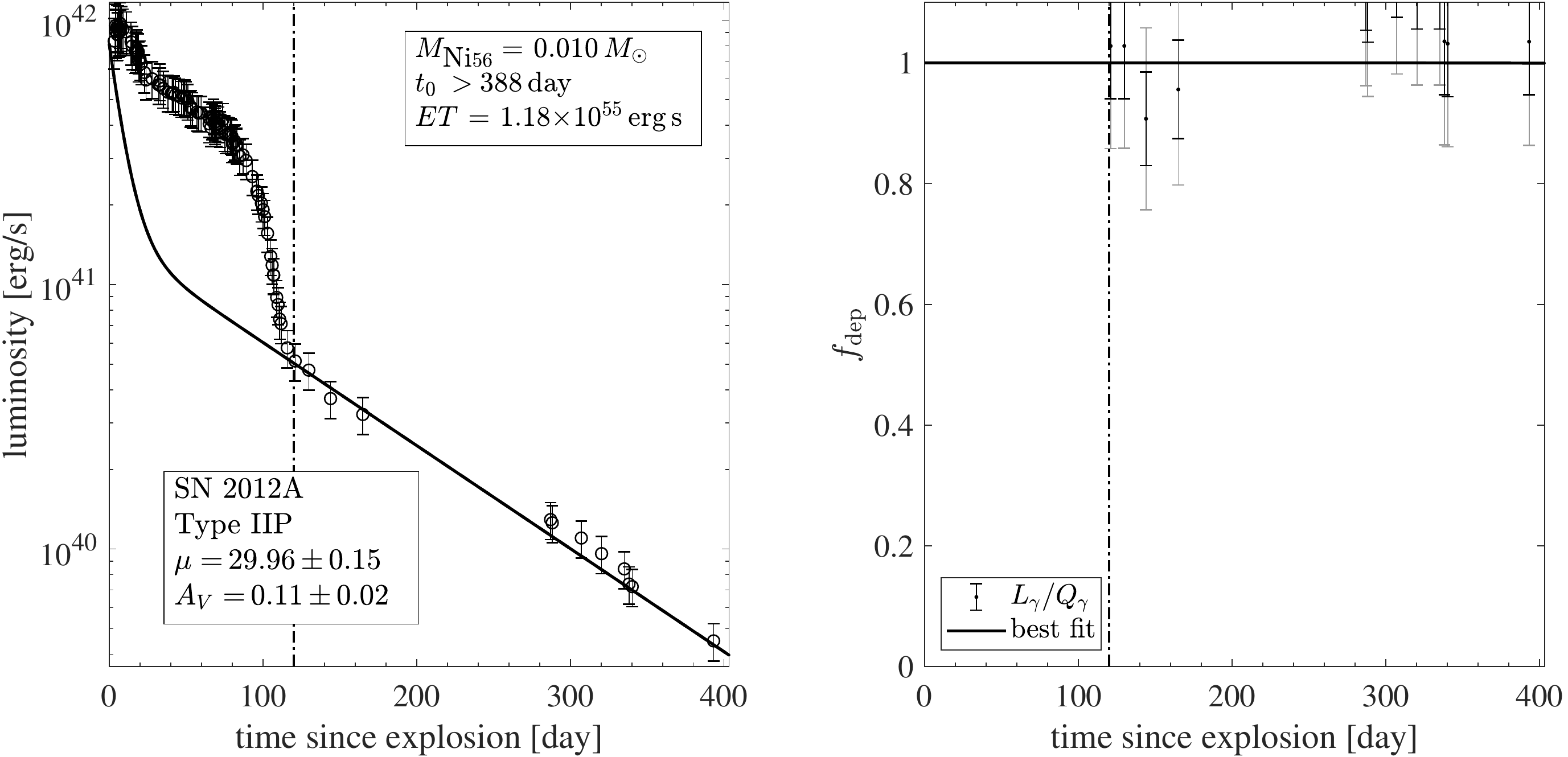}
								
	\vspace{0.25 cm}
	\includegraphics[width=0.67\textwidth]{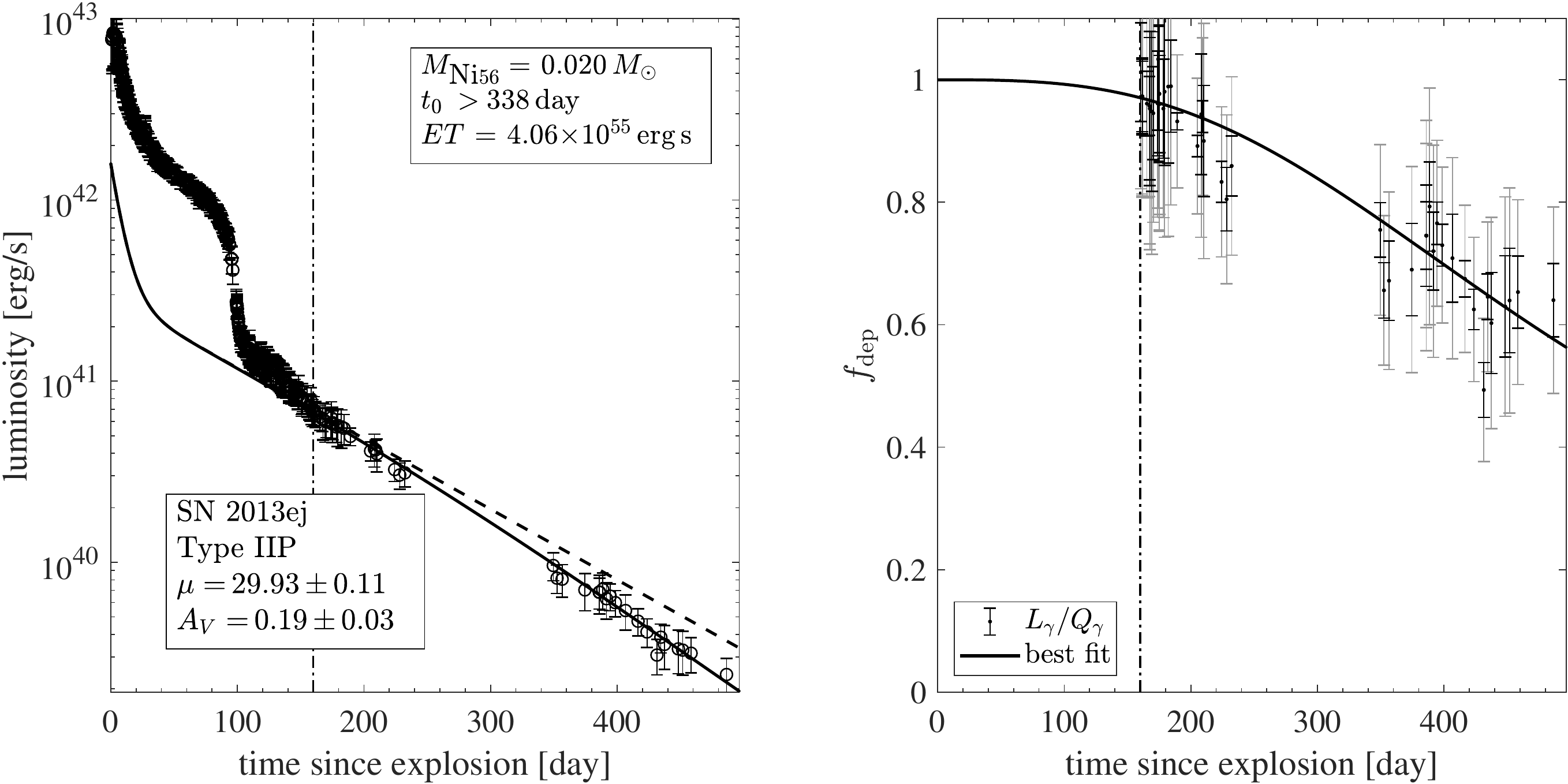}
	
	\vspace{0.25 cm}	
	\includegraphics[width=0.67\textwidth]{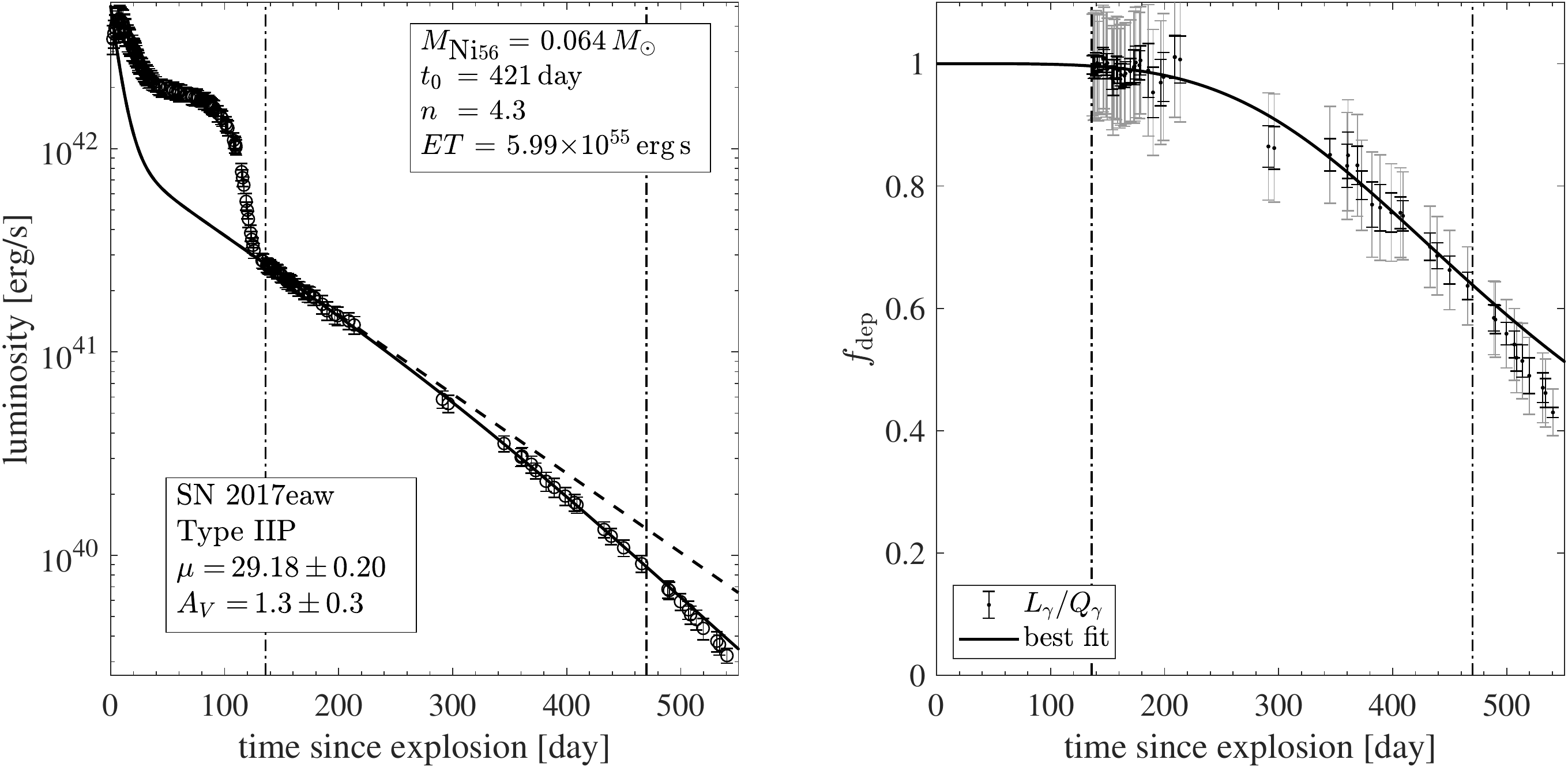}
								
	\vspace{0.25 cm}
	\includegraphics[width=0.67\textwidth]{figures/1987A-eps-converted-to.pdf}
	\caption{(continued) Same as Figure~\ref{fig:fits} for the full SNe sample.}	
\end{figure*}
\FloatBarrier

\onecolumn
\section{Model parameters and $ \gamma $-ray simulation results}
\label{tab:models}
A summary of the models that were used in our $\gamma $-ray analysis and along with the results of the calculations is provided in Table~\ref{tab:modelparam}. We provide for each model the ejecta mass, the total kinetic energy of the ejecta, the amount of $^{56}$Ni mixing as defined in Equation~\eqref{eq:mixing}, and the ratio of the mass to the square root of kinetic energy, Equation~\eqref{eq:alpha}.
\begin{ThreePartTable}
	\begin{TableNotes}
	\item [a] $ ^{56} $Ni mixing, see Equation \eqref{eq:mixing}.
	\item [b] The ratio of the ejecta mass to the square root of the kinetic energy (Equation \ref{eq:alpha}).
	\item [c] The references are: 1. \cite{Yoon2019} 2. \cite{dessart2016inferring} 3. \cite{blinnikov1998comparative} 4. \cite{sukhbold2016core} 5. \cite{blinnikov2000radiation} 6. \cite{utrobin2005supernova} 7. \cite{dessart2019supernovae} 8. \cite{hillier2019photometric} 9. \cite{utrobin2009high}.
	\end{TableNotes}

	\begin{longtable}{lcccccccc}
	Name & Type & $ M_\text{ej} $ $(M_\odot)$ & $ E_\text{kin} $ ($ 10^{51} \,\text{erg}$) & $ ^{56} $Ni mixing\tnote{a}&  $\alpha$\tnote{b}& simulation $ t_0 $ & simulation $ n $ & ref.\tnote{c}\\ \midrule
	\endfirsthead
	Name & Type & $ M_\text{ej} $ $[M_\odot]$ & $ E_\text{kin} $ [$ 10^{51} \,\text{erg}$] & $ ^{56} $Ni mixing\tnote{a}&  $\alpha$\tnote{b}& simulation $ t_0 $ & simulation $ n $ & ref.\tnote{c}\\ \midrule
	\endhead
	\endfoot
	\insertTableNotes  
	\endlastfoot
		
	HE3.87\textunderscore fm0.15\textunderscore E1.0  & SE &  2.4   &  0.67  &  0.087  &  2.93  & 132 & 4.5 & 1 \\
	HE3.87\textunderscore fm0.5\textunderscore E1.0   & SE&  2.4   &  0.67  &  0.28  &  2.93  & 123 & 3.2 & 1 \\
	HE3.87\textunderscore fm5.0\textunderscore E1.0   & SE&  2.4   &  0.67  &  0.50  &  2.93  & 114 & 2.1 & 1 \\
	HE3.87\textunderscore fm0.15\textunderscore E1.5  & SE&  2.4   &  1.17   &  0.087  &  2.22  & 97 & 4.5 & 1 \\
	HE3.87\textunderscore fm0.5\textunderscore E1.5   & SE&  2.4   &  1.17   &  0.28  &  2.22  & 93 & 3.1 & 1 \\
	HE3.87\textunderscore fm5.0\textunderscore E1.5   & SE&  2.4   &  1.17   &  0.50  &  2.22  & 84 & 2.1 & 1 \\
	HE3.87\textunderscore fm0.15\textunderscore E1.8  & SE&  2.4   &  1.47   &  0.087  &  1.98  & 87 & 4.6 & 1 \\
	HE3.87\textunderscore fm0.5\textunderscore E1.8   & SE&  2.4   &  1.47   &  0.28  &  1.98  & 81 & 3.3 & 1 \\
	HE3.87\textunderscore fm5.0\textunderscore E1.8   & SE&  2.4   &  1.47   &  0.50  &  1.98  & 73 & 2.1 & 1 \\
	CO3.93\textunderscore fm0.15\textunderscore E1.0  & SE&  2.49  &  0.53  &  0.087  &  3.42  & 187 & 3.5 & 1 \\
	CO3.93\textunderscore fm0.5\textunderscore E1.0   & SE&  2.49  &  0.53  &  0.28  &  3.42  & 152 & 2.4 & 1 \\
	CO3.93\textunderscore fm5.0\textunderscore E1.0   & SE&  2.49  &  0.53  &  0.50  &  3.42  & 133 & 1.6 & 1 \\
	CO3.93\textunderscore fm0.15\textunderscore E1.55  & SE&  2.49  &  0.99  &  0.087  &  2.5   & 112 & 3.8 & 1 \\
	CO3.93\textunderscore fm0.5\textunderscore E1.5   & SE&  2.49  &  0.99  &  0.28  &  2.5   & 96 & 2.8 & 1 \\
	CO3.93\textunderscore fm5.0\textunderscore E1.5   & SE&  2.49  &  0.99  &  0.50  &  2.5   & 83 & 1.9 & 1 \\
	CO3.93\textunderscore fm0.15\textunderscore E1.8  & SE&  2.49  &  1.27   &  0.087  &  2.21  & 90 & 4.3 & 1 \\
	CO3.93\textunderscore fm0.5\textunderscore E1.8   & SE&  2.49  & 1.27   &  0.28  &  2.21  & 82 & 2.8 & 1 \\
	CO3.93\textunderscore fm5.0\textunderscore E1.8   & SE&  2.49  & 1.27   &  0.50  &  2.21  & 72 & 1.9 & 1 \\
	he4p96Ax1  & SE&  3.54  &  1.25  &  0.14  & 3.16  & 133 & 3.3 & 2 \\
	he4p96Ax2  & SE&  3.62  &  1.29  &  0.3  &  3.18  & 122 & 2.6 & 2 \\
	he4p96Bx1  & SE&  3.61  &  2.49  &  0.088  &  2.29  & 88 & 3.6 & 2 \\
	he4p96Bx2  & SE&  3.63  &  2.49  &  0.22  &  2.3  & 84 & 3.1 & 2 \\
	he5p1Gx2  & SE&  3.65  &  5.4  &  0.13  &  1.57  & 63 & 3.2 & 2 \\
	he6p5Ax1  & SE&  4.98  &  1.26  &  0.2  &  4.43  & 182 & 3.1 & 2 \\
	he6p5Ax2  & SE&  4.97  &  1.26  &  0.35  &  4.43  & 168 & 2.5 & 2 \\
	he6p5Bx1  & SE&  4.95  &  2.43  &  0.12  &  3.18  & 119 & 3.7 & 2 \\
	he6p5Bx2  & SE&  4.98  &  2.43  &  0.27  &  3.19  & 114 & 3 & 2 \\
	he6p5Gx1  & SE&  5.14  &  5.29  &  0.087  &  2.24  & 79 & 3.7 & 2 \\
	he6p5Gx2  & SE&  5.18  &  5.3  &  0.2  &  2.25  & 77 & 3.3 & 2 \\
	3p0Ax1  & SE&  1.73  &  1.25  &  0.11  &  1.55  & 70 & 3.1 & 2 \\
	3p0Ax2  & SE&  1.72  &  1.24  &  0.19  &  1.54  & 66 & 2.8 & 2 \\
	3p0Bx2  & SE&  1.73  &  2.5  &  0.14  &  1.09  & 47 & 3.1 & 2 \\
	3p0Cx1  & SE&  1.71  &  0.62  &  0.15  &  2.17  & 100 & 2.8 & 2 \\
	3p0Cx2  & SE&  1.71  &  0.62  &  0.26  &  2.16  & 92 & 2.4 & 2 \\
	3p65Ax1  & SE&  2.23  &  1.24  &  0.14  &  2  & 84 & 3.1 & 2 \\
	3p65Ax2  & SE&  2.22  &  1.22  &  0.25  &  2.01  & 79 & 2.9 & 2 \\
	1993J\textunderscore 13C & SE & 2.26 & 1.32 & 0.12 & 1.97 & 57 & 2.5 & 3 \\
	s16.2 & II-pec & 11.9 & 0.79 & 0.072 & 13.3 & 683 & 3.7 & 4 \\
	s18.2 & II-pec& 13.2 & 1.25 & 0.065 & 11.8 & 608 & 3.9 & 4 \\
	s19.2 & II-pec& 13.8 & 0.77 & 0.062 & 15.7 & 765 & 3.7 & 4 \\
	s20.2 & II-pec& 14.2 & 1.75 & 0.06 & 10.8 & 488 & 3.9 & 4 \\
	1987A\textunderscore 14E1 & II-pec& 14.7 & 1.06 & 0.26 & 14.2 & 583 & 2.9 & 5 \\
	1987A\textunderscore M18 & II-pec& 18.1 & 1.50 & 0.21 & 14.8 & 645 & 2.5 & 6 \\
	a2 & Type II-pec & 14.3 & 0.47 & 0.3 & 20.9 & 1059 & 2.3 & 7 \\
	a2vth & Type II-pec & 14.3 & 0.47 & 0.3 & 20.9 & 1057 & 2.3 & 7 \\
	a3 & Type II-pec & 13.5 & 0.87 & 0.29 & 14.5 & 705 & 2.4 & 7 \\
	a3m1 & Type II-pec & 12.5 & 0.87 & 0.17 & 13.5 & 652 & 3 & 7 \\
	a3m2 & Type II-pec & 12.5 & 0.87 & 0.096 & 13.5 & 680 & 3.2 & 7 \\
	a3ni & Type II-pec & 13.4 & 0.87 & 0.31 & 14.4 & 715 & 2.3 & 7 \\
	a4 & Type II-pec & 13.2 & 1.26 & 0.29 & 11.8 & 573 & 2.4 & 7 \\
	a4he & Type II-pec & 13.2 & 1.25 & 0.3 & 11.8 & 564 & 2.2 & 7 \\
	a4ni3 & Type II-pec & 12.6 & 1.26 & 0.31 & 11.2 & 514 & 2.5 & 7 \\
	a5 & Type II-pec & 13.1 & 2.46 & 0.3 & 8.36 & 410 & 2.3 & 7 \\
	15M\textunderscore  mdot1p5 & II & 12.1 & 1.23 & 0.12 & 10.9 & 533 & 3.4 & 8 \\
	15M\textunderscore  mdot3p0\textunderscore e1pm0  & II & 13.2 & 1.19 & 0.49 & 12.1 & 386 & 2.3 & 8 \\
	15M\textunderscore  mdot3p0\textunderscore e1pm1  & II & 11.1 & 1.28 & 0.48 & 9.84 & 340 & 2.2 & 8 \\
	15M\textunderscore  mdot3p0\textunderscore e3pm1  & II & 11.3 & 1.30 & 0.48 & 9.91 & 330 & 2.3 & 8 \\
	15M\textunderscore  mdot3p0\textunderscore e5pm1  & II & 11.6 & 1.36 & 0.48 & 9.96 & 327 & 2.3 & 8 \\
	15M\textunderscore  mdot3p0\textunderscore e5pm2  & II & 11.1 & 1.28 & 0.48 & 9.82 & 335 & 2.2 & 8 \\
	15M\textunderscore  mdot3p0\textunderscore e7pm1  & II & 12.1 & 1.28 & 0.48 & 10.7 & 343 & 2.4 & 8 \\
	2004et & II & 22.9 & 2.23 & 0.017 & 15.3 & 638 & 4 & 9 \\
	\end{longtable}
\label{tab:modelparam}
\end{ThreePartTable}


\bsp	
\label{lastpage}
\end{document}